\documentclass[aps,prc,nofootinbib,twoside,twocolumn]{revtex4-2}
\usepackage{mathpazo}
\usepackage[T1]{fontenc}
\usepackage[latin9]{inputenc}
\usepackage[a4paper]{geometry}
\usepackage{color}
\usepackage{verbatim}
\usepackage{amsmath}
\usepackage{amssymb}
\usepackage{graphicx}

\makeatletter

\def\@oddhead{\rightmark \hfill Core-corona procedure and microcanonical hadronization\hfill \thepage}
\def\@evenhead{\thepage \hfill Klaus WERNER\hfill}
\def\fnum@table{\tablename~{\bf\thetable}}
\def\fnum@figure{\figurename~{\bf\thefigure}}
\def\tablename{\footnotesize{\bf Table}}
\def\figurename{\footnotesize{\bf Figure}}

\usepackage{dcolumn}

\def\citet{\cite}

\AtBeginDocument{
  
}

\usepackage{babel}

\makeatother

\usepackage{babel}
\begin{document}

\title{Core-corona procedure and microcanonical hadronization to understand
strangeness enhancement in proton-proton and heavy ion collisions
in the EPOS4 framework}

\author{K. Werner}

\affiliation{SUBATECH, Nantes University - IN2P3/CNRS - IMT Atlantique, Nantes, France }

\begin{abstract}
The multiplicity dependence of multistrange hadron yields in proton-proton
and lead-lead collisions at several TeV allows one to study the transition
from very big to very small systems, in particular, concerning collective
effects. I investigate this, employing a core-corona approach based
on new microcanonical hadronization procedures in the EPOS4 framework,
as well as new methods allowing one to transform energy-momentum flow
through freeze-out surfaces into invariant-mass elements. I try to
disentangle effects due to ``canonical suppression'' and ``core-corona
separation'', which will both lead to a reduction of the yields at
low multiplicity.\\

\end{abstract}

\maketitle


\section{Introduction}


An enhancement of multistrange hadrons in relativistic heavy ion
collisions compared with proton-proton scattering is one of the oldest
signals proposed to detect the creation of a ``quark-gluon plasma''
\cite{Koch:1986}. It was observed first (as expected) in heavy
ion collisions \cite{WA97:1999,NA49:2002,NA57:2004,STAR:2008,ALICE:2013-PbPb-Xi-Oga}
but, unexpectedly, such ``signals'' have also been detected in proton-protons
collisions, where the enhancement concerns high-multiplicity compared
with low-multiplicity events \cite{ALICE:2016-pp-Ks-Lda-Xi-Oga}. Even
more, when one plots ratios of multistrange hadrons to pions, as
a function of the multiplicity $dn/d\eta$ at rapidity zero, for both
proton-proton (at 7 TeV) and heavy ions (PbPb at 2.76 TeV), one observes
a unique curve, increasing monotonically from low-multiplicity proton-proton
up to central PbPb. Also shown in Ref. \cite{ALICE:2016-pp-Ks-Lda-Xi-Oga}:
the standard Monte Carlo event generators do not really describe the
data.

During the past five years, the EPOS4 project was developed (see first
publications \cite{werner:2023-epos4-overview,werner:2023-epos4-heavy,werner:2023-epos4-smatrix}),
being an attempt to construct a realistic model for describing relativistic
collisions of different systems (from proton-proton to nucleus-nucleus)
at energies from several GeV per nucleon up to several TeV. So the
model should be able to deal with strangeness enhancement, but not
only. It is a ``general purpose'' approach that is meant to describe
any observable. The results shown in this paper are only a very small
fraction of simulation results, since as a first test of the new approach
simulations were performed at low energies, i.e., 7.7, 11.5, 14.5, 19.6,
27, 39, 62.4, 130, and 200 GeV / nucleon, and as well at high energies,
i.e., 2.76, 5.02, 7, 8, 13 TeV / nucleon, for different systems,
looking for yields, spectra, and flow, for light and heavy flavor.
All this is done with the same code version, namely EPOS4.0.0, which
is at this moment being published via a dedicated web page. No attempt
has been made to ``fit'' one particular curve for one particular
system, the idea is more to see to what extent one understands the enormous
amount of data accumulated over the past two decades.

A fundamental ingredient of the EPOS4 approach is the observation
that multiple (nucleonic or partonic) scatterings must happen in parallel,
and not sequentially, based on very elementary considerations concerning
time-scales. To take this into account, EPOS4 brings together ancient
knowledge about S-matrix theory (to deal with parallel scatterings)
and modern concepts of perturbative QCD and saturation, going much
beyond the usual factorization approach. The parallel scattering principle
requires sophisticated Monte Carlo techniques, inspired by those used
in statistical physics to investigate the Ising model.

In the EPOS4 approach, one distinguishes ``primary scatterings''
and ``secondary scatterings''. The former refer
to the above-mentioned parallel scatterings with the initial nucleons
(and their partonic constituents) being involved, happening at very
high energies instantaneously. The theoretical tool is S-matrix theory,
using a particular form of the proton-proton scattering S-matrix (
Gribov-Regge approach \cite{Gribov:1967vfb,Gribov:1968jf,GribovLipatov:1972,Abramovsky:1973fm}),
which can be straighforwardly generalized to be used for nucleus-nucleus
collisions. Although these basic assumptions are not only well-motivated,
but also very simple and transparent, the practical application is
complicated. This is due to the fact that when developing matrix elements
in terms of multiple scattering diagrams, the large majority of the
diagrams cancel when it comes to inclusive cross sections. But in
EPOS4 one keeps all contributions since one wants to go beyond inclusive
cross sections (important when studying high multiplicity events).
The challenge for EPOS4 is to use the full parallel scattering scenario,
but in a smart way such that for inclusive cross section the cancellations
actually work. This is the new part in EPOS4, strongly based on an
interplay between parallel scatterings and saturation. This is discussed
in detail in separate publications \cite{werner:2023-epos4-overview,werner:2023-epos4-heavy,werner:2023-epos4-smatrix}.
Our S-matrix approach has a modular structure, it is based on so-called
``Pomerons'' (representing elementary parton-parton scatterings),
and these Pomerons are constructed based on ``parton ladders'',
which makes the link to perturbative QCD. Also, this part has been
completely redone, with particular care concerning the role of heavy
flavor \cite{werner:2023-epos4-heavy}.

In this paper, I want to focus on the secondary scatterings, and
in particular on the hadronization of ``plasma droplets''. In Sec .
\ref{=======parallel-scattering-scenario=00003D00003D=======},
I summarize briefly the ``parallel scattering approach'' for primary
scatterings. In Sec . \ref{=======the-role-of-core=======},
I discuss how the EPOS4 multiple-scattering diagrams translate
into particle production, which means first the production of so-called
prehadrons,which may originate from Pomerons or from remnants. Based
one these prehadrons, a core-corona procedure will be employed, which
allows one to identify the ``core'', which will then be treated as a
fluid which evolves and eventually decays into hadrons. In Sec.
\ref{=======microcanonical-hadronization=======},
I discuss the main topic of this paper, namely the hadronization
of the core.

\section{Parallel scattering scenario in EPOS4 \label{=======parallel-scattering-scenario=00003D00003D=======}}

Let me briefly summarize the EPOS4 approach, considering first high-energy collisions, where a parallel scattering approach for
primary scatterings is mandatory, in AA collisions but also concerning
multiple partonic scatterings in pp reactions. 
In Refs. \cite{werner:2023-epos4-overview,werner:2023-epos4-heavy,werner:2023-epos4-smatrix},
it is shown in detail how such a ``parallel scattering scheme'' can
be constructed rigorously based on S-matrix theory, which will be
sketched in the following.

\begin{figure}[h]
\centering{}\includegraphics[scale=0.21]
{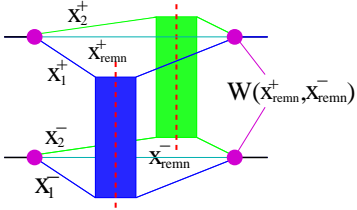}
\caption{Double scattering, each box representing a cut Pomeron $G$ (single
inelastic scattering). \label{two-boxes}}
\end{figure}

The starting point is the elastic-scattering T-matrix $T$ for pp
scattering, expressed as a product of ``elementary'' T-matrices
$T_{\mathrm{Pom}}$ representing parton-parton scattering via Pomeron
exchange. The expression can be easily generalized for AA scattering.
In both cases, pp or AA, this formalism allows a strict parallel scattering
picture. The precise content of the Pomerons will be discussed later.
The connection with inelastic scattering provides the optical theorem,
using so-called cutting rules, which allows one to express the total cross
section (which by definition adds up all inelastic processes) in terms
``cut Pomerons'' $G$, as shown for the case of two inelastic scatterings
in Fig. \ref{two-boxes}. 
The light-cone momentum fractions $x^{+}$ and $x^{-}$ are shared
between the two Pomerons and the remnants ($W$ in the plot represents
some vertex function). Each cut Pomeron $G$ represents a squared
amplitude of a single inelastic scattering.

\begin{figure}[h]
\centering{}%
\begin{minipage}[c]{0.2\columnwidth}%
\noindent \begin{center}
\includegraphics[scale=0.2]
{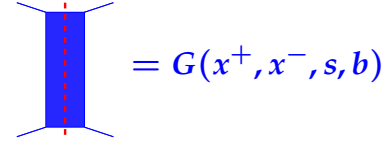} 
\par\end{center}%
\end{minipage}%
\begin{minipage}[c]{0.2\columnwidth}%
\noindent \begin{center}
\textcolor{blue}{\large{}{}{} 
\[
\boldsymbol{\boldsymbol{=G(x^{+},x^{-},s,b)}}
\]
}{\large{}{} } 
\par\end{center}%
\end{minipage}$\qquad\qquad$\caption{The cut Pomeron $G$. \label{box-equal-G}}
\end{figure}

The important new issue in Ref. \cite{werner:2023-epos4-overview,werner:2023-epos4-heavy,werner:2023-epos4-smatrix}
is the understanding of how energy conservation ruins factorization
(which strictly speaking makes the model inapplicable), and how to
solve the problem via an appropriate definition of $G$. The cut Pomeron,
representing a single inelastic scattering is the fundamental quantity
in the EPOS formalism. For the moment, one considers the Pomeron as
a ``box'' (the precise internal structure will be discussed later),
with two external legs representing two incoming partons carrying
light-cone momentum fractions $x^{+}$ and $x^{-}$, so $G=G(x^{+},x^{-},s,b)$,
with the energy squared $s$, and the impact parameter $b$, see Fig.
\ref{box-equal-G}. 

Let me define the ``Pomeron energy fraction'' $x_{\mathrm{PE}}=x^{+}x^{-}=M_{\mathrm{Pom}}^{2}/s,$
with $M_{\mathrm{Pom}}$ being the transverse mass of the Pomeron,
which is the crucial variable characterizing cut Pomerons: the bigger
$x_{\mathrm{PE}}$, the bigger the Pomeron's invariant mass and the
number of produced particles. Large invariant masses also favor high-$p_{t}$ jet production.

Let me consider a AA collision (including pp as a special case). I
define, for a given cut Pomeron connected to projectile nucleon $i$
and target nucleon $j$, a ``connection number'' $N_{\mathrm{conn}}=(N_{\mathrm{P}}+N_{\mathrm{T}})/2,$
with $N_{\mathrm{P}}$ being the number of Pomerons connected to \emph{i,
}and with $N_{\mathrm{T}}$ being the number of Pomerons connected
to \emph{j. }The case $N_{\mathrm{conn}}=1$ corresponds to an isolated
Pomeron, which may take all the energy of the initial nucleons, whereas
in case of $N_{\mathrm{conn}}>1$ the energy will be shared. To quantify
the effect of the energy sharing, one defines $f^{(N_{\mathrm{conn}})}(x_{\mathrm{PE}}$)
to be the inclusive $x_{\mathrm{PE}}$ distribution, i.e. the probability
that a single Pomeron carries an energy fraction $x_{\mathrm{PE}}$,
for Pomerons with given values of $N_{\mathrm{conn}}$. The main problem
which ruins factorization is the fact that the distribution for $N_{\mathrm{conn}}>1$
will be deformed with respect to the $N_{\mathrm{conn}}=1$ case,
due to energy sharing. I define the corresponding ``deformation
function'' $R_{\mathrm{deform}}(x_{\mathrm{PE}})$ as a ratio of
$f^{(N_{\mathrm{conn}})}(x_{\mathrm{PE}}$) over $f^{(1)}(x_{\mathrm{PE}}$).
I also use the notation $R_{\mathrm{deform}}^{(N_{\mathrm{conn}})}(x_{\mathrm{PE}})$
to undeline its $N_{\mathrm{conn}}$ dependence. As shown in Refs. \cite{werner:2023-epos4-overview,werner:2023-epos4-smatrix},
this function can be calculated and tabulated. As discussed in very
much detail in Ref. \cite{werner:2023-epos4-heavy}, one also calculates
and tabulate some function $G_{\mathrm{QCD}}(Q^{2},\,x^{+},x^{-},s,b)$,
which contains as a basic element a cut parton ladder based on 
Dokshitzer-Gribov-Lipatov-Altarelli-Parisi (DGLAP)
parton evolutions \cite{GribovLipatov:1972,AltarelliParisi:1977,Dokshitzer:1977}
from the projectile and target side, with an elementary QCD cross
section in the middle, $Q^{2}$ being the low virtuality cutoff in
the DGLAP evolution. The latter is usually taken to be constant and
of the order of $1\,$GeV, whereas here one allows any value. With all this
preparation, one is now able to connect $G$ (used in the multi-Pomeron
diagrams) and $G_{\mathrm{QCD}}$ [which contains all the 
perturbative QCD (pQCD) diagrams],
as follows: 
For each cut Pomeron, for given $x^{\pm}$, $s$, $b$, and $N_{\mathrm{conn}}$,
one postulates: 
\begin{equation}
G(\!\,x^{+}\!\!,x^{-}\!\!,s,b)\!=\frac{n}{R_{\mathrm{deform}}^{(N_{\mathrm{conn}})}(x_{\mathrm{PE}})}G_{\mathrm{QCD}}(Q_{\mathrm{sat}}^{2}\,,\,x^{+}\!\!,x^{-}\!\!,s,b),\label{fundamental-epos4-equation}
\end{equation}
with
\begin{equation}
 G \; \mathrm{independent}\; \mathrm{of} \;  N_{\mathrm{conn}}.
\end{equation}
But $Q_{\mathrm{sat}}^{2}$ does depend on $N_{\mathrm{conn}}$ (and also on $x^{\pm}$). 
The symbol $n$ is a factor not depending on $x^{\pm}$. This is the
fundamental equation in the new EPOS4 approach. As shown in Refs. \cite{werner:2023-epos4-overview,werner:2023-epos4-smatrix},
this deformation in the denominator of Eq. (\ref{fundamental-epos4-equation})
corrects perfectly the deformation of the $x_{\mathrm{PE}}$ distribution,
the only effect of energy sharing is a modification of $Q_{\mathrm{sat}}^{2}$,
and the latter will only affect low-$p_{t}$ processes. In this way,
one recovers factorization, and binary scaling in AA (with $R_{AA}=1$)
at large transverse momenta for inclusive cross sections (but only
for them). In other words, at large $p_{t}$, the full multiple-scattering
machinery produces exactly the same result as compared with simply taking
one single Pomeron (single scattering), which allows one to define
and compute (in the EPOS framework) parton distribution function,
tabulate them, and then compute inclusive jet cross sections based
on these (using standard procedures).

\begin{figure}[h]
\centering{}\includegraphics[scale=0.22]
{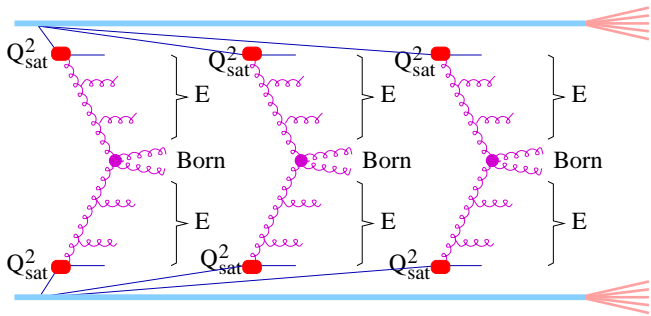}
\caption{Rigorous parallel scattering scenario, for $n=3$ parallel scatterings,
including non-linear effects via saturation scales. The red symbols
should remind one that the parts of the diagram representing nonlinear
effects are replaced by simply using saturation scales. \label{saturation-three-pom}}
\end{figure}

\begin{figure}[h]
\centering{}\includegraphics[scale=0.22]
{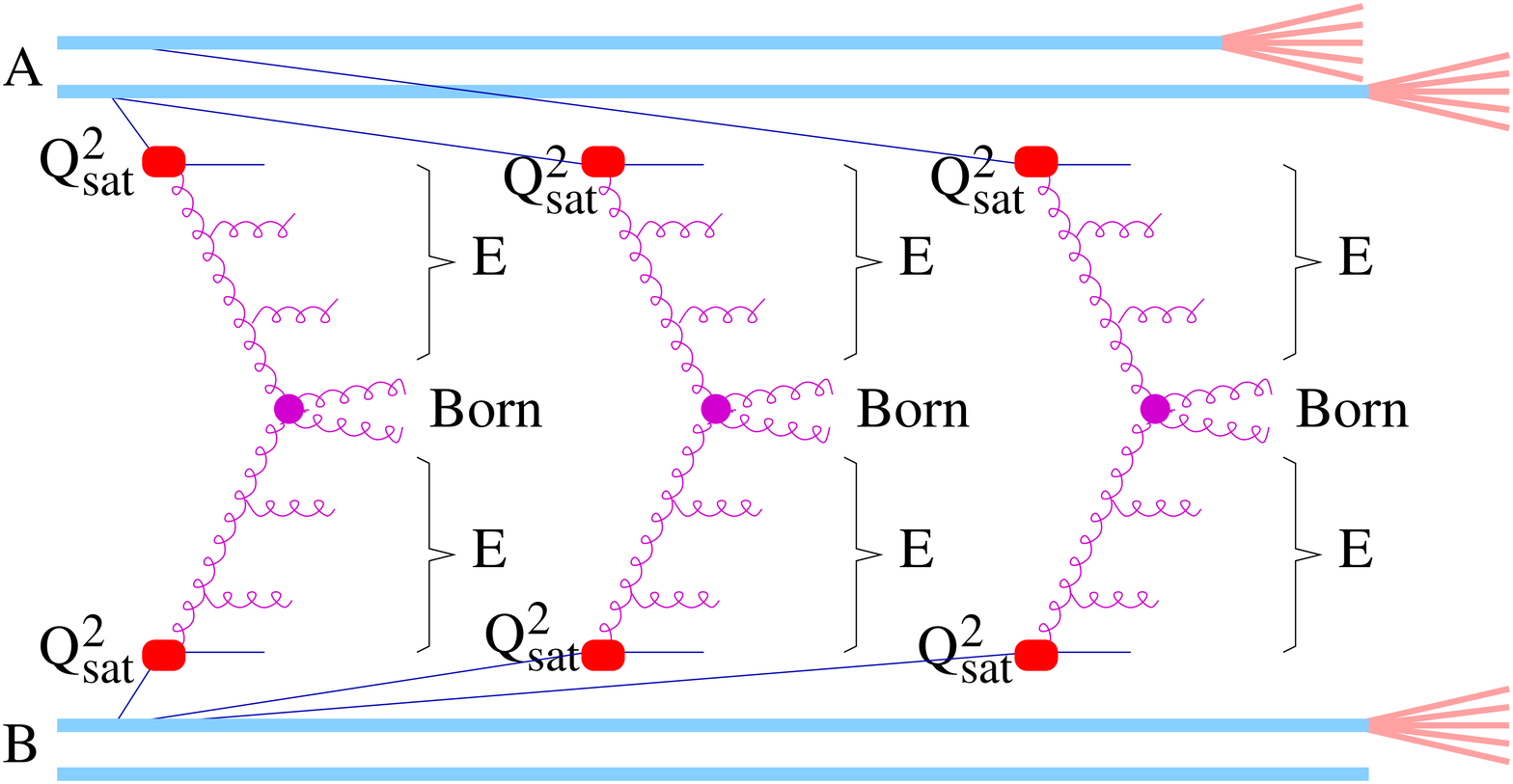}
\caption{Rigorous parallel scattering scenario, for $n=3$ parallel scatterings
for a collision of a nucleus $A$ with a nucleus $B$ , including
non-linear effects via saturation scales. \label{saturation-three-pom-1}}
\end{figure}

But such a ``shortcut'' (also referred to as ``EPOS4 factorization
mode'') is only possible for inclusive cross sections, and only at
high $p_{t}$. But this represents only a very small fraction of all
possible applications, and there are very interesting cases outside
the applicability of that approach. A prominent example, one of the
highlights of the past decade in our domain, concerns collective phenomena
in small systems. It has been shown that high-multiplicity pp events
show very similar collective features as earlier observed in heavy
ion collisions \cite{CMS:2010ifv}. High multiplicity means automatically
``multiple parton scattering''. As discussed earlier, this means
that one has to employ the full parallel scattering machinery developed
earlier, based on S-matrix theory. One cannot use the usual parton
distribution functions (representing the partonic structure of a fast
nucleon), one has to treat the different scatterings (happening in
parallel) individually, and for each scattering, one has a parton evolution according
to some evolution function $E$ (representing the partonic structure
of a fast parton), as sketched in Fig. \ref{saturation-three-pom}.
One still has DGLAP evolution, for each of the scatterings, but one
introduces saturation scales. But, most importantly, these scales are
not constants, they depend on the number of scatterings, and they
depend as well on $x^{+}$ and $x^{-}$. An example of a 
multiple-scattering AA configuration is shown in Fig. \ref{saturation-three-pom-1}.
The diagrams shown in Figs. \ref{saturation-three-pom} and \ref{saturation-three-pom-1}
are meant to be ``symbolic'', the real structure is somewhat more
complex. Also for simplicity, one consider only gluons in the diagrams.
I also do not consider (for simplicity) timelike parton emissions,
but in the real EPOS4 simulations they are of course taken care of,
based on angular ordering. For a complete description, see Ref. \cite{werner:2023-epos4-heavy}.

But all this ``multiple-scattering discussion'' is not the full
story. The S-matrix part concerns ``primary scatterings'', happening
instantaneously at $t=0$. As a result, in the case of a large number
of Pomerons, one has correspondingly a large number of ``prehadrons'',
and based on these, ``secondary interactions'' occur: fluid formation
and decay and hadronic rescatterings. This will be discussed in the
next section.

\section{The role of core, corona, and remnants \label{=======the-role-of-core=======}}

In this section, I will discuss briefly how the EPOS4 multiple scattering
diagrams translate into particle production, which means first the
production of so-called prehadrons. As discussed below, our Pomerons
are mapped into kinky relativistic strings, where string decay traditionally
produces string segments which correspond to hadrons. But one considers
the possibility of having a dense environment, and here the string
segments cannot ``evolve'' into hadrons. So one uses the term ``prehadrons''
for these segments, and they either ``fuse'' to produce the core,
or become hadrons if they escape the core. A similar argument is used
for excited remnants, which may decay into hadrons, but again this
may happen in a dense area. In Ref. \cite{werner:2023-epos4-heavy}, one
discusses the different types of Pomerons and their relation with pQCD,
including technical details for the computations, and one also discusses
how the pQCD diagrams translate into kinky strings. String decay into
segments (now called prehadrons) is disussed in detail in Ref. \cite{Drescher:2000ha}.
Based on these prehadrons, a core-corona procedure will be employed,
which allows identifying the ``core'', which will then be treated
as a fluid that evolves and eventually decays into hadrons, which
still may collide with each other.

\begin{figure}[h]
\noindent \centering{}\includegraphics[scale=0.2]
{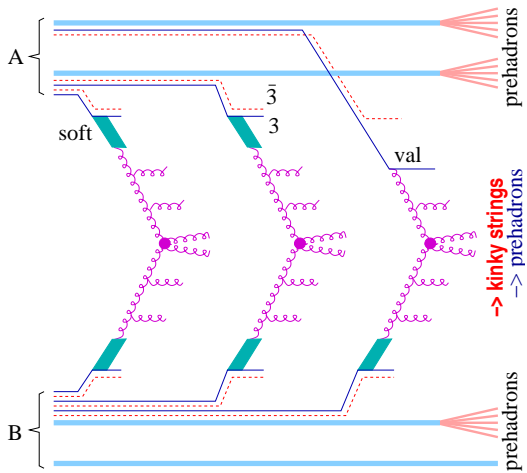}
\caption{Multiple scattering configuration of two colliding nuclei $A$ and
$B$ at LHC energies, each nucleus being composed of two nucleons,
with three scatterings (from three cut Pomerons). Dark blue lines
mark active quarks, red dashed lines active antiquarks, and light
blue thick lines projectile and target remnants. One of the target
nucleons is just a spectator.\label{partonic-configuration}}
\end{figure}

In Fig. \ref{partonic-configuration}, I consider an example of a
multiple cut Pomeron (i.e. multiple scattering) configuration of two
colliding nuclei $A$ and $B$ at Large Hadron Collider (LHC) energies
(several TeV), each
nucleus composed of two nucleons (just for illustration). Dark blue
lines mark active quarks, red dashed lines active antiquarks, and
light blue thick lines projectile and target remnants (nucleons minus
the active (anti)quarks). I have chosen an example with two scatterings
of ``sea-sea'' type, and one of ``val-sea'' type (see Ref. \cite{werner:2023-epos4-heavy}).
One considers each incident nucleon as a reservoir of three valence
quarks plus quark-antiquark pairs. The ``objects'' which represent
the ``external legs'' of the individual scatterings are colorless: 
quark-antiquark pairs in most cases as shown in the figure,
but one may as well have quark-diquark pairs or even antiquark-antidiquark
pairs \textendash{} in any case, a $3$ and a $\bar{3}$ color representation.
In general, each of the horizontal gluon lines would develop the so-called
\begin{figure}[h]
\noindent \centering{}\includegraphics[scale=0.2]
{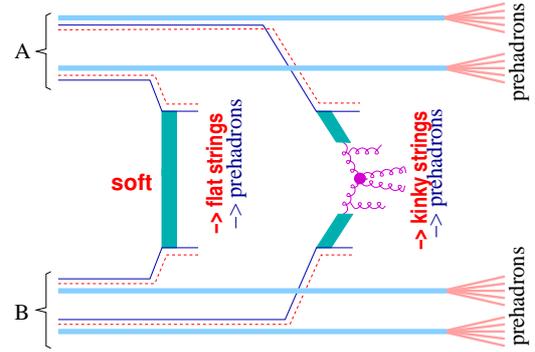}
\caption{As Fig. \ref{partonic-configuration}, but for lower energy.\label{partonic-configuration-1}}
\end{figure}
timelike cascade, which is not considered here for the simplicity
of the discussion. I also consider only gluons here, in general,
there are as well $g\to q+\bar{q}$ splittings. This multiple-scattering
picture works perfectly also at lower energies 
[BNL Relativistic Heavy Ion Collider (RHIC)].
With decreasing
energy, it becomes simply more and more likely that the Pomerons in
Fig. \ref{partonic-configuration} are replaced by purely soft ones,
as indicated in Fig. \ref{partonic-configuration-1}. Also the Pomerons
get less energetic, producing fewer particles.

For a given diagram as for example Fig. \ref{partonic-configuration},
one constructs the corresponding color flow diagram (two arrows per
gluon, forward and backward arrows for quarks and antiquarks). One
then follows the color flow arrows, for example starting with a $\bar{3}$
being an external leg on the projectile side, until one finds a $3$
which is in our example an external leg on the target side. This defines
a chain of partons. Assuming that the external legs are simply quarks
and antiquarks, one gets six chains of the type $\bar{q}-g-g-...-g-q$,
mapped (in a unique fashion) to kinky strings, where each parton corresponds
to a kink. This is explained in detail in Ref. \cite{werner:2023-epos4-heavy}.
The kinky strings are then decayed into prehadrons.

A second source of particle production are the remnants. In case of
multiple scattering as in Fig. \ref{partonic-configuration}, the
projectile and target remnants remain colorless, but they may
change their flavor content during the multiple collisions. The quark-antiquark
pair ``taken out'' for a collision (the ``external
legs'' for the individual collisions), may be $u$-$\bar{s}$,
then the remnant for an incident proton has flavor $uds$. In addition,
the remnants get massive, much more than a simply resonance excitation.
One may have remnants with masses of $10\,\mathrm{GeV/c^{2}}$ or more,
which contribute significantly to particle production (at large rapidities).

\begin{figure}[h]
\centering{}(a)\hspace*{4.2cm}(b)\hspace*{5.2cm}\\
 \includegraphics[scale=0.22]
{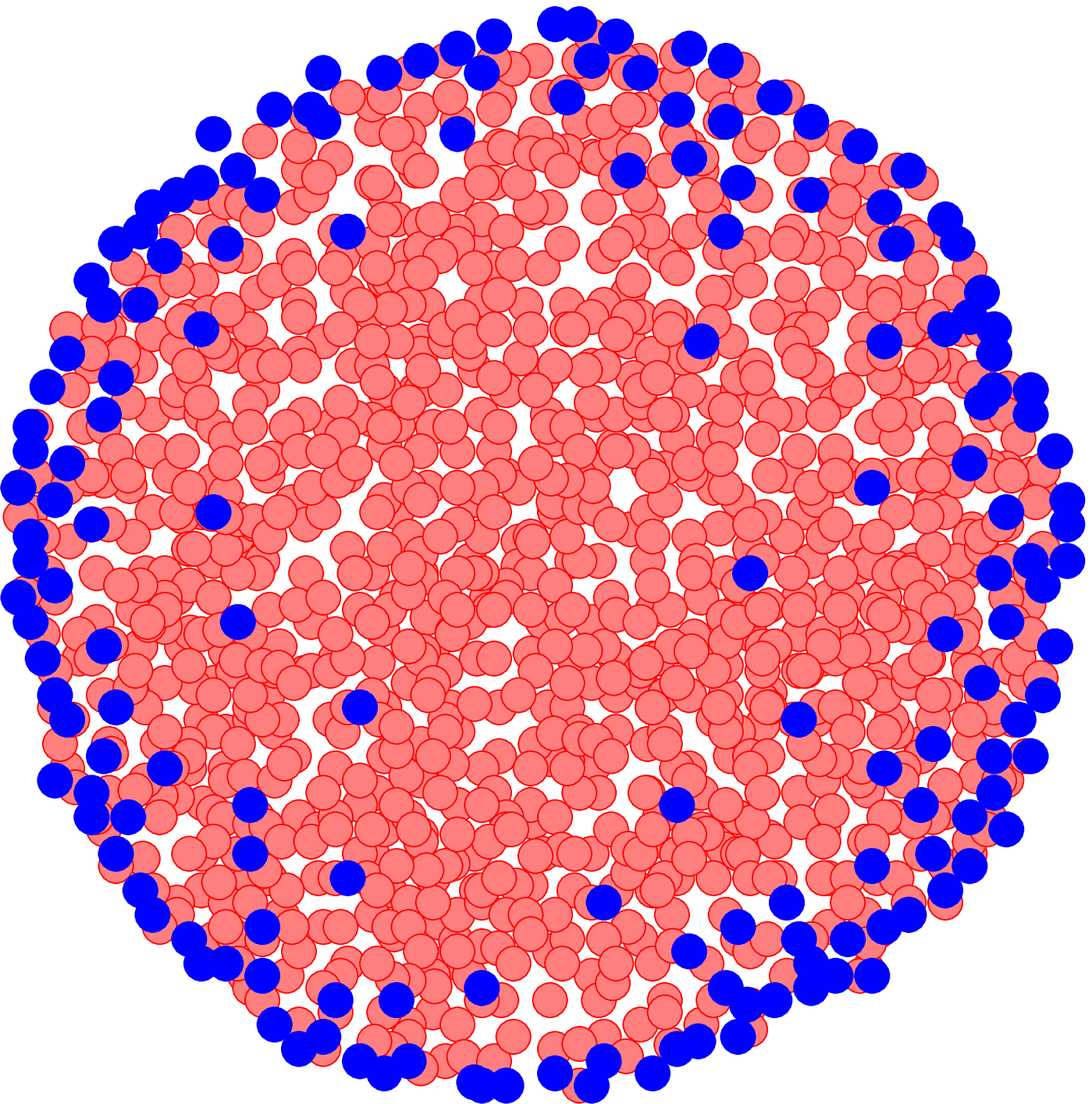}
$\qquad\qquad$\includegraphics[scale=0.22]
{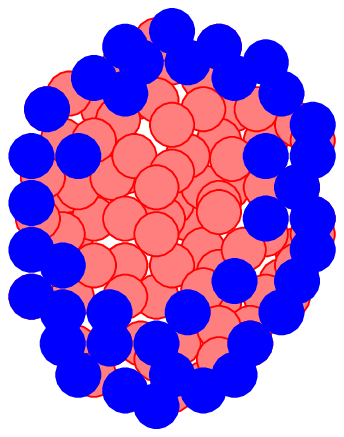}
\caption{Sketch of the core-corona separation for a ``big'' (a) and a ``small''
(b) system. The dots are prehadrons in the transverse plane, red refers
to core, blue to corona. \label{sketch-core-corona}}
\end{figure}

In the above discussion, I mentioned the production of ``prehadrons''
from strings and from remnant decays. Based on these prehadrons, one
employs a so-called core-corona procedure (introduced in Ref. \cite{Werner:2007bf},
updated in Ref. \cite{Werner:2013tya}), at some given (early) proper-time
$\tau_{0}$, see Fig. \ref{sketch-core-corona}. 
The method is used for all systems, big ones (as central AA) or small
ones (peripheral AA, or even pp scattering). One considers all prehadrons
at $\tau_{0}$, marked as dots in Fig. \ref{sketch-core-corona}.
For each 
prehadron, one computes 
\begin{equation}
p_{t}^{\mathrm{new}}=p_{t}
-0.25 \big(f_{\mathrm{EL1}}\!+\!Z(f_{\mathrm{EL2}}\!-\!f_{\mathrm{EL1}})\big) p_{t}^{g_{\mathrm{EL}}} \!\!
\int_{\gamma} \rho^{h_{\mathrm{EL}}}\, dL,
\end{equation}
where $\gamma$ is the trajectory of the prehadron moving out of the system, 
and $Z=(N_{\mathrm{part}}-2)/N_{\mathrm{part}}^{\mathrm{max}}$ the ``centrality'' 
based on the number of participants. 
I define ``cone prehadrons'' as prehadrons within a jet cone radius $R$ 
with respect to a parton with $p_t\ge 1\,$GeV/c, considering all the partons 
which constitute the kinky string being the origin of the prehadron in question.
All others are ``noncone prehadrons''. 
The jet cone radius is defined as 
$R^2=(\Delta\eta)^{2}+(\Delta\phi)^{2}$,
with $\Delta\eta$ and $\Delta\phi$ being respectively the difference
in pseudorapidity and azimuthal angle, with respect to the parton.
I use jet cone radii of 0.3 (pp) and 0.2 (PbPb). I use different sets of parameters 
$\{f_{\mathrm{EL1}},f_{\mathrm{EL2}},g_{\mathrm{EL}},h_{\mathrm{EL}}\}$ 
for noncone and cone prehadrons, namely, 
$\{0.1,-,0.5,0.375\}$ (pp cone), $\{0.4,-,0,1\}$ (pp noncone), $\{0.3,0.3,0.5,0.375\}$ (PbPb cone), $\{0.5,0.7,0,1\}$ (PbPb noncone).
The prehadron in question is then considered to be of ``core'' or ``corona'' type, based on the following criteria: \begin{itemize} \item For $p_{t}^{\mathrm{new}}\le 0$, the prehadron is considered to be a ``core prehadron''. \item For $p_{t}^{\mathrm{new}}>0$, the prehadron escapes, it is called ``corona prehadron''. \end{itemize}
The core prehadrons constitute ``bulk matter'' and will be treated
via hydrodynamics. The corona prehadrons become simply hadrons and
propagate with reduced energy (due to the energy loss). In Fig. \ref{sketch-core-corona},
I distinguish between big and small systems. Actually, the method
used is the same, but there are relatively more core prehadrons in
the big systems. Corona particles are either very energetic (then
they move out even from the center), or they are close to the surface,
or one has a combination of both.

The core-corona procedure is a crucial element in the EPOS4 approach,
which allows soft and hard elements to emerge naturally from a unique
formalism (rather than constructing a two-component approach). It
is also important to understand that a hard process (in the usual
terminology), which is in the EPOS4 framework a high
transverse momentum pQCD process, will also contribute
to low-$p_{t}$ particle production, since only string breaks close
to the the high-$p_{t}$ kink will produce high-$p_{t}$ particles.

The importance of the core-corona picture is also emphasized in Refs. \cite{Becattini:2008ya},
\cite{Aichelin:2010ns}, and \cite{Yuuka:2021}.

Let me investigate the core-corona effects in EPOS4 first for proton-proton
collisions. One wants to understand the relative importance of the core
part, and one also wants to know which fraction is coming from remnant
decay.
\begin{figure}[h]
\centering{}\includegraphics[bb=20bp 25bp 520bp 790bp,clip,scale=0.42]
{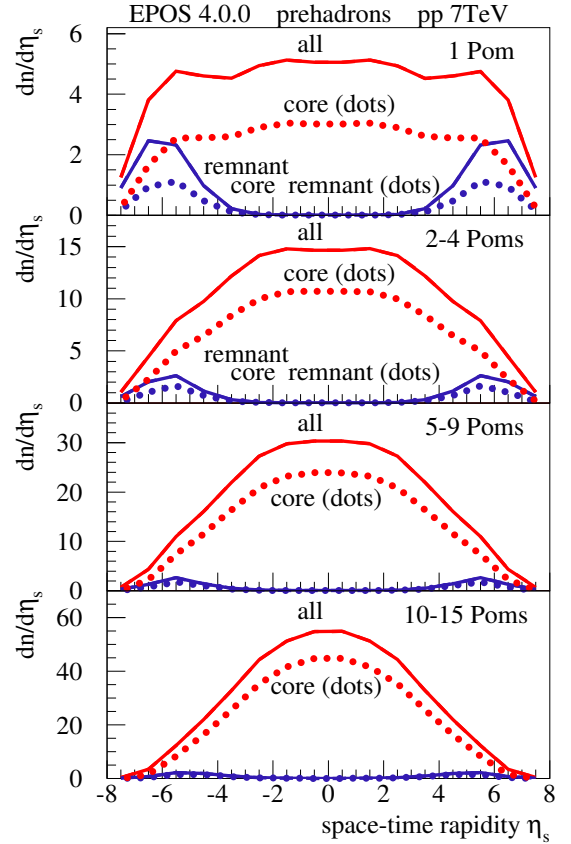}
\caption{The prehadron yield as a function of space-time rapidity, for different
Pomeron numbers in proton-proton collisions at 7 TeV. The curves refer
to all prehadrons (red full), all core prehadrons (red dotted), prehadrons
from remnant decay (blue full), and core prehadrons from remnant decay
(blue dotted). \label{prehadrons-pp}}
\end{figure}
In Fig. \ref{prehadrons-pp}, I show results for different event
classes (defined via Pomeron numbers) in proton-proton collisions
at 7 TeV. I show in each case four different curves: all prehadrons
(red full), all core prehadrons (red dotted), prehadrons from remnant
decay (blue full), and core prehadrons from remnant decay (blue dotted).
Not too surprisingly, remnant contributions show up in general at
large rapidities, but in all cases they do contribute to the core.
Comparing the red full and dotted curves, one sees that the core fraction
is in all cases substantial, most significantly for the event class
``10-15 Pomerons'', but even for small Pomeron numbers, the core
contribution remains important. As a small side remark: it is often
said that ``collectivity'' is seen in high multiplicity pp scattering.
In our EPOS4 analysis, it is even essential in minimum bias collisions
(with an average Pomeron number of around 2), very strongly supported
by data.

\begin{figure}[h]
\centering{}\includegraphics[bb=20bp 25bp 520bp 790bp,clip,scale=0.41]
{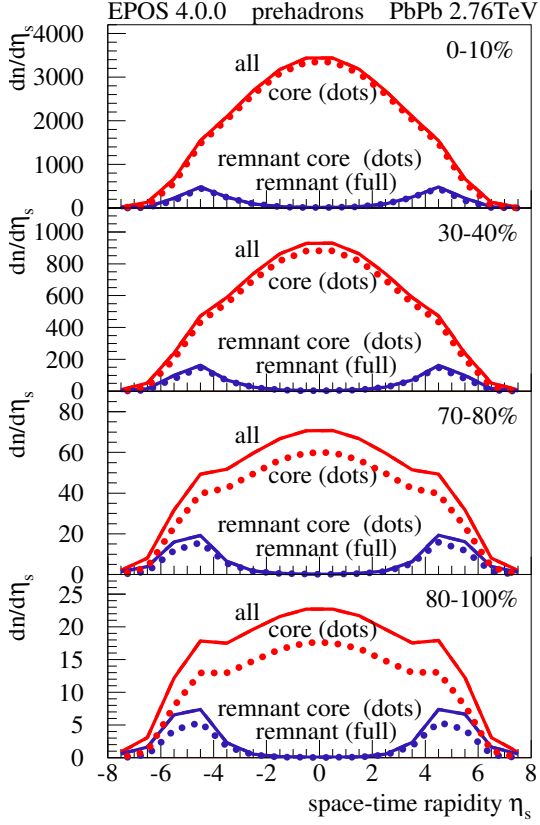}
\caption{The prehadron yield as a function of space-time rapidity, for different
centralities in PbPb collisions at 2.76 TeV. The curves refer to all
prehadrons (red full), all core prehadrons (red dotted), prehadrons
from remnant decay (blue full), and core prehadrons from remnant decay
(blue dotted).\label{prehadrons-aa}}
\end{figure}

Let me now turn to AA scattering, to understand the relative importance
of the core part, and of the fraction coming from remnant decay. In
Fig. \ref{prehadrons-aa}, I show results for different centralities
in PbPb collisions at 2.76 TeV, namely (from top to bottom): 0-10\%,
30-40\%, 70-80\%, and 80-100\% (based on the distribution of the impact
parameter). Also here, the remnant contributions show up preferentially
at large rapidities, and in all cases, they do contribute to the core.
Comparing the red full and dotted curves, one sees that the core fraction
(ratio of $dn/d\eta_{s}$ of the core contribution over all) is in
all cases substantial: 0.97 for 0-10\%, 0.95 for 30-40\%, 0.85 for
70-80\%, and 0.77 for 80-100\%. One also sees that this ``core dominance''
extends over a wide rapidity range.

Having identified core pre-hadrons, one computes the corresponding energy-momentum
tensor $T^{\mu\nu}$ and the flavor flow vector at some position $x$
at initial proper time $\tau=\tau_{0}$ as 
\begin{equation}
T^{\mu\nu}(x)=\sum_{i}\frac{p_{i}^{\mu}p_{i}^{\nu}}{p_{i}^{0}}g(x-x_{i})
\end{equation}
and 
\begin{equation}
N_{q}^{\mu}(x)=\sum_{i}\frac{p_{i}^{\mu}}{p_{i}^{0}}\,q_{i}\,g(x-x_{i}),
\end{equation}
with $q_{i}\in\{u,d,s\}$ being the net flavor content and $p_{i}$
the four-momentum of prehadron $i$. The function $g$ is a Gaussian
smoothing kernel, in Milne coordinates given as $g(x)=h(\eta/d_{\Vert})h(x/d_{\bot})h(y/d_{\bot})$,
with $h(s)=\frac{1}{\sqrt{2\pi}}\exp(-\frac{1}{2}s^{2})$, with parameters
$d_{\Vert}$ and $d_{\bot}$. 

The parameter $d_{\Vert}$ is given as
$\min(1.0,r(s))$, with $r(s)=\max(0.5,y(s)/y(s_{\mathrm{ref}}))$,
where $y(s)$ is the rapidity of the projectile nucleons in the nucleon-nucleon center-of-mass 
for a squared nucleon-nucleon energy $s$, and where $s_{\mathrm{ref}}=(2.76\mathrm{TeV)^{2}}$ is some reference value.
The parameter $d_{\bot}$ is given as $0.3+0.3Z$ for PbPb and as $0.3-0.02\min(N_{\mathrm{Pom}}-1,12)$ for pp,
where $Z=(N_{\mathrm{part}}-2)/N_{\mathrm{part}}^{\mathrm{max}}$ is the centrality based on the number of participants and $N_{\mathrm{Pom}}$ the number of (cut) Pomerons (characterizing the event activity in pp).

The Lorentz transformation of $T^{\mu\nu}$ into the
comoving frame provides the energy density $\varepsilon$ and the
flow velocity components $v^{i}$, which will be used as the initial
condition for a hydrodynamical evolution \cite{Werner:2013tya}. This
is based on the hypothesis that equilibration 
\begin{figure}
\centering{}\includegraphics[bb=15bp 30bp 540bp 790bp,clip,scale=0.41]
{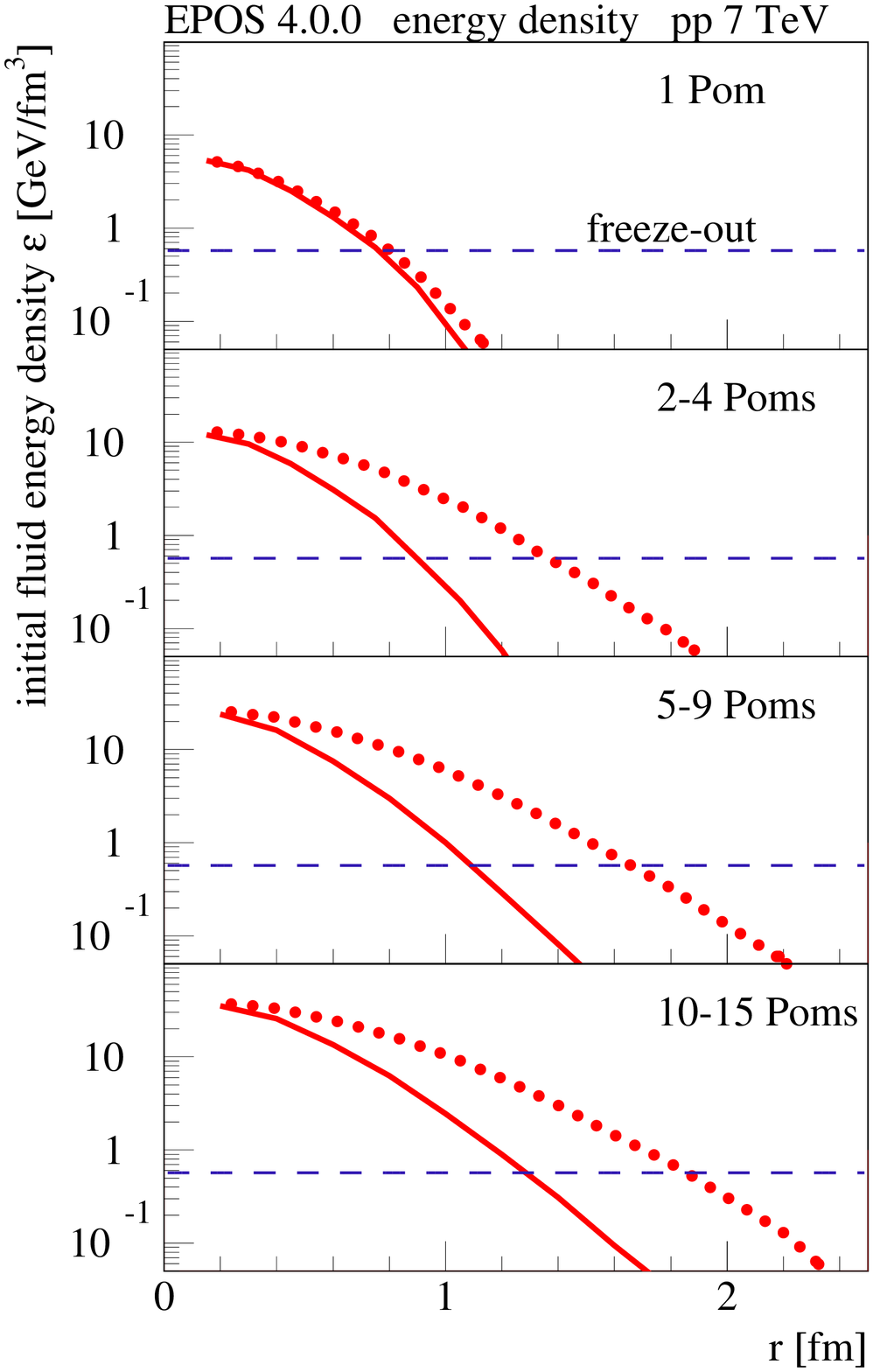} 

\centering{}\includegraphics[bb=15bp 30bp 540bp 790bp,clip,scale=0.41]
{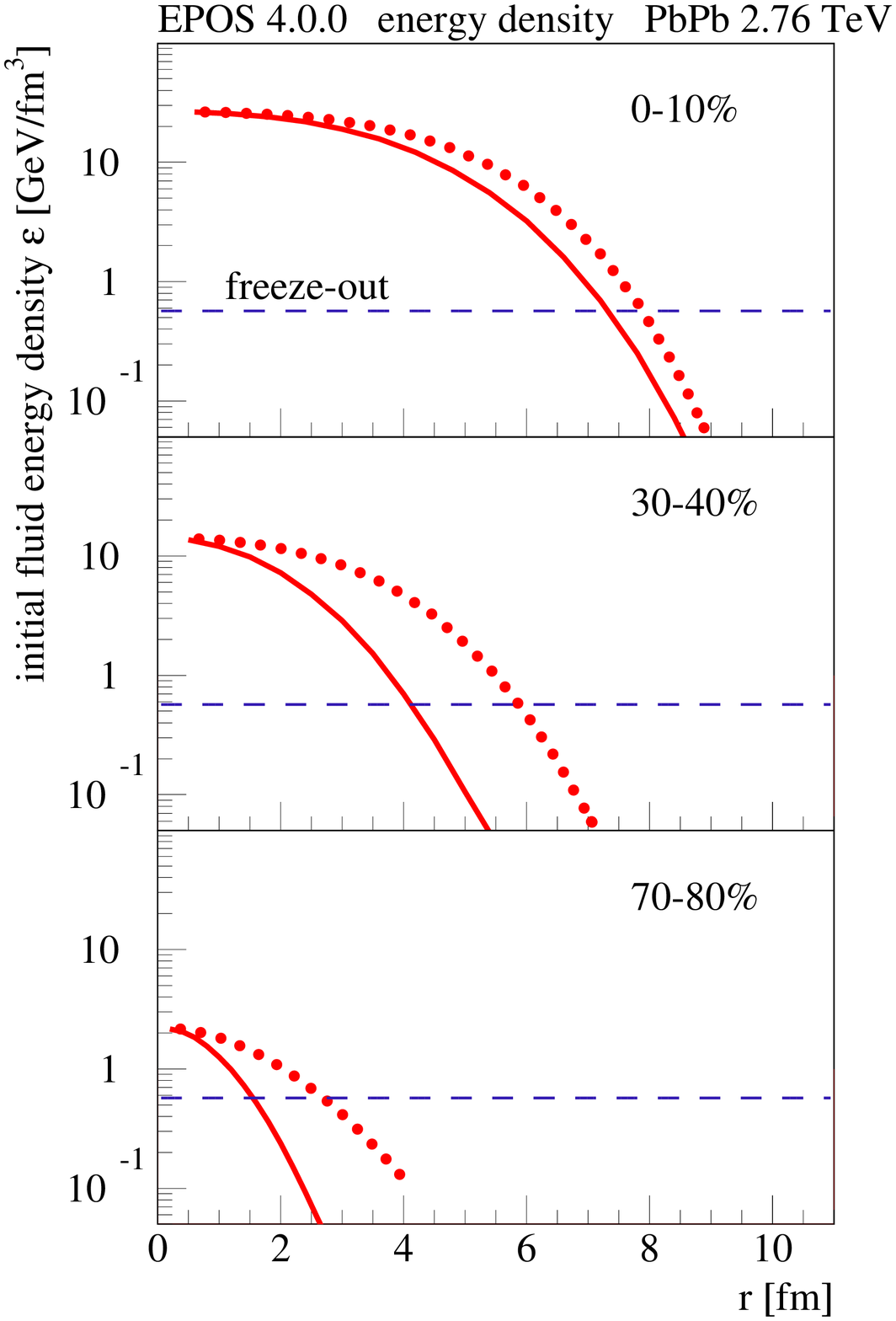}
\caption{Energy density at the initial proper-time $\tau_{0}$ as a function
of the transverse coordinate $r$. The full red lines correspond to
an azimuthal angle $\phi=0$, and the dotted red lines to $\phi=\pi/2$.
The blue dashed lines represent the freeze-out energy density. I
show results for different event classes (defined via Pomeron numbers)
in pp collisions at 7 TeV (upper plot) and for different centralities
(defined via impact parameter) in PbPb collisions at 2.76 TeV (lower
plot). \label{energy-density}}
\end{figure}
happens rapidly and affects essentially the space components of the
energy-momentum tensor. In Fig. \ref{energy-density} (upper plot),
I show the energy density at the initial proper-time $\tau_{0}$
as a function of the transverse coordinate $r$ for different event
classes (defined via Pomeron numbers) in pp collisions at 7 TeV. The
values of $\tau_{0}$ is $0.4\,$fm/c in EPOS4.0.0. I also indicate
in the figure the freeze-out energy density (blue dashed line). For
each event, one determines (based on the energy density distribution)
the event plane angle $\psi$ and rotate the system accordingly (to
have after rotation event plane angles zero). The plots in Fig. \ref{energy-density}
represent averages over such rotated events, the full lines correspond
to azimuthal angles $\phi=0$, the dotted lines to $\phi=\pi/2$.
The difference between the two lines reflects the azimuthal asymmetry.
As already seen in Fig. \ref{prehadrons-pp}, even in the case of
a single Pomeron, there is some core production, and one gets actually
an energy density of several $\mathrm{GeV/fm^{3}}$, but the radial
extension is very small, and the lifetime (before freeze-out) very
small. For large Pomeron numbers, the energy densities and the radial
extensions get bigger, but the latter remain of the order of 1 $\mathrm{fm/c}$.

In Fig. \ref{energy-density} (lower plot), I show the energy density
at the initial proper-time $\tau_{0}$ as a function of the transverse
coordinate $r$ for different centralities (defined via impact parameter)
in PbPb collisions at 2.76 TeV. The values of $\tau_{0}$ are between
$2\,$fm/c and $1\,$fm/c from central to peripheral in EPOS4.0.0.
Also here one does the rotations according to the event plane angles,
as discussed above for pp scattering, before taking event averages.
Based on these, the full lines correspond to azimuthal angles $\phi=0$,
the dotted line to $\phi=\pi/2$. As already seen in Fig. \ref{prehadrons-aa},
even in peripheral collisions, there is some core production, and
one gets actually an energy density of about $\mathrm{2\,GeV/fm^{3}}$
for 70-80\% centrality, but the radial extension is small, and the
lifetime as well.

It follows a viscous~hydrodynamic~expansion. Starting from the initial
proper time $\tau_{0}$, the core part of the system evolves according
to the equations of relativistic viscous hydrodynamics \cite{Werner:2013tya,Karpenko_2014},
where one uses presently $\eta/s=0.08$. The ``core-matter'' hadronizes
on some hyper-surface defined by a constant energy density $\epsilon_{H}$
(presently $0.57\mathrm{\,GeV/fm^{3}}$). In earlier versions \cite{Werner:2011-hydro-pp-900GeV},
one used a so-called Cooper-Frye procedure. This is problematic in
particular for small systems: not only do energy and flavor conservation
become important, but one also encounters problems due to the fact that
one gets small ``droplets'' with huge baryon chemical potential, with
strange results for heavy baryons. In EPOS4, one systematically
uses microcanonical hadronization, with major technical improvements
compared to earlier versions, as I will discuss below.

\section{Microcanonical hadronization of the core \label{=======microcanonical-hadronization=======}}

In EPOS4, one systematically uses micro-canonical hadronization.
Due to several technical improvements, our procedures work for all
kinds of systems (big and small ones), and one uses it for proton-proton
as well as heavy ion scattering. But not only for technical reasons.
The usual Cooper-Frye procedure amounts to a smooth transition from
a fluid to a particle description. However, one has a violently expanding
system into the vacuum, where around some critical energy density
the system goes very quickly from one state (plasma) to a very different
one (hadrons), in a complicated fashion. So one assumes \textendash{}
whatever the precise mechanism might be \textendash{} that the system
decays into multi-hadron states, randomly. And the most random way
corresponds to maximizing the entropy, which amounts to the micro-canonical
ensemble. Being a crucial element of the new EPOS4 approach, I will
discuss this in the following.

It is important to note that 
\begin{itemize}
\item there is no need to match the dynamical part of hydro evolution, one
considers a sudden statistical decay 
\item one has full energy and flavor conservation, which is important for
small systems 
\item the procedure is extremely fast, and can be used for ``big systems''
and in particular study the limiting case of ``very large plasma
droplets'' (infinite volume limit). 
\end{itemize}

\subsection{Infinite volume limit \label{-------infinite-volume-limit-------}}

Let me first look at the infinite volume limit, which is the grand
canonical ensemble. Here, for single particle spectra (particle species
$k$), one has the distribution%
\begin{equation}
f_{k}=\frac{g_{k}V}{(2\pi\hbar)^{3}}\,\exp\left(-\frac{E_{k}}{T}\right),\label{GraCan distribution}
\end{equation}
with degeneracy $g_{k}$, volume $V$, energy $E_{k}$ and temperature
$T$. The average energy is 
\begin{equation}
\left\langle E\right\rangle =\sum_{k}\frac{g_{k}V}{(2\pi\hbar)^{3}}\int_{0}^{\infty}E_{k}\exp\left(-\frac{E_{k}}{T}\right)4\pi p^{2}dp,
\end{equation}
Changing variables via $E_{k}dE_{k}=pdp$,%
{} and using 
\begin{equation}
K_{1}(z)=z\int_{1}^{\infty}\exp(-zx)\sqrt{x^{2}-1}dx,
\end{equation}
\begin{equation}
3\,K_{2}(z)=z^{2}\int_{1}^{\infty}\exp(-zx)\sqrt{x^{2}-1}^{3}dx,
\end{equation}
one gets 
\begin{equation}
\left\langle E\right\rangle =\sum_{k}\frac{4\pi g_{k}V}{(2\pi\hbar)^{3}}m_{k}^{2}T\left(3TK_{2}(\frac{m_{k}}{T})+m_{k}K_{1}(\frac{m_{k}}{T})\right).\label{GraCan average}
\end{equation}
This formula allows one to compare micro-canonical and grand-canonical
decay using the same average energy (this will be done later). Generating
hadrons according to Eq. (\ref{GraCan distribution}), onev may compute
the total energy $E$ of the produced particles, and then compare
with the average energy $\left\langle E\right\rangle $, Eq. (\ref{GraCan average}).
\begin{figure}[h]
\centering{}\includegraphics[bb=50bp 60bp 520bp 550bp,scale=0.24]
{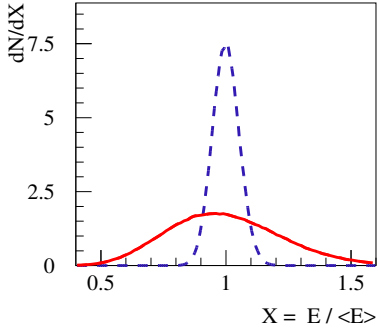}
\caption{The distribution of $X=E/\left\langle E\right\rangle $, i.e. the
ratio of final energy over initial energy, for a temperature of 130
MeV and volumes $V=50\,$fm$^{\boldsymbol{3}}$ (red curve) and $V=1000\,$fm$^{\boldsymbol{3}}$
(blue dashed curve).\label{GraCan energy}}
\end{figure}
In Fig. \ref{GraCan energy}, I plot the distribution of $E/\left\langle E\right\rangle $,
for a temperature of 130 MeV and two values for the volume $V$, namely
V=50 fm$^{\boldsymbol{3}}$ and V=1000 fm$^{\boldsymbol{3}}$. This
shows the amount of violation of energy conservation, which increases
with decreasing volume.

\subsection{Microcanonical hadronization of plasma droplets{\Large{}{} \label{-------Microcanonical-hadronization-of-plasma-droplets-------}}}

Let me now turn to micro-canonical decay of a ``plasma droplet''
of volume $V$ and energy $E$ (also referred to as mass $M$). Here,
all the states are equally probable, so the weight of a decay of the
droplet in its center-of-mass system (CMS into $n$ hadrons with momenta
$\vec{p}_{i}$ is given as

\begin{equation}
dP\boldsymbol{=}C_{\textrm{vol}}\,C_{\textrm{deg}}\,C_{\textrm{ident}}\,d\Phi_{\mathrm{NRPS}},\label{microcanonical-law-1}
\end{equation}
with the constants (the indices referring to volume, degeneracy,
and identical particles) 
\begin{equation}
C_{\textrm{vol}}=\frac{V^{n}}{(2\pi\hbar)^{3n}}\,\,,\,\,\,\,C_{\textrm{deg}}=\prod_{i=1}^{n}g_{i}\,\,,\,\,\,\,C_{\textrm{ident}}=\prod_{\alpha\in\mathcal{S}}{\frac{1}{n_{\alpha}!}}\,\,,\label{microcanonical-law-2}
\end{equation}
and with 
\begin{equation}
d\Phi_{\mathrm{NRPS}}=\delta(E-\Sigma E_{i})\,\delta(\Sigma\vec{p}_{i})\,\prod_{A}\delta_{Q_{A},\Sigma q_{A\,i}}\,\prod_{i=1}^{n}d^{3}p_{i}\,,\label{microcanonical-law-3}
\end{equation}

\noindent with $E_{i}=\sqrt{m_{i}^{2}+p_{i}^{2}}$ being the energy
and $\vec{p}_{i}$ the 3-momentum of particle $i$. Here, $C_{\textrm{deg}}$
accounts for degeneracies ($g_{i}$ is the degeneracy of particle
$i$), and $C_{\textrm{ident}}$ accounts for the occurrence of identical
particles ($n_{\alpha}$ is the number of particles of species $\alpha$).
The term $\delta_{Q_{A},\Sigma q_{A\,i}}$ reflects conservation laws
(baryon ($A=B$), electric charge ($A=C$) and strangeness ($A=S$).
It should be noted that one has to use the ``non-relativistic phase
space (NRPS)'' $d\Phi_{\mathrm{NRPS}}$ rather than the Lorentz-invariant
phase space (LIPS) $d\Phi_{\mathrm{LIPS}}$ used for the decay
of massive particles, where asymptotic states are defined over an
infinitely large volume, whereas here one considers a finite volume
(see discussion in Ref. \cite{Becattini_2004}). But of course, one uses
the relativistic expressions $E_{i}=\sqrt{p_{i}^{2}+m_{i}^{2}}$ for
hadron energies.

Numerical procedures to deal with these n-body phase-space expressions
are quite involved, and have a long history. Cerulus and Hagedorn
\cite{Hagedorn:1958} provided in 1958 a way to evaluate the NRPS
integrals $\Phi_{\mathrm{NRPS}}=\int d\Phi_{\mathrm{NRPS}}$, which
they used to compute particle production in high energy collisions.
Shortly after it was proposed to better use a covariant formula, referred
to as the Lorentz-invariant phase-space (LIPS) integral. The corresponding
algorithms for Monte Carlo applications were discussed by James \cite{James:1968}
in 1968 and heavily used for computing particle production. Being still
relevant concerning the decay of a quark gluon plasma, the NRPS integrals
and the corresponding Hagedorn prescriptions were used for Monte Carlo
realizations (Werner and Aichelin \cite{Werner:1995} in 1994, Becattini
and Ferroni \cite{Becattini_2004} in 2004) of particle production
in heavy ion collisions. In 2012 (Bignamini et al. \cite{Bignamini_2012}),
it was proposed to use the fact that the non-relativistic phase-space
(NRPS) element is up to a factor equal to the Lorentz invariant 
phase-space (LIPS) element, which is much easier to handle for $n$ not
too large. 

The Cerulus-Hagedorn method employed in Ref. \cite{Werner:1995}
is not very efficient and needs enormous computing efforts at intermediate
$n$ (around $n=10$) and at very large $n$, whereas the LIPS
method employed in Ref. \cite{Bignamini_2012} works well at small $n$,
but gets very time consuming at large $n$.

I will employ a hybrid method, i.e. an improved (and very efficient)
Cerulus-Hagedorn method for large $n$, and the LIPS method for small
$n$, such that one has finally very efficient procedures for any
$n$, even for very big values (for very big plasma droplets).

Let me first discuss the ``improved Cerulus-Hagedorn method''. Cerulus
and Hagedorn \cite{Hagedorn:1958} proposed to write the phase-space
integral as 
\begin{equation}
\Phi_{\mathrm{NRPS}}=(4\pi)^{n}\int\prod_{i=1}^{n}p_{i}^{2}\,\delta(E-\sum_{i=1}^{n}E_{i})\,W(p_{1},\ldots,p_{n})\prod_{i=1}^{n}dp_{i},\label{NRPS-integral}
\end{equation}
with $p_{i}=\left|\vec{p}_{i}\right|$, and with the ``random walk
function'' $W$ (also called angular integral) 
\begin{equation}
W(p_{1},\ldots,p_{n}):=\frac{1}{(4\pi)^{n}}\int\delta\big(\sum_{i=1}^{n}p_{i}\vec{u}_{i}\big)\prod_{i=1}^{n}d\Omega_{i}\,,\label{angular-integral}
\end{equation}
with $\vec{u}_{i}=\vec{p}_{i}/p_{i}$ and $\Omega_{i}$ referring
to the corresponding solid angle. The crucial point is an efficient
calculation of the angular integral $W$. It may be written as (omitting
the arguments $p_{1},\ldots,p_{n}$) 
\begin{equation}
W=\frac{1}{(4\pi)^{n}}\frac{1}{(2\pi)^{3}}\int\int e^{-i\vec{\lambda}\Sigma p_{j}\vec{u}_{j}}\,\prod_{j=1}^{n}d\Omega_{j}\,d^{3}\lambda,\label{13a-1}
\end{equation}
which leads to 
\begin{equation}
W=\int_{0}^{\infty}F(\lambda)\,d\lambda,
\end{equation}
with the integrand being 
\begin{equation}
F(\lambda)=\frac{\lambda^{2}}{2\pi^{2}}\,\prod_{j=1}^{n}\frac{\sin p_{j}\lambda}{p_{j}\lambda}=\frac{\lambda^{2}}{2\pi^{2}}\,\prod_{j=1}^{n}\frac{\sin\pi{a}_{j}}{\pi{a}_{j}},
\end{equation}
with $a_{j}=p_{j}\frac{\lambda}{\pi}$. One may use Euler's product
formula to get 
\begin{equation}
\prod_{j=1}^{n}\frac{\sin\pi{a}_{j}}{\pi{a}_{j}}=\prod_{j=1}^{n}\prod_{k=1}^{\infty}\left(1-\frac{a_{j}^{2}}{k^{2}}\right),
\end{equation}
which is for large $n$ approximately equal to 
\begin{equation}
\exp\left(-{\frac{\pi^{2}}{6}}\sum_{j=1}^{n}a_{j}^{2}\right)
\end{equation}
(having used $\sum_{k=1}^{\infty}\frac{1}{k^{2}}=\frac{\pi^{2}}{6}$).%
{} This means that the expression $\prod_{j=1}^{n}\{\sin p_{j}\lambda/p_{j}\lambda\}$
is well approximated by 
\begin{equation}
\exp\left(-P^{2}\lambda^{2}\right),\quad\mathrm{with}\;P=\sqrt{\frac{1}{6}\sum_{j=1}^{n}p_{j}^{\:2}}.
\end{equation}
The crucial point is the following: although this ``approximation''
is not precise enough to be used directly, one may use the fact that
one has a rough estimate of the integrand $F(\lambda)$ to make a
variable transformation which allows one to make a very accurate and efficient
numerical integration. Thanks to the approxmation $F(\lambda)\approx\frac{\lambda^{2}}{2\pi^{2}}\exp\left(-P^{2}\lambda^{2}\right)$,
one knows that 
\begin{equation}
F_{0}(\lambda)=F(\lambda)\times\exp\left(P^{2}\lambda^{2}\right)
\end{equation}
is a slowly varying function of $\lambda$ (polynomial dependence).
The angular integral may then be written as 
\begin{equation}
W=\int_{0}^{\infty}F(\lambda)\,d\lambda=\frac{1}{P}\int_{0}^{\infty}F_{0}\left(\frac{x}{P}\right)\,\times\exp\left(-x^{2}\right)\,dx.
\end{equation}
So one has an integral of the form $\int f(x)e^{-x^{2}}dx$, with
a well behaved function $f$, which allows one to evaluate the integral
with very high precision by using the Gauss-Hermite method, i.e., 
\[
W\approx\frac{1}{P}\sum_{k=1}^{K}w_{j}^{GH}F_{0}\left(\frac{x_{j}^{GH}}{P}\right),
\]
with nodes and weights $x_{j}^{GH}$ and $w_{j}^{GH}$ found in text
books. With only six nodes one gets excellent results.

Having solved the problem of computing $W$, the next step is to write
$d\Phi_{\mathrm{NRPS}}$ in terms of independent variables, which
can be done (see Ref. \cite{Werner:1995} and also Appendix \ref{=======Coordinate-transformations-for-NRPS=======})
as 
\begin{equation}
d\Phi_{\mathrm{NRPS}}=dr_{1}...dr_{n-1}\frac{(4\pi)^{n}\,T^{n-1}}{(n-1)!}\prod_{i=1}^{n}p_{i}\,E_{i}\,W(p_{1},...,p_{n}),\label{dPhi-NRPS-rvariables-1}
\end{equation}
with $n-1$ independent variables $r_{i}$ defined in $[0,1]$, and
with $z_{i}=r_{i}^{1/i}$, ~$x_{i}=z_{i}x_{i+1}$, ~$s_{i}=x_{i}T$,
~$t_{i}=s_{i}-s_{i-1}$, ~$E_{i}=t_{i}+m_{i}$, and $T=M-\sum_{i=1}^{n}m_{i}$,
where a configuration is defined by the number of hadrons, their species
(only configurations respecting the conservation laws $\Sigma q_{A\,i}=Q_{A}$
are considered), and their momenta. In addition to $r_{i}$, one
needs $2n$ more variables, to take care of the random orientations
in space of the momentum vectors as $(x,y,z)$, with $x=\sqrt{1-z^{2}}\cos(2\pi u)$,
$y=\sqrt{1-z^{2}}\sin(2\pi u)$, and $z=2w-1$, characterized in terms
of two independent variables $u,w\in[0,1]$ (for each of the $n$
hadrons).

However, this does not work for small $n$ (below 30), because in
that case the angular integral $W$ cannot be calculated reliably,
as discussed earlier. Here, one uses the LIPS method \cite{James:1968,Bignamini_2012}.
The NRPS and LIPS integrands are very similar, the only difference
is an additional $1/2E_{i}$ factor in the LIPS case. So one uses 
\begin{equation}
d\Phi_{\mathrm{NRPS}}=d\Phi_{\mathrm{LIPS}}\times\prod_{i=1}^{n}2E_{i}\:,\label{dPhi-LIPS-rvariables-0}
\end{equation}
with (see appendix \ref{=======LIPS-method-for-small-n=======})
\begin{align}
d\Phi_{\mathrm{LIPS}} & =dr_{1}...dr_{n-2}\frac{\pi(2\pi)^{n-2}}{M}\:\prod_{i=1}^{n-1}p(M_{i+1};M_{i},m_{i+1})\nonumber \\
 & \qquad\times\Big(M-\sum_{j=1}^{n}m_{j}\Big)^{n-2}\,\frac{1}{(n-2)!},\label{dPhi-LIPS-rvariables-1}
\end{align}
with $n-2$ independent variables $r_{i}$ defined in $[0,1]$, and
with $x_{i+1}=x'_{i}$, $x'_{i}=z_{i}x'_{i+1}$, and $(z_{i})^{i}=r_{i},$
and 
\begin{equation}
M_{i}=\sum_{j=1}^{i}m_{i}+x_{i}\left(M-\sum_{j=1}^{n}m_{j}\right),
\end{equation}
and with 
\begin{equation}
p(m;m_{a},m_{b})=\sqrt{\left\{ \frac{1}{2m}\left(m^{2}+m_{a}^{2}-m_{b}^{2}\right)\right\} ^{2}-m_{a}^{2}}.
\end{equation}
These formulas correspond to a sequence of successive two-body decays
$M=M_{n}\to M_{n-1}+m_{n},$ $M_{n-1}\to M_{n-2}+m_{n-1}$, etc, where
the decays are considered in the CMS system of the decaying mass $M_{i}$
(the only way to get simple formulas), with the decay products having
random orientations $(x,y,z)$ in space, with $x=\sqrt{1-z^{2}}\cos(2\pi u)$,
$y=\sqrt{1-z^{2}}\sin(2\pi u)$, and $z=2w-1$, characterized in terms
of two independent variables $u,w\in[0,1]$ (for each of the $n-1$
decays). At the end of the procedure \textendash{} starting from the
last decay, going backward untill the first decay \textendash{} one
has to perform a sequence of Lorentz boosts to have finally the momenta
in the CMS frame of the decaying droplet. This will become time-consuming
for large $n$ (>50), whereas for smaller $n$ the procedure is very
fast. \\

So one has excellent methods for computing $d\Phi_{\mathrm{NRPS}}$,
for large $n$ [based on Eq. (\ref{dPhi-NRPS-rvariables-1})], and
for small $n$ [based on Eqs. (\ref{dPhi-LIPS-rvariables-0},\ref{dPhi-LIPS-rvariables-1})],
and fortunately around $n\approx30-50$ both methods work, so finally
one has very fast procedures for any size of droplets. So one is ready
to employ Markov chain methods to generate configurations according
to the weights $d\Phi_{\mathrm{NRPS}}$. A configuration of $n$ hadrons
is characterized by their particle species and momenta, the latter
ones expressed in terms of $k$ independent variables (all of them
defined in $[0,1]$). Based on the above discussion, in the case of
the improved Cerulus-Hagedorn method [using Eq. (\ref{dPhi-NRPS-rvariables-1})],
one has $k=n-1+2n$, in case of the LIPS method [using Eqs.
(\ref{dPhi-LIPS-rvariables-0},\ref{dPhi-LIPS-rvariables-1})], one
has $k=n-2+2(n-1)$. For details concerning the Markov chain methods,
see Appendix \ref{=======Sampling-via-Markov=======},
where one discusses in particular the (new) algorithm to define the proposal
matrix in the Markov chain.\\

\begin{figure}[h]
\centering{}\includegraphics[bb=20bp 20bp 400bp 595bp,clip,scale=0.69]
{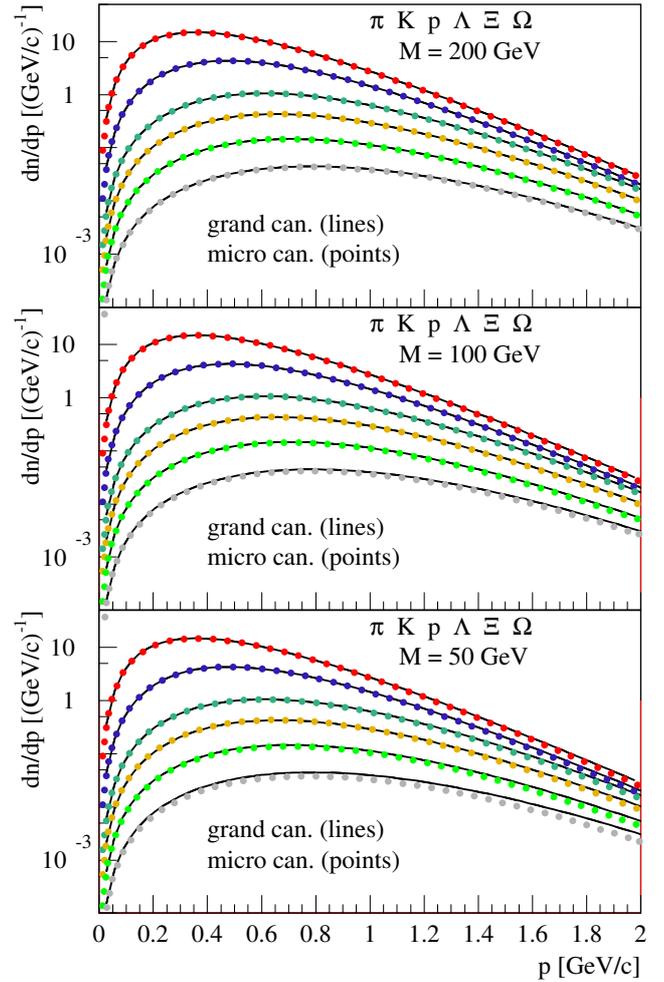}
\caption{Momentum distributions of pions (red), kaons (blue), protons (green),
lambdas (yellow), $\Xi$ baryons (light green), and $\Omega$ baryons
(grey), where one counts particles and antiparticles. I show results
for masses $M=200$ GeV, $M=100$ GeV, and $M=50$ GeV. I show results
for microcanonical particle production (points) compared with the grand
canonical one (lines) \label{fig: MiCan/GraCan 1}. }
\end{figure}
\begin{figure}[h]
\centering{}\includegraphics[bb=20bp 20bp 400bp 595bp,clip,scale=0.69]
{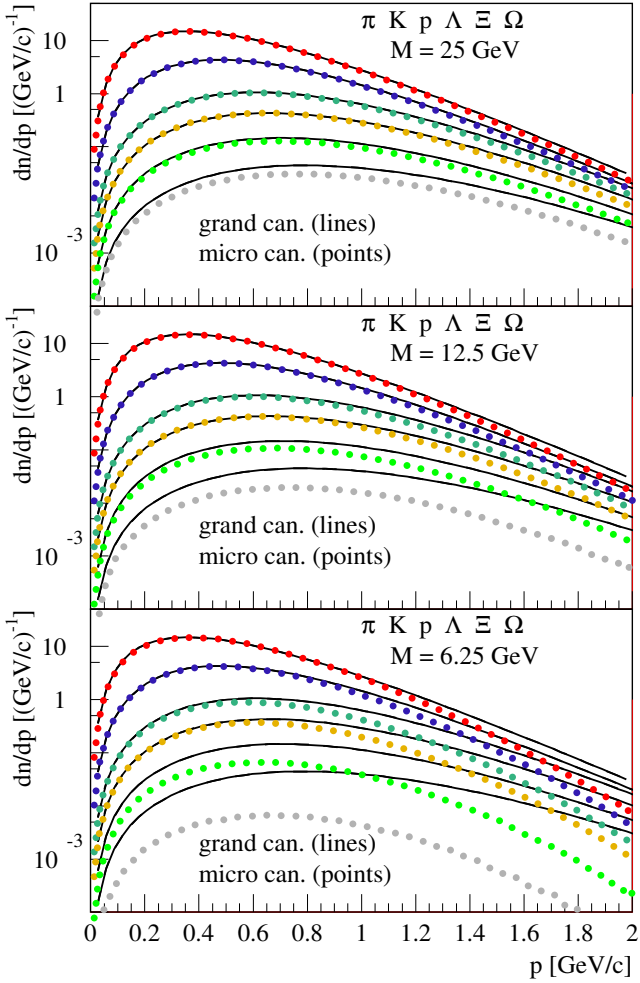}
\caption{Same as Fig. \ref{fig: MiCan/GraCan 1}, but for masses $M=25$ GeV,
$M=12.5$ GeV, and $M=6.25$ GeV\label{fig: MiCan/GraCan 2}. }
\end{figure}
Having a reliable method for very large $n$ is in particular useful
in order to investigate the convergence towards the infinite volume
limit. I will first \textendash{} also as a check that the procedures
work \textendash{} compare our microcanonical results with grand-canonical
decay, knowing that the latter is the limit of the former for large
droplets. I consider a ``complete'' set of hadrons, compatible
with a recent PDG (particle data group) list. I consider the decay of droplets
of different invariant masses $M$ (corresponding to the total energy
$E$ in the above formulas). The volume is chosen such that one has
always the same energy density, namely $M/V=\varepsilon_{\mathrm{FO}}=0.57\,\mathrm{GeV}/\mathrm{fm}{}^{3}$,
which corresponds to a temperature of $T=167\,$MeV.

In Figs. \ref{fig: MiCan/GraCan 1} and \ref{fig: MiCan/GraCan 2}, I
show momentum distributions of pions (red), kaons (blue), protons
(green), lambdas (yellow), $\Xi$ baryons (light geen), and $\Omega$
baryons (grey), where one counts particles and antiparticles. I show
results for different masses $M$, namely $M=200$ GeV, $M=100$ GeV,
$M=50$ GeV, $M=25$ GeV, $M=12.5$ GeV, and $M=6.25$ GeV. I show results
for microcanonical particle production (points) compared with the grand
canonical one (lines). The grand canonical results are all taken for
the same volume, namely $V_{0}=M_{0}/\varepsilon_{\mathrm{FO}}$,
with $M_{0}=50$ GeV, to have always the same reference curve,
therefore one multiplies the microcanonical results with $M_{0}/M$,
to have the correct normalization.

Looking at the $M=200\mathrm{\,GeV}$ results (upper plot in Fig.
\ref{fig: MiCan/GraCan 1}), one sees no difference between the microcanonical
and the grand canonical results, which means that this system ( $M=200\mathrm{\,GeV}$
and $V=M/\varepsilon_{\mathrm{FO}}$,) correspond already to the ``infinite
volume limit''. Considering $M=100\mathrm{\,GeV}$ and $M=50\mathrm{\,GeV}$
(middle and lower plot in Fig. \ref{fig: MiCan/GraCan 1}), one sees
slight deviations of the microcanonical curves from the grand canonical
ones, for heavy particles ($\Xi$ baryons and $\Omega$ baryons) and
at large $p_{t}$. For pions and kaons, the two scenarios are still
identical.

Looking at the plots in Fig. \ref{fig: MiCan/GraCan 2}, representing
$M=25\mathrm{\,GeV}$, $M=12.5\mathrm{\,GeV}$, and $M=6.25\mathrm{\,GeV}$,
one sees increasing differences between microcanonical and the grand
canonical results, the biggest ones for heavy particles, in particular
the $\Omega$ baryons. Finally not only the shape of the distributions
is affected, but even the absolute yield. In case of $M=6.25\mathrm{\,GeV}$,
one observes a strong $\Omega$ baryon suppression in the microcanonical
compared with the grand canonical case.

So far one considers the decay of a static plasma droplet of
given mass $M$ and volume $V$ (with $M/V=\varepsilon_{\mathrm{FO}}$),
according to the microcanonical probability distribution, whereas
in reality, one has an expanding fluid, where fluid cells pass eventually
below the critical energy density $\varepsilon_{\mathrm{FO}}$. I
discuss in the following how to treat such a case.

\subsection{Flow through hypersurface elements \label{-------Flow-through-hypersurface-elements-------}}

\begin{figure}[h]
\centering{}(a)$\qquad\qquad\qquad\qquad$(b)$\qquad\qquad$\\
 \includegraphics[scale=0.2]
{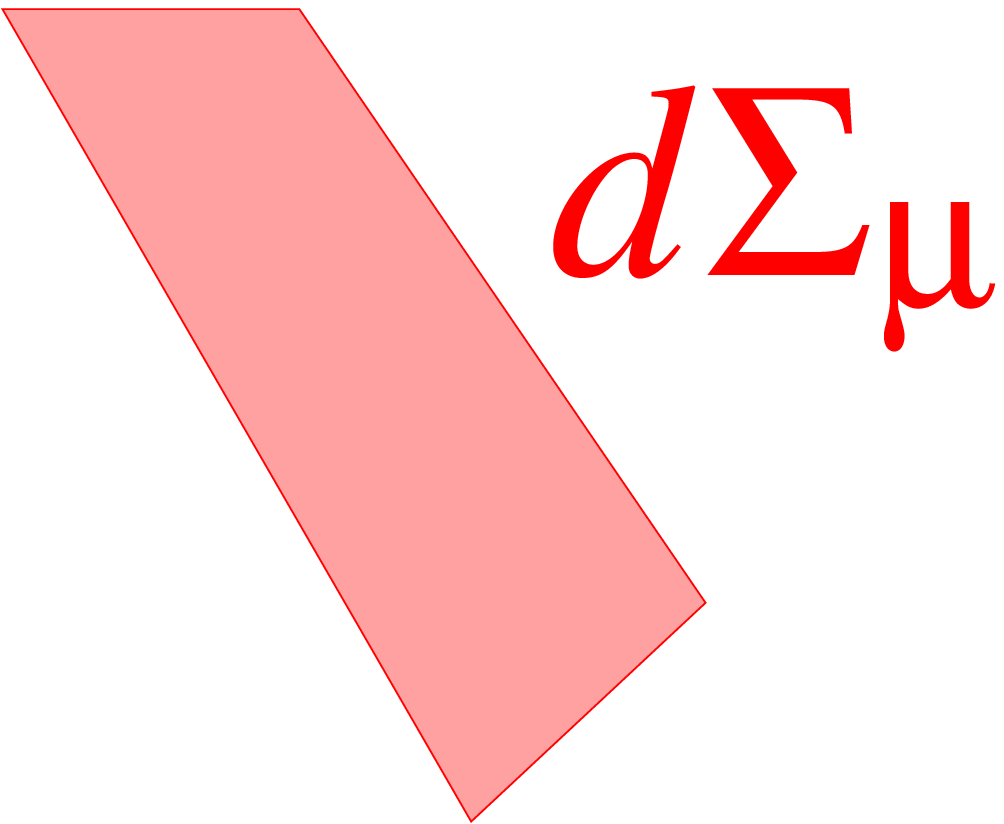} 
$\qquad\qquad$\includegraphics[scale=0.2]
{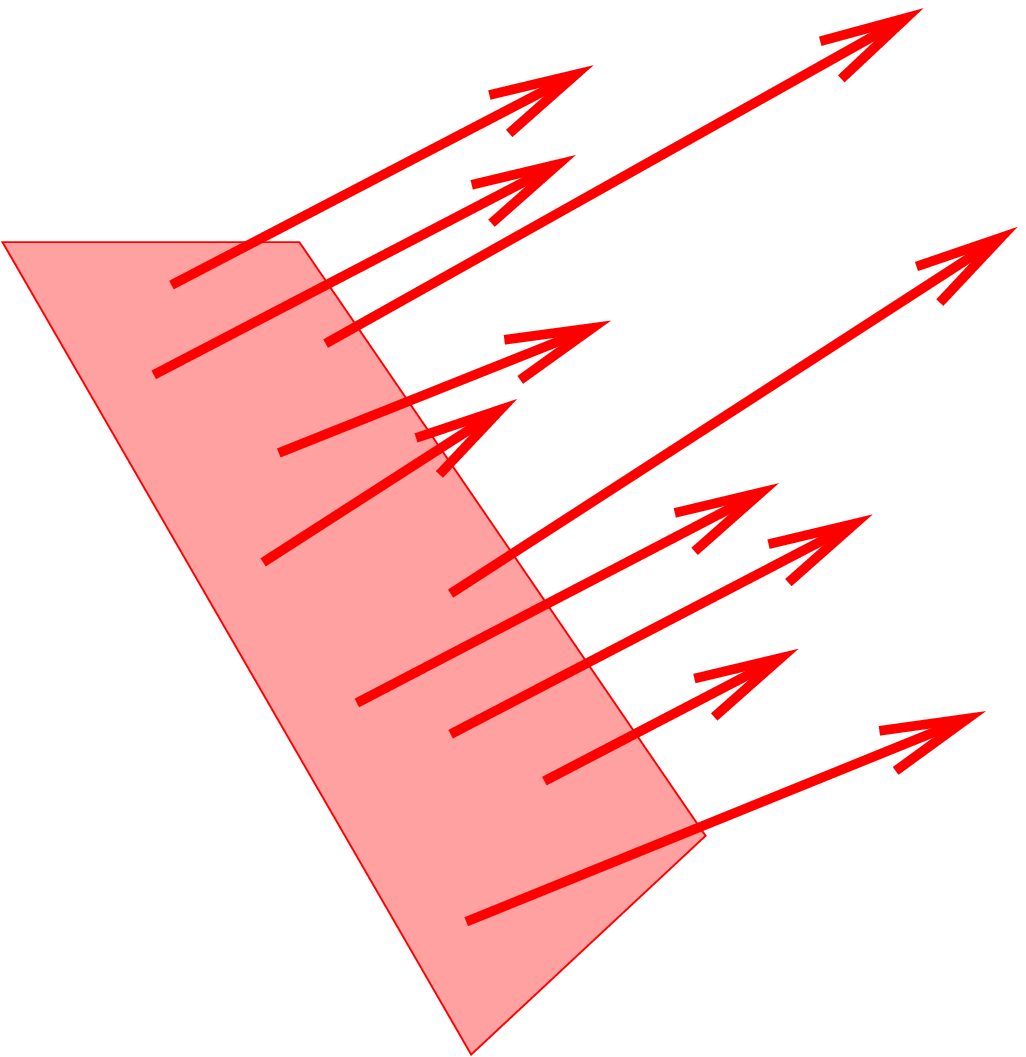}\\
 (c)$\qquad\qquad$\\
 $\qquad\qquad$\includegraphics[scale=0.2]
{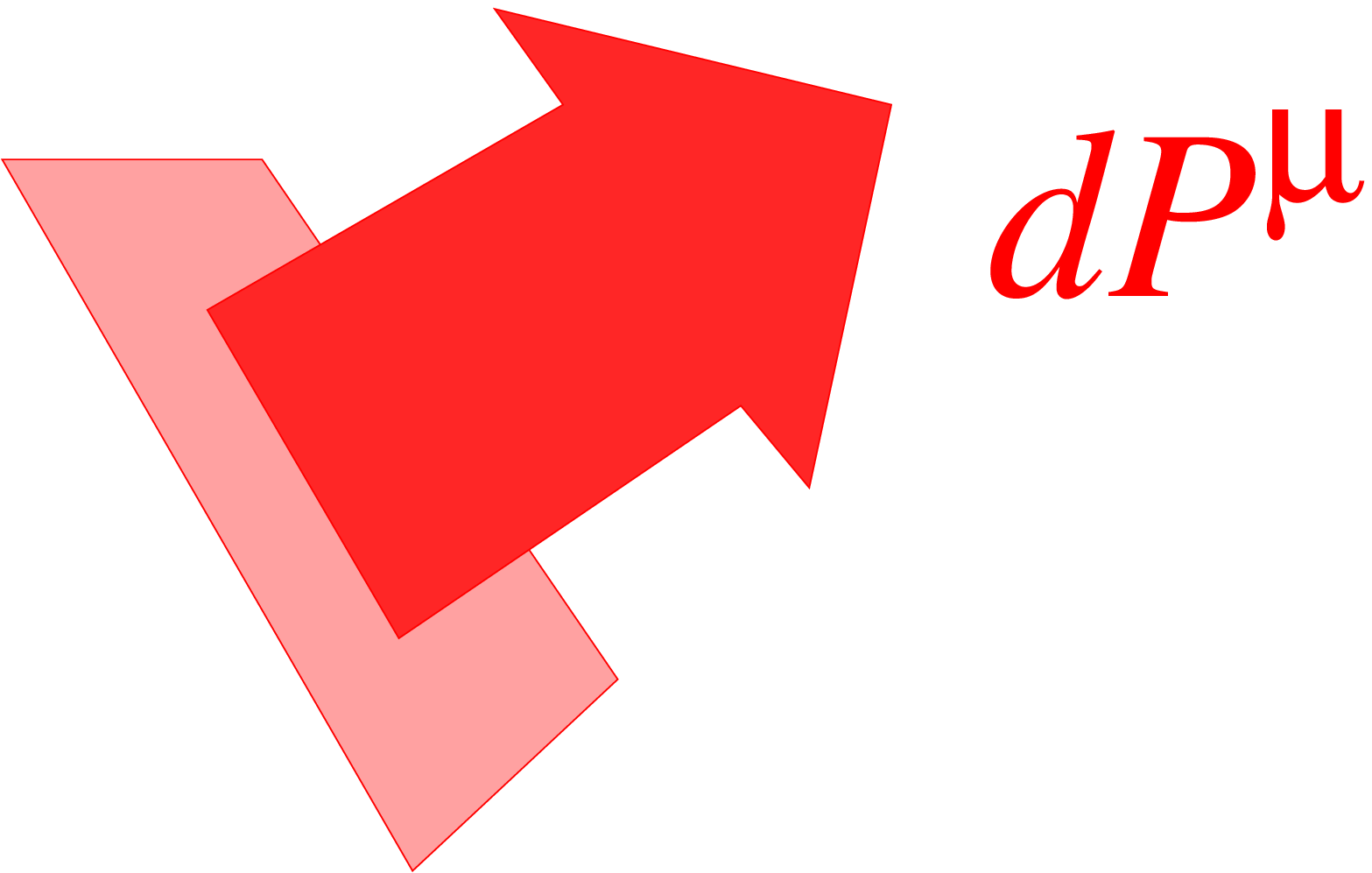}
\caption{(a) FO hypersurface element. (b) Cooper-Frye hadronization as particle
flow through FO hypersurface element. (c) The flow of energy-momentum
$dP^{\mu}$ through the surface element $d\Sigma_{\mu}$. \label{Cooper-Frye-hadronization}}
\end{figure}

Let me assume that one has an expanding fluid, characterized by an
energy-momentum tensor $T^{\mu\nu}$ and some vector $J_{A}^{\mu}$
representing the current of conserved quantities $A$ (charge, baryon
number, strangeness). The energy-momentum tensor can be expanded in
the case of a viscous fluid in terms of the energy density $\varepsilon$,
the flow vector $u^{\mu}$ (four-velocity of fluid cell), the shear
stress tensor, and bulk pressure. Given the space-time dependence
of the energy density, one may define a ``freeze-out (FO) hypersurface''
via 
\begin{equation}
\varepsilon(\tau,\eta,r,\varphi)=\varepsilon_{\mathrm{FO}},
\end{equation}
such that a point on the FO hypersurface can be expressed as $x^{\mu}(\tau,\eta,r,\varphi)$
with 
\begin{equation}
x^{0}=\tau\cosh\eta,\:x^{1}=r\cos\varphi,\:x^{2}=r\sin\varphi,\:x^{3}=\tau\sinh\eta,\label{milne}
\end{equation}
with for example $r=r(\tau,\eta,\varphi)$. A hypersurface element
[sketched in Fig. \ref{Cooper-Frye-hadronization}(a)] may then be
defined as 
\begin{equation}
d\Sigma_{\mu}=\varepsilon_{\mu\nu\kappa\lambda}\frac{\partial x^{\nu}}{\partial\tau}\frac{\partial x^{\kappa}}{\partial\varphi}\frac{\partial x^{\lambda}}{\partial\eta}\,d\tau\,d\varphi\,d\eta.
\end{equation}
One uses Milne coordinates and the corresponding natural frame (see
appendix \ref{=======Hyper-surfaces-and-Milne=======})
for all vector and tensor representations.

\begin{figure}[h]
\centering{} \includegraphics[scale=0.25]
{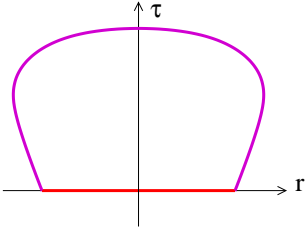}
\caption{Closed hypersurface $\Sigma_{\mathrm{ini}}$ (red) plus $\Sigma_{\mathrm{FO}}$
(magenta). \label{closed-hypersurface} }
\end{figure}

Cooper-Frye hadronization amounts to calculating the production of
particles with four-momenta $p^{\mu}$ according to 
\begin{equation}
E\frac{dn}{d^{3}p}=\int d\Sigma_{\mu}p^{\mu}f_{T}(u^{\nu}p_{\nu}),
\end{equation}
with $f_{T}(E)$ being ``thermal'' (grand canonical) distribution
at a given temperature. As sketched in Fig. \ref{Cooper-Frye-hadronization}(b),
the Cooper-Frye procedure amounts to considering particle flow through
FO hypersurface elements. Here one is going to proceed differently. Rather
than considering particle flow, one considers energy-momentum flow.
As sketched in Fig. \ref{Cooper-Frye-hadronization}(c), one defines
the flow of energy-momentum $dP$ and the flow of conserved charges
$dQ_{A}$ through the surface element $d\Sigma$ as 
\begin{eqnarray}
dP^{\mu} & \boldsymbol{=} & T^{\mu\nu}d\Sigma_{\nu},\\
dQ_{A} & \boldsymbol{=} & J_{A}^{\nu}d\Sigma_{\nu},
\end{eqnarray}
with $A\in\{C,B,S\}$, corresponding electric charge, baryon number
and strangeness. Momentum and charges are conserved, i.e., $\int_{\Sigma}dP^{\mu}=0$
and $\int_{\Sigma}dQ_{A}=0$ for a closed surface $\Sigma$, and as
a consequence, if one starts at some given proper time $\tau_{\mathrm{ini}}$,
one gets 
\begin{eqnarray}
\int_{\Sigma_{\mathrm{FO}}}dP^{\mu} & \boldsymbol{=} & P_{\mathrm{ini}}^{\mu},\\
\int_{\Sigma_{\mathrm{FO}}}dQ_{A} & \boldsymbol{=} & Q_{A\,\mathrm{ini}},
\end{eqnarray}
provided the hypersurface $\Sigma_{\mathrm{ini}}$ corresponding to
$t_{\mathrm{ini}}$ plus the FO hypersurface $\Sigma_{\mathrm{FO}}$
represent a closed hypersurface, as sketched in Fig. \ref{closed-hypersurface}.
This means, when integrating $dP^{\mu}$ and $dQ_{A}$ over the complete
FO hypersurface, one recovers completely the initial energy-momentum
and conserved charges, and one therefore considers $dP^{\mu}$ and $dQ_{A}$
as the ``basic objects'' which allow one to realize particle
production by respecting the conservation of energy-momentum and of
conserved charges.

\begin{figure}[h]
\centering{} \includegraphics[scale=0.4]
{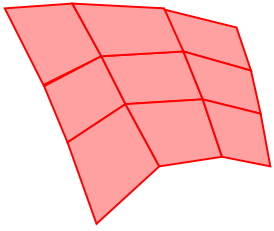}
\caption{Small hypersurface elements $\Delta\Sigma^{(K)}$ covering the whole
hypersurface $\Sigma_{\mathrm{FO}}$.\label{small-hypersurface-elements} }
\end{figure}

In practice, one discretizes the hypersurface, and consider small (but
finite) elements $\Delta\Sigma^{(K)}$ which cover the whole hypersurface
$\Sigma_{\mathrm{FO}}$, as sketched in Fig. \ref{small-hypersurface-elements},
 to be explained in the following. Before defining hypersurface and
hypersurface elements, the hydrodynamic evolution is completed, and
all relevant quantities (energy density, flow velocity, current of
conserved charges, ...) are known on a four-dimensional grid $(\tau_{i},\eta_{j},r_{k},\varphi_{n})$,
with constant step sizes $\Delta a=a_{i+1}-a_{i}$ for the four variables
$a\in\{\tau,\eta,r,\varphi\}$. 

The ranges for $r$ and $\eta$ are
adapted to the system size (PbPb is obviously bigger than pp) and
the energy (for LHC one needs larger $\eta$ ranges than for RHIC energies).
First the transverse  size $R$ is determined (based on the core distribution in space), event-by-event, 
such that the freeze-out surface will be covered, with $R$ from 3 fm (peripheral)  to 10 fm (central) for PbPb at 2.76 TeV, and with $R=2.4$ for pp at 7 TeV. 
Then a longitudinal width (in space-time rapidity) $W$ is determined, as $21.2\, r(s)$,
with $r(s)=y(s)/y(s_{\mathrm{ref}})$,
where $y(s)$ is the rapidity of the projectile nucleons in the nucleon-nucleon center-of-mass 
for a squared nucleon-nucleon energy $s$, and where $s_{\mathrm{ref}}=(2.76\mathrm{TeV)^{2}}$ is some reference value.
Then one defines: $\Delta r=R / 74$, $c\Delta\tau=2R/50 $, $\Delta\eta=W/26 $, $\Delta\varphi=2\pi/60$ (EPOS4.0.0 default).

The hypersurface defined by $\varepsilon(\tau,\eta,r,\varphi)=\varepsilon_{\mathrm{FO}}$
is computed based on the knowledge of the energy density on the grid
and is given in terms of four-vectors $x=x^{(i,j,n)}$ depending on
three indices $i$, $j$, $n$ as
\begin{align}
 & x^{0}=\tau_{i}\cosh\eta_{j},\:x^{1}=r_{i,j,n}\cos\varphi_{n},\label{milne-1}\\
 & x^{2}=r_{i,j,n}\sin\varphi_{n},\:x^{3}=\tau_{i}\sinh\eta_{j},\label{milne-2}
\end{align}
with $r_{i,j,n}$ being the root of $f(r)=\varepsilon(\tau_{i},\eta_{j},r,\varphi_{n})-\varepsilon_{\mathrm{FO}}$,
for fixed $i$, $j$, $n$ computed via linear interpolation from
the known values of $f(r_{k})$ and $f(r_{k+1})$, where $k$ has
been determined such that $f(r_{k})$ and $f(r_{k+1})$ have opposite
sign (and $[r_{k},r_{k+1}]$ contains the root). Knowing the hypersurface
given as $x=x^{(i,j,n)}$, one defines hypersurface elements $\Delta\Sigma^{(i,j,n)}$
as 
\begin{equation}
\Delta\Sigma_{\mu}^{(i,j,n)}=\varepsilon_{\mu\nu\kappa\lambda}\left\{ \frac{\partial x^{\nu}}{\partial\tau}\frac{\partial x^{\kappa}}{\partial\varphi}\frac{\partial x^{\lambda}}{\partial\eta}\right\} _{x=x^{(ijn)}}\,\Delta\tau\,\Delta\varphi\,\Delta\eta,
\end{equation}
where the partial derivatives are obtained from Eq. (\ref{milne})
and $\Delta\tau$, $\Delta\varphi$, and $\Delta\eta$ are the step
sizes of our grid. Having determined $r_{i,j,n}$ via interpolation,
one employs the same interpolation procedure to construct the energy-momentum
tensor $T^{\mu\nu}$and the current of conserved charges $J_{A}^{\nu}$
corresponding to $r_{i,j,n}$, from the known values of these quantities
at grid points $(\tau_{i},\eta_{j},r_{k},\varphi_{n})$ and $(\tau_{i},\eta_{j},r_{k+1},\varphi_{n})$
and use then for the interpolated values the symbols $T^{(i,j,n)\,\mu\nu}$
and $J_{A}^{(i,j,n)\,\nu}$. To simplify the following discussion,
one introduces an integer $K=K(i,j,n)$, being uniquely related to the
triplet $(i,j,n)$, so one uses the quantities $\Delta\Sigma_{\mu}^{(K)}$,
$T^{(K)\,\mu\nu}$ and $J_{A}^{(K)\,\nu}$. One then computes the energy-momentum
flow vector $\Delta P^{(K)\,\mu}=T^{(K)\,\mu\nu}\Delta\Sigma_{\nu}^{(K)}$
and the flow of conserved charges $\Delta Q_{A}^{(K)}=J_{A}^{(K)\,\nu}\Delta\Sigma_{\nu}^{(K)}$.

For each hypersurface element $\Delta\Sigma^{(K)}$ and its associated
flow vector $\Delta P^{(K)}$, one defines an invariant mass element
\begin{equation}
\Delta M^{(K)}=\sqrt{\Delta P^{(K)}\cdot\Delta P^{(K)}},
\end{equation}
where ``$\cdot$'' refers to a product of four-vectors. One also
defines an associated flow four-velocity $U$ as

\begin{equation}
U^{(K)}=\Delta P^{(K)}/\Delta M^{(K)}.
\end{equation}
The four-velocity $U^{\boldsymbol{\mu}}$ is NOT equal to the fluid
velocity $u^{\mu}$, this would only be true in case of zero pressure.

If one was only interested in global particle yields, one might sum
up all the masses to get the total effective invariant mass: 
\begin{equation}
M_{\mathrm{eff}}=\sum_{K=1}^{K_{\mathrm{max}}}\Delta M^{(K)},\label{meff}
\end{equation}
sum up the charges as
\begin{equation}
Q_{\mathrm{eff}\,A}=\sum_{K=1}^{K_{\mathrm{max}}}\Delta Q_{A}^{(K)},
\end{equation}
and decay this object according to microcanonical hadronization, as
discussed in the previous section. But in that case, one has no information
transverse momentum and rapidity dependencies of particle production.

\begin{figure}[h]
\centering{} \includegraphics[scale=0.4]
{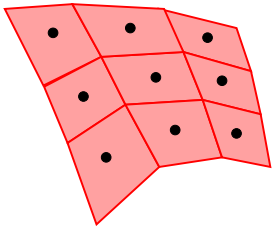}
\caption{For all hypersurface elements (their centers, black points), the coordinates
$(\tau,\eta,r,\varphi)$ are known.\label{hypersurface-element-coordinates} }
\end{figure}

But one can do better. One still computes $M_{\mathrm{eff}}$ and performs
the microcanonical decay, as discussed above. But one actually has
much more information than just the masses. One knows in addition for
each hypersurface element $\Delta\Sigma^{(K)}$ the associated four-velocity
$U^{(K)}$, representing the energy-momentum flow through the surface
element, and one knows all its coordinates, see Fig. \ref{hypersurface-element-coordinates}.
Since one counts the hypersurface elements $\Delta\Sigma^{(K)}$ as
$K=1,2,3...K_{\mathrm{max}}$, one may define a probability law $P$
on this set $\{1,2,3,...,K_{\mathrm{max}}\}$ as 
\begin{equation}
P(K)=\frac{\Delta M^{(K)}}{\sum_{K}\Delta M^{(K)}},
\end{equation}
representing the relative weight of a particular hypersurface element.
After having done the microcanonical decay, one may consider $P(K)$
as the weight of having produced a particle at a hypersurface position
given by the coordinates $\tau_{K},\varphi_{K},r_{K},\eta_{K}$, corresponding
to the hypersurface element $\Delta\Sigma^{(K)}$. In other words,
one generates the coordinates $\tau_{K},\varphi_{K},r_{K},\eta_{K}$
of the produced hadrons according to the law $P(K)$. In addition,
one may consider $P(K)$ also as the weight of having a four-velocity
$U^{(K)}$, and so one generates not only the position according to
$P(K)$, but also the four-velocity $U^{(K)}$, which one then uses
to Lorentz boost the particles into the lab frame (remembering that
the microcanonical decay is done in the center-of-mass frame of an
effective invariant mass).%
{} In other words, one gives back the flow, which has been taken out to
construct the effective invariant mass.

In the actual realization of the generation of a set of particles
from the Monte Carlo procedure (microcanonical decay plus subsequent
sampling based on $P(K)$ and the following boosts of the generated particles according to $U^{(K)}$)
one will not have 100\%
energy-momentum conservation, so one repeats 
the sampling based on $P(K)$ and the following boosts
several times, to take the ``best'' selection, giving an accuracy
on the 1\% level.

\begin{figure}[h]
\centering{} \includegraphics[bb=30bp 50bp 700bp 570bp,clip,scale=0.27]
{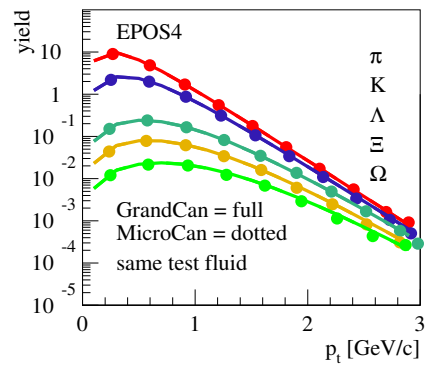}
\caption{Particle yields in $p_{t}$ bins of width 0.2 GeV/c, computed based
on ``microcanonical decay + boost'' (dotted lines) and based on
``grand canonical decay + boost'' (full lines), for the same test
fluid, see text.\label{check-grand-canonical} }
\end{figure}

To test the above procedure, one considers a realistic expanding fluid
and the corresponding hypersurface created in a single AuAu simulation
at 200 AGeV with an impact parameter of 8 fm, considering a limited
$\tau$ range (two steps). Transverse flow velocities on the freeze-out
surfacte reach up to 0.45$c$. The total effective mass is around
160 GeV, so this should correspond already to the infinite-volume
limit. One therefore computes particle yields in two ways: using the
above method of microcanonical decay plus Lorentz boosts according
to $U^{(K)}$, based on Monte Carlo simulations, and using the grand
canonical method by doing a three-dimensional (3D) momentum integration of Eq. (\ref{GraCan distribution}),
including as well Lorentz boosts according to $U^{(K)}$. The results
are shown in Fig. \ref{check-grand-canonical},
where I plot transverse momentum distributions for different hadron
species, from top to bottom: $\pi$, $K$, $\Lambda$, $\Xi$, and
$\Omega$ (since one is just interested in comparing two methods,
one computes yields in $p_{t}$ bins of width $0.2$ GeV/c, not divided
by bin size). As expected, one observes indeed the same result (which
is also a good check of the numerics, since the two methods are competely
independent of each other).

At very high energies, the above procedure may be problematic, since
the FO hypersurface may extend over a wide $\eta$ range, and it is
not obvious if it is correct to consider the whole surface as one
object with one single effective invariant mass. One therefore has
the possibility (and will use it) to split the surface into several
pieces corresponding to intervals in $\eta$ with width $\Delta\eta$,
as $[\eta_{i}-\Delta\eta/2,\eta_{i}+\Delta\eta/2]$, with $\eta_{i+1}-\eta_{i}=\Delta\eta$.
One may then do the sum as in Eq. (\ref{meff}), but just summing the
hypersurface elements within $I(\eta_{i})=[\eta_{i}-\Delta\eta/2,\eta_{i}+\Delta\eta/2]$,
i.e. 
\begin{equation}
\Delta M_{\mathrm{eff}}(\eta_{i})=\sum_{\eta(K)\in I(\eta_{i})}\Delta M^{(K)},\label{meff-1}
\end{equation}
[and correspondingly for $Q_{\mathrm{eff}\,A}(\eta_{i})$] where $\eta(K)$
is the $\eta$ value associated with $K$. Rather than a single effective
mass, one has now a discrete set of masses $\Delta M_{\mathrm{eff}}(\eta_{i})$,
which one may decay independently in their center-of-mass frames (in
the same way as dicussed for $M_{\mathrm{eff}}$. One also notes the
rapidities $y_{i}$ corresponding to $\Delta M_{\mathrm{eff}}(\eta_{i})$,
which allows one to finally boost the particles into the lab frame. It
should be noted that the numerical values of $y_{i}$ are very close
to $\eta_{i}$.

One has to do this splitting carefully. If the $\Delta\eta$ interval
is chosen to be very small, one may create artificially small masses,
with the corresponding suppression of heavy mass hadrons. In this
sense, the width $\Delta\eta$ is a physical parameter, measuring
the width where hypersurface elements are causally connected. I will
come back to this point later.

Let me finally address a technical issue. When constructing objects
for given $\eta$ intervals, after summing $\Delta Q_{A}^{(K)}$ elements
within this interval, one may end up with non-integer values, which
is modified by taking the next-closest integer instead. The violation
of the global conservation laws even at RHIC energies is small (less
than 1\%), so for the moment no correction is applied. This might
be changed in the future, but in particular at low energies there
are even more important issues to be considered (breakdown of the
parallel scattering scheme, first of all). But it is planned to reconsider
these issues in a future work dedicated to lower energies.

\subsection{Core hadronization in pp scattering {\Large{}{}\label{-------core-hadronization-in-pp-scattering-------}}}

I will apply the procedures discussed in the last section to investigate
proton-proton scattering at LHC energies (I will show results for
a center-of-mass energy of 7 TeV), and I will try to understand what
kind of effective masses are produced, and where they are produced
in space and time.

\begin{figure}[h]
\begin{centering}
(a) 2 Pomerons $\tau_{0}$$\qquad\qquad$(b) 2 Pomerons $\tau_{1}$\\
 \includegraphics[bb=0bp 0bp 567bp 520bp,clip,scale=0.21]
{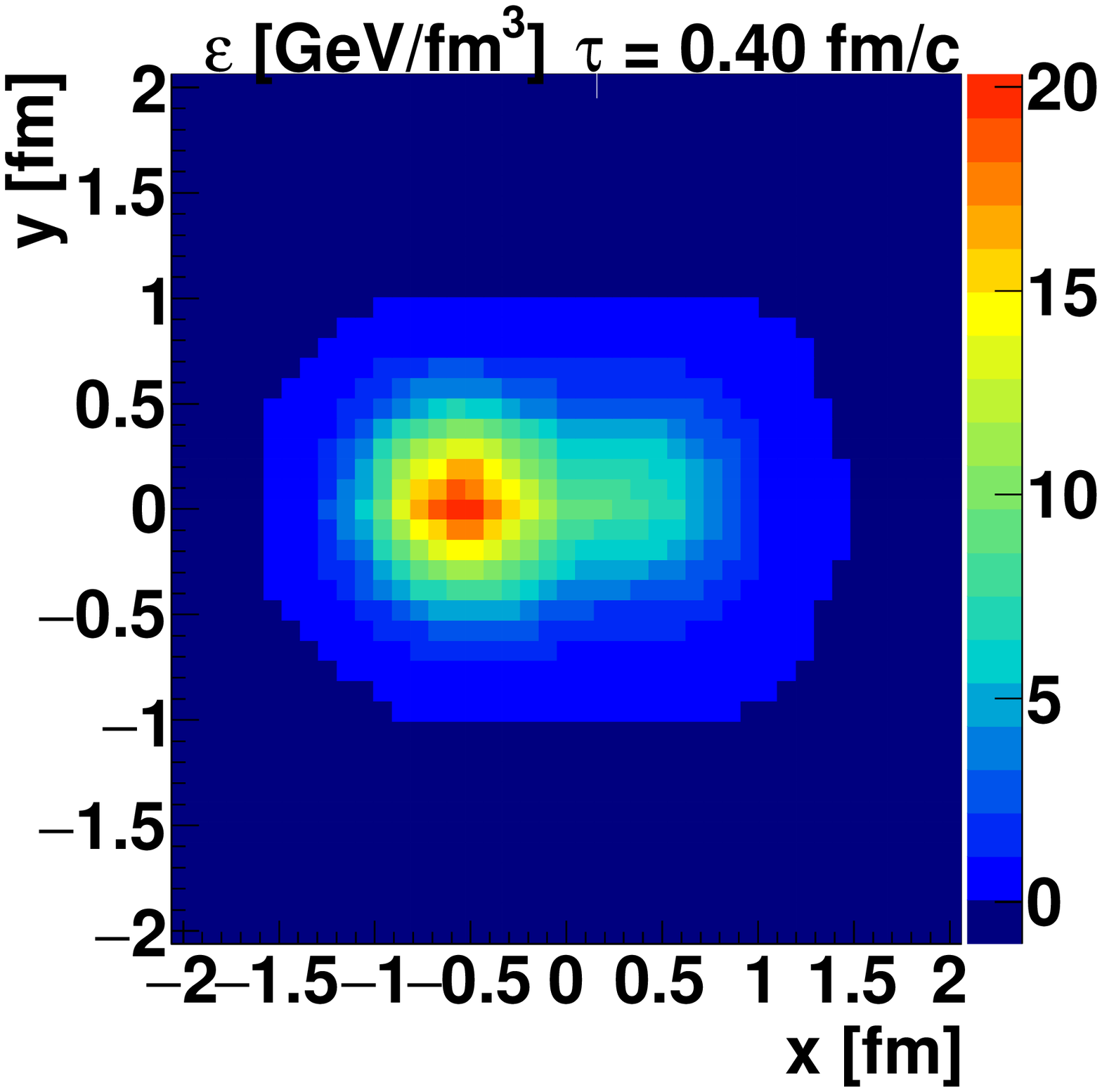}
\includegraphics[bb=0bp 0bp 567bp 520bp,clip,scale=0.21]
{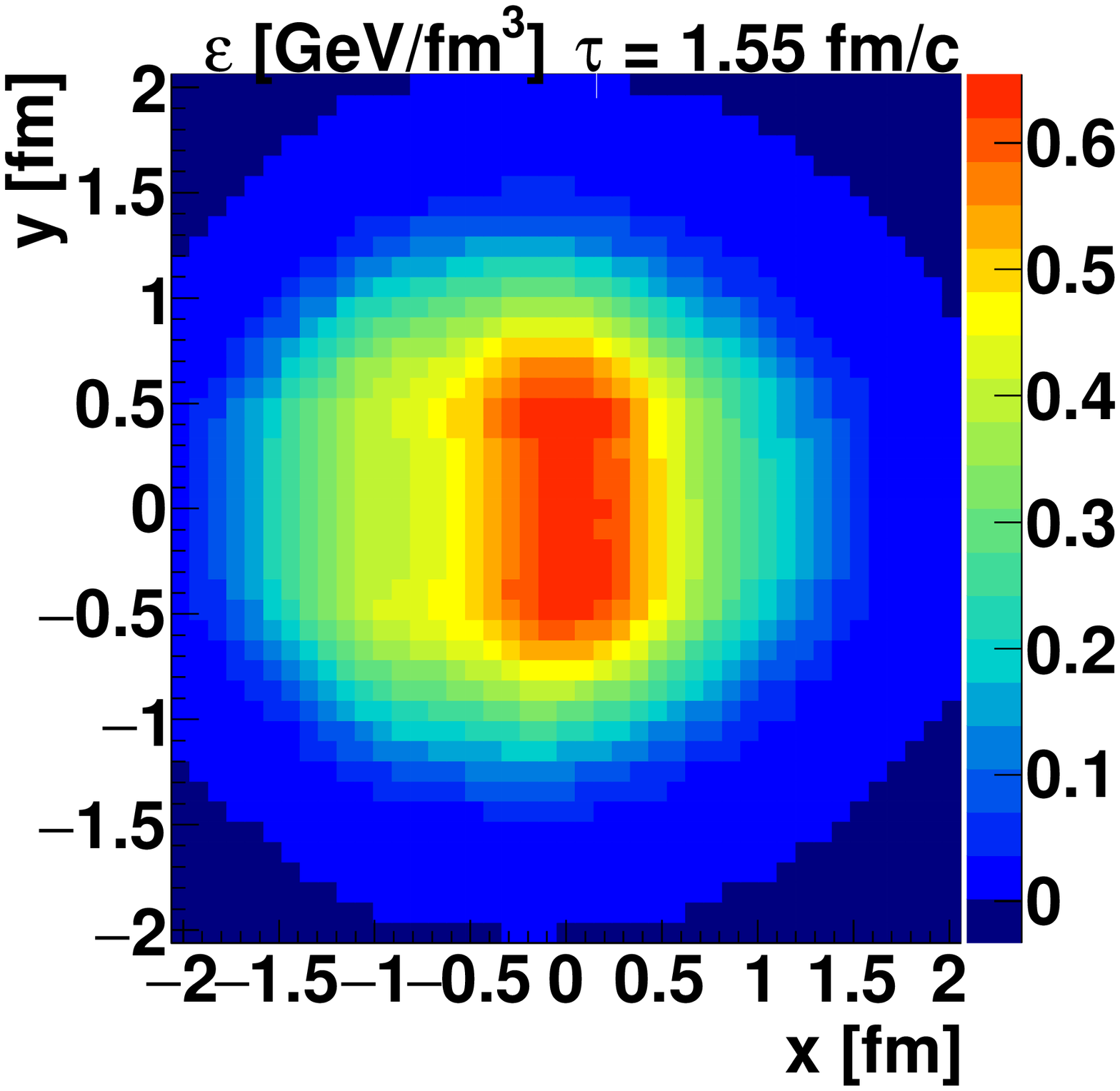}\\
 (c) 6 Pomerons $\tau_{0}$$\qquad\qquad$(d) 6 Pomerons $\tau_{1}$\\
 \includegraphics[bb=0bp 0bp 567bp 520bp,clip,scale=0.21]
{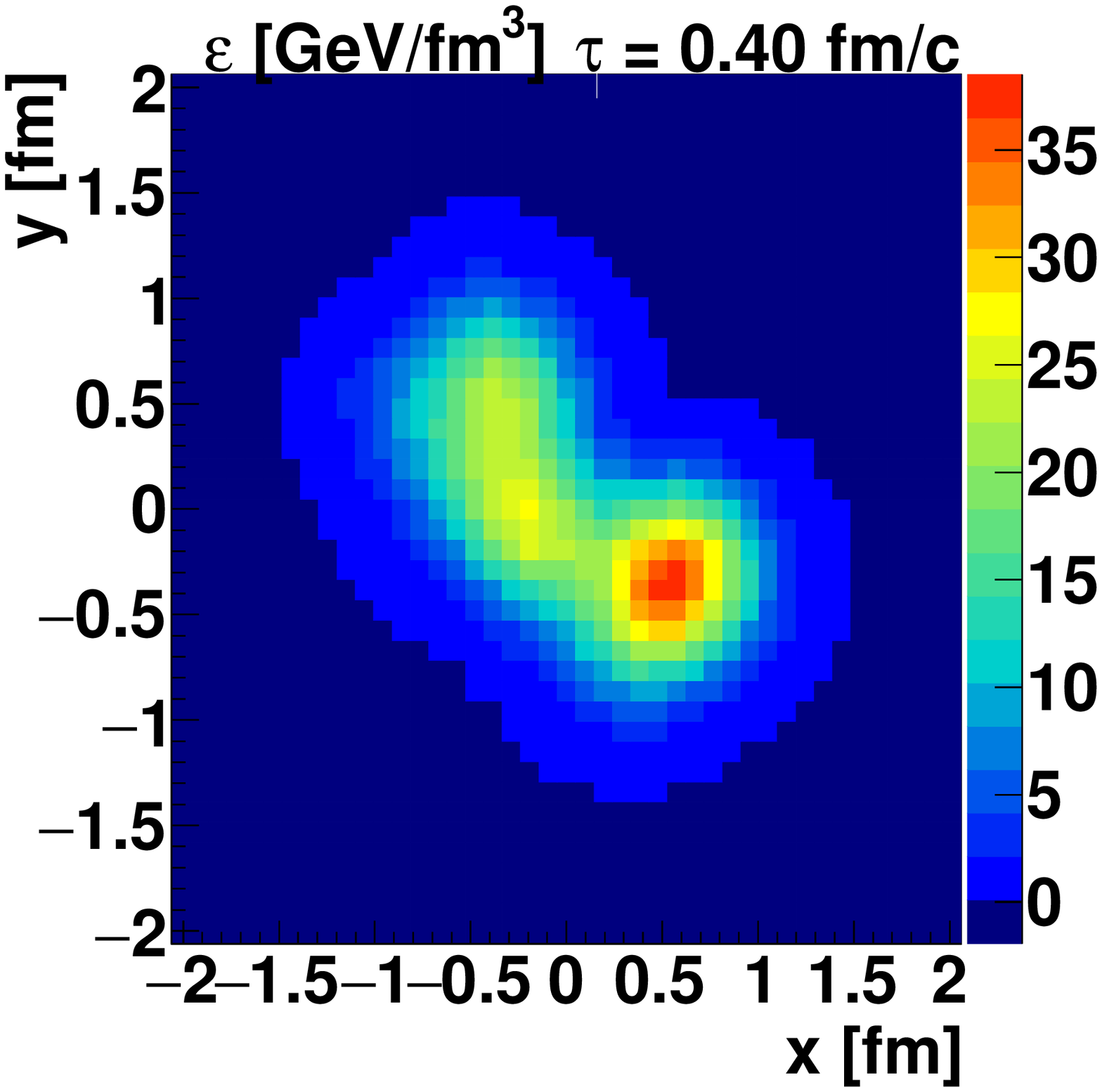}
\includegraphics[bb=0bp 0bp 567bp 520bp,clip,scale=0.21]
{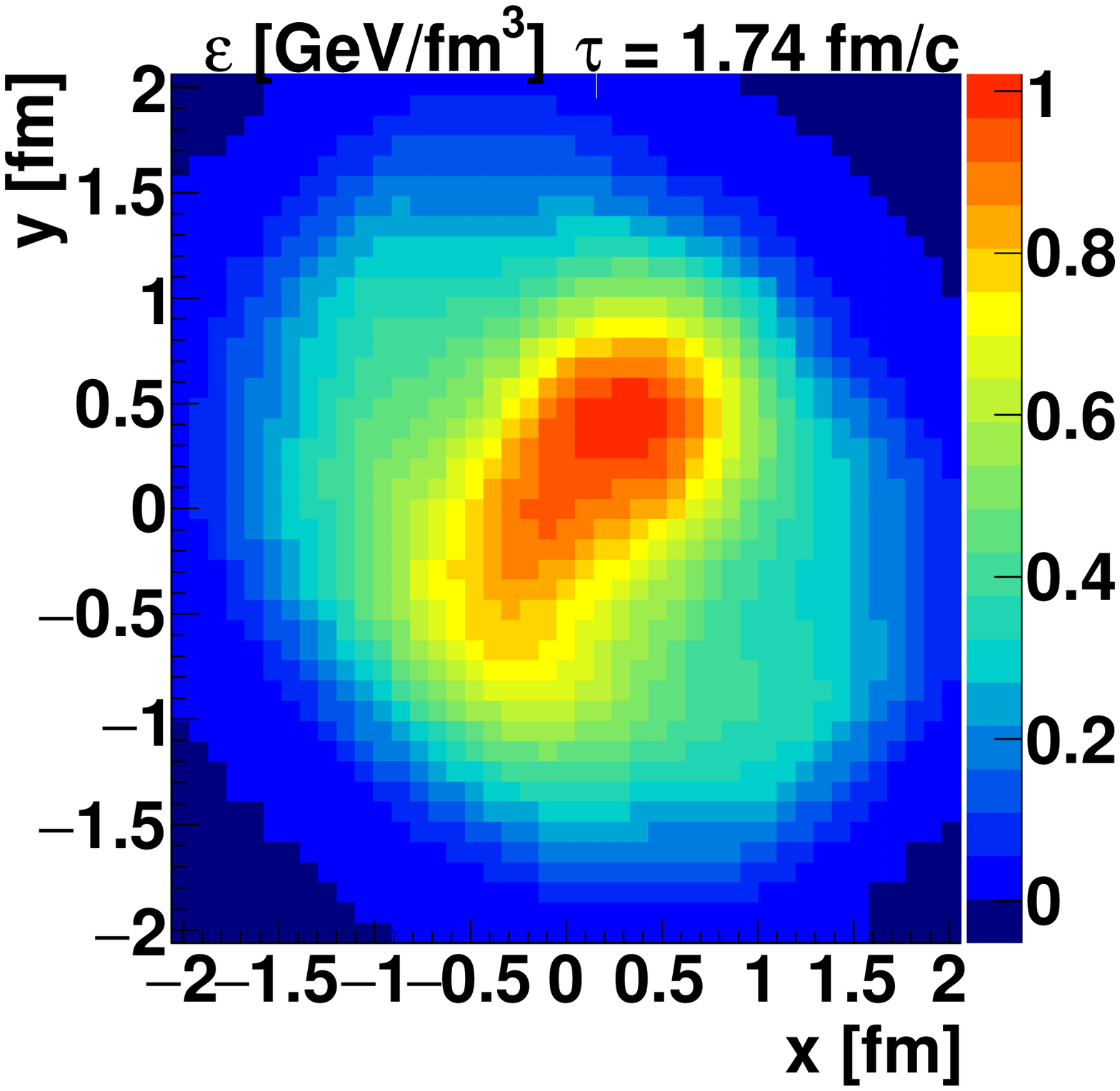}\\
 (e) 12 Pomerons $\tau_{0}$$\qquad\qquad$(f) 12 Pomerons $\tau_{1}$\\
 \includegraphics[bb=0bp 0bp 567bp 520bp,clip,scale=0.21]
{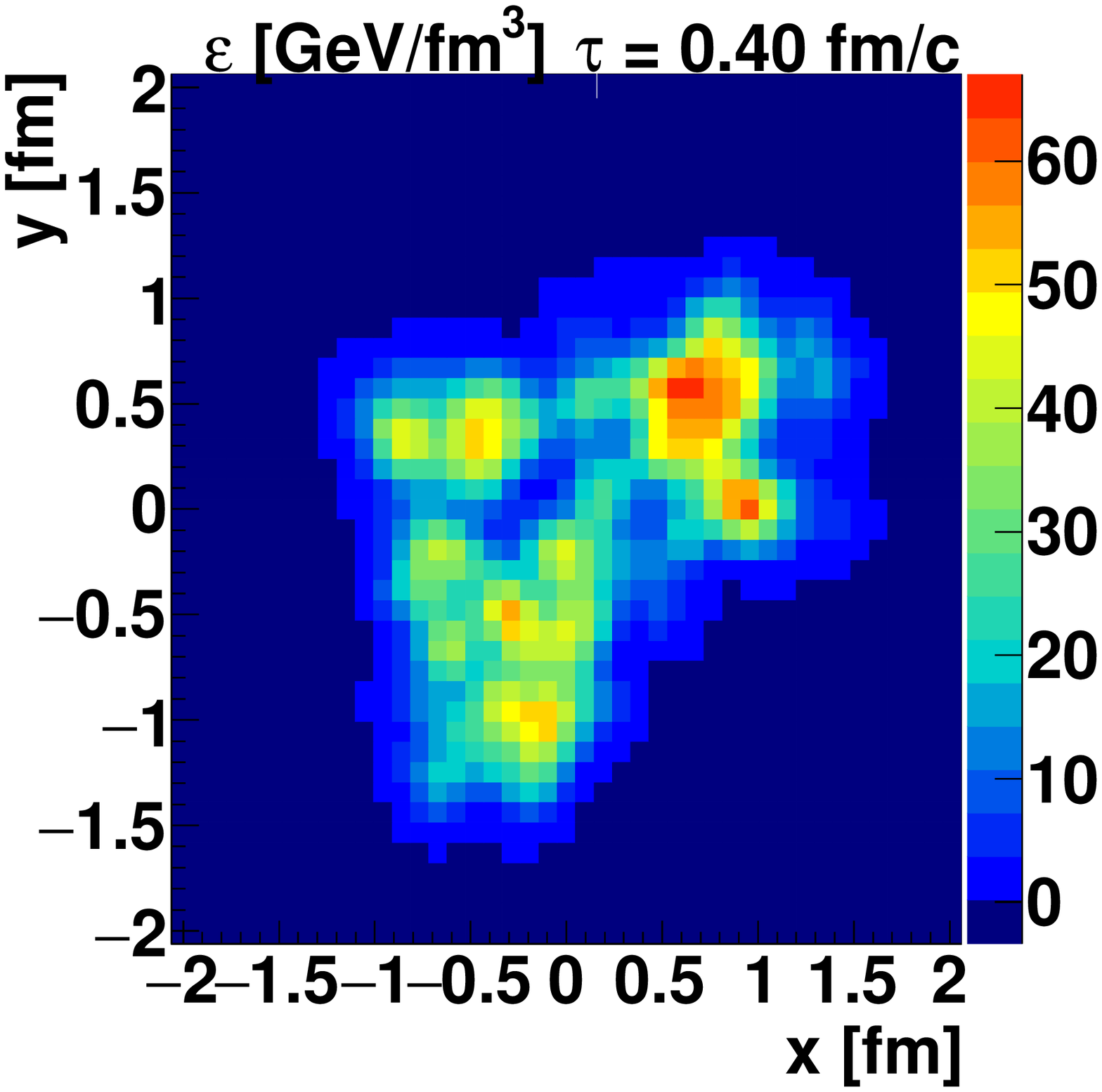}
\includegraphics[bb=0bp 0bp 567bp 520bp,clip,scale=0.21]
{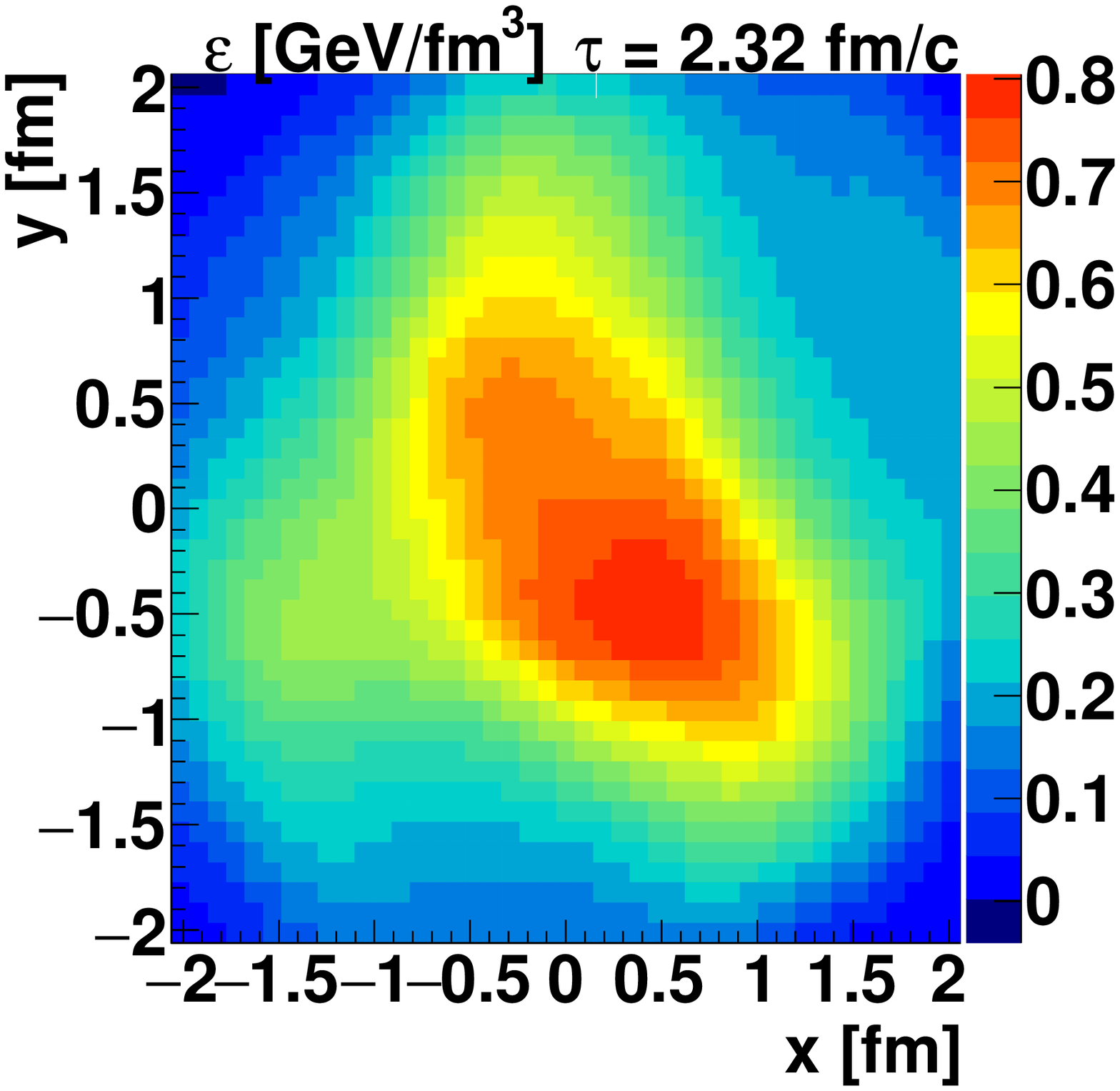} 
\par\end{centering}
\centering{} \caption{Energy density in the transverse plane ($x,y)$ for proton-proton
scattering involving (from top to bottom) 2, 6, and 12 Pomerons. The
left column represents the start time $\tau_{0}$ (of the hydro evolution),
and the right column a later time $\tau_{1}$, close to final freeze-out.
\label{Energy-density-2-6-12-Pomerons}}
\end{figure}

Let me consider (randomly chosen, but typical) proton-proton scattering
events involving 2 Pomerons, 6 Pomerons, and 12 Pomerons. In Fig.
\ref{Energy-density-2-6-12-Pomerons}, I plot for the three cases
the energy density in the transverse plane ($x,y)$. I consider in
each case two snapshots, namely at the start time of the hydro evolution
$\tau_{0}=0.40\,\mathrm{fm/c}$ (left column) and a later time $\tau_{1}$
(different values, $1.5-2.5\,\mathrm{fm/c}$) close to final freeze-out
(right column). In all cases, the initial distributions have elongated
shapes (just by accident, due to the random positions of interacting
partons). One can clearly see that the final distributions are as
well elongated, but perpendicular to the initial ones, as expected
in a hydrodynamical expansion.

\begin{figure}[h]
\begin{centering}
\includegraphics[scale=0.4]
{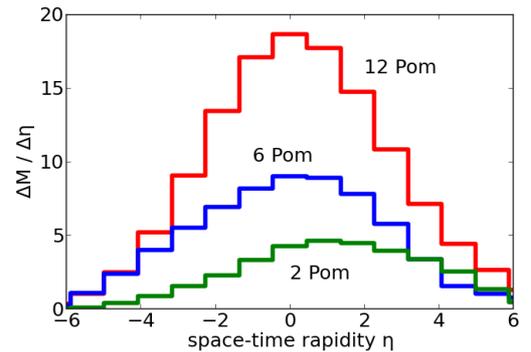} 
\par\end{centering}
\centering{} \caption{Mass distribution $\Delta M/\Delta\eta$ as a function of the space-time
rapidity $\eta$, for an event with 2 Pomerons, an event with 6 Pomerons,
and an event with 12 Pomerons. \label{fig: Mass distribution}}
\end{figure}

As explained in the last section, one computes effective mass elements
representing the momentum flow through surface elements. It is in
particular useful to split the $\eta$ range into intervals $[\eta_{i}-\Delta\eta/2,\eta_{i}+\Delta\eta/2]$,
with $\eta_{i+1}-\eta_{i}=\Delta\eta$, and sum up the corresponding
hypersurface elements to get a set of effective masses $\Delta M_{\mathrm{eff}}(\eta_{i})$.
In this way, one obtains an ``effective mass distribution'' $\Delta M/\Delta\eta$,
which one may plot as a function of $\eta$, as shown in Fig. \ref{fig: Mass distribution}
for the above-mentioned events with 2 Pomerons (corresponding roughly
to minimum bias), events with 6 Pomerons (roughly 3 times minimum
bias), and events with 12 Pomerons, (roughly 6 times minimum bias).
As expected, the $dM/d\eta$ values increase with increasing Pomeron
number, somewhat less than linear.

The total energy (of the core part) may be estimated as 
\begin{equation}
E=\sum\frac{\Delta M}{\Delta\eta}(\eta_{i})\,\cosh(\eta_{i})\Delta\eta,
\end{equation}
using the fact that $y_{i}$ is very close to $\eta_{i}$. One gets
for the three cases: 
\begin{equation}
E=\left\{ \begin{array}{cc}
573\,\mathrm{GeV} & \,\,\,(2\,\mathrm{Pom})\\
1843\,\mathrm{GeV} & \,\,\,(6\,\mathrm{Pom})\\
2225\,\mathrm{GeV} & \,\,\,(12\,\mathrm{Pom})
\end{array}\right.
\end{equation}
These numbers should be compared with the total available energy, which
is $7000$~GeV. This shows that the total energies of the core part
in all three cases represent only a relatively small fraction of the
total energy, but the observable effects are big! One may now compute
the total effective masses as 
\begin{equation}
M_{\mathrm{eff}}=\sum\frac{\Delta M}{\Delta\eta}(\eta_{i})\Delta\eta,
\end{equation}
and one gets for the three cases: 
\begin{equation}
M_{\mathrm{eff}}=\left\{ \begin{array}{cc}
30.6\,\mathrm{GeV} & \,\,\,(2\,\mathrm{Pom})\\
61.4\,\mathrm{GeV} & \,\,\,(6\,\mathrm{Pom})\\
115.4\,\mathrm{GeV} & \,\,\,(12\,\mathrm{Pom})
\end{array}\right.
\end{equation}
These numbers are much smaller than the energies $E$, the latter
ones containing the kinetic energy of the longitudinal expansion.
So the masses $M$ are the relevant quantities concerning the production
yields for the different particle species, and these masses will be used
in the micro-canonical decay procedures.

The effective masses $\Delta M_{\mathrm{eff}}(\eta_{i})$ are sums
of small hypersurface elements $\Delta\Sigma^{(K)}$, each one corresponding
to a set of coordinates $\tau,\eta,r,\varphi$, which allows one to plot
projections of these coordinates into the $\eta-\tau$ plane, as done
\begin{figure}[h]
\centering{}\includegraphics[bb=20bp 25bp 520bp 330bp,scale=0.4]
{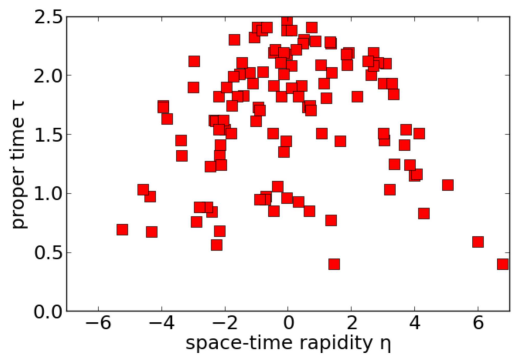}
\caption{Hypersurface element positions projected into the $\eta-\tau$ plane.\label{Hypersurface-element-positions} }
\end{figure}
in Fig. \ref{Hypersurface-element-positions} for our 12-Pomeron example.
This shows that the hypersurface is quite extended in space and time.
In general, hadronization at large space-time rapidities (corresponding
to large rapidities) happens early, the latest hadronization takes
place around $\eta=0$, the extension in time is around 2 fm/c, the
width in $\eta$ is around 10 units.

\begin{figure}[h]
\centering{}\includegraphics[scale=0.23]
{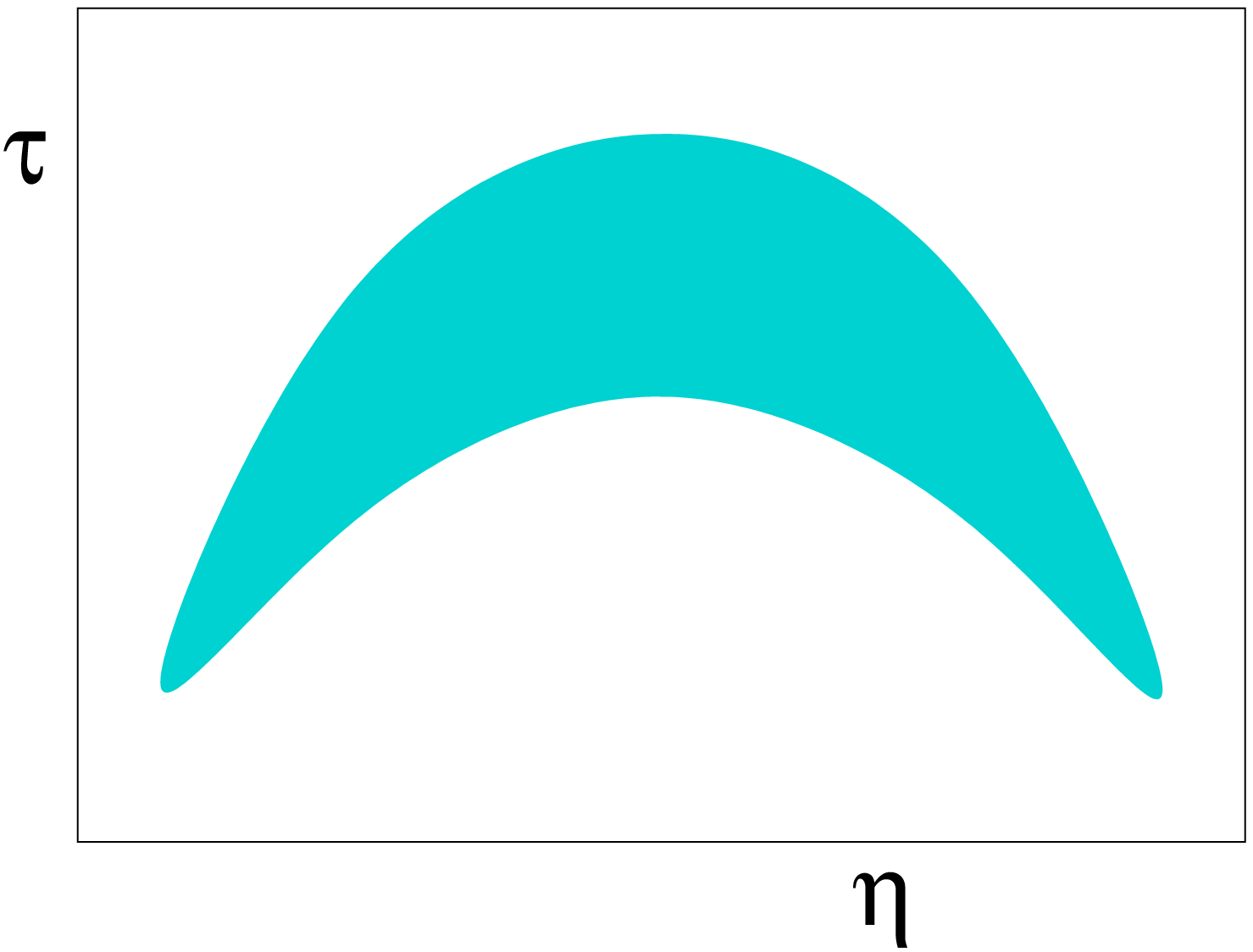}
\includegraphics[scale=0.23]
{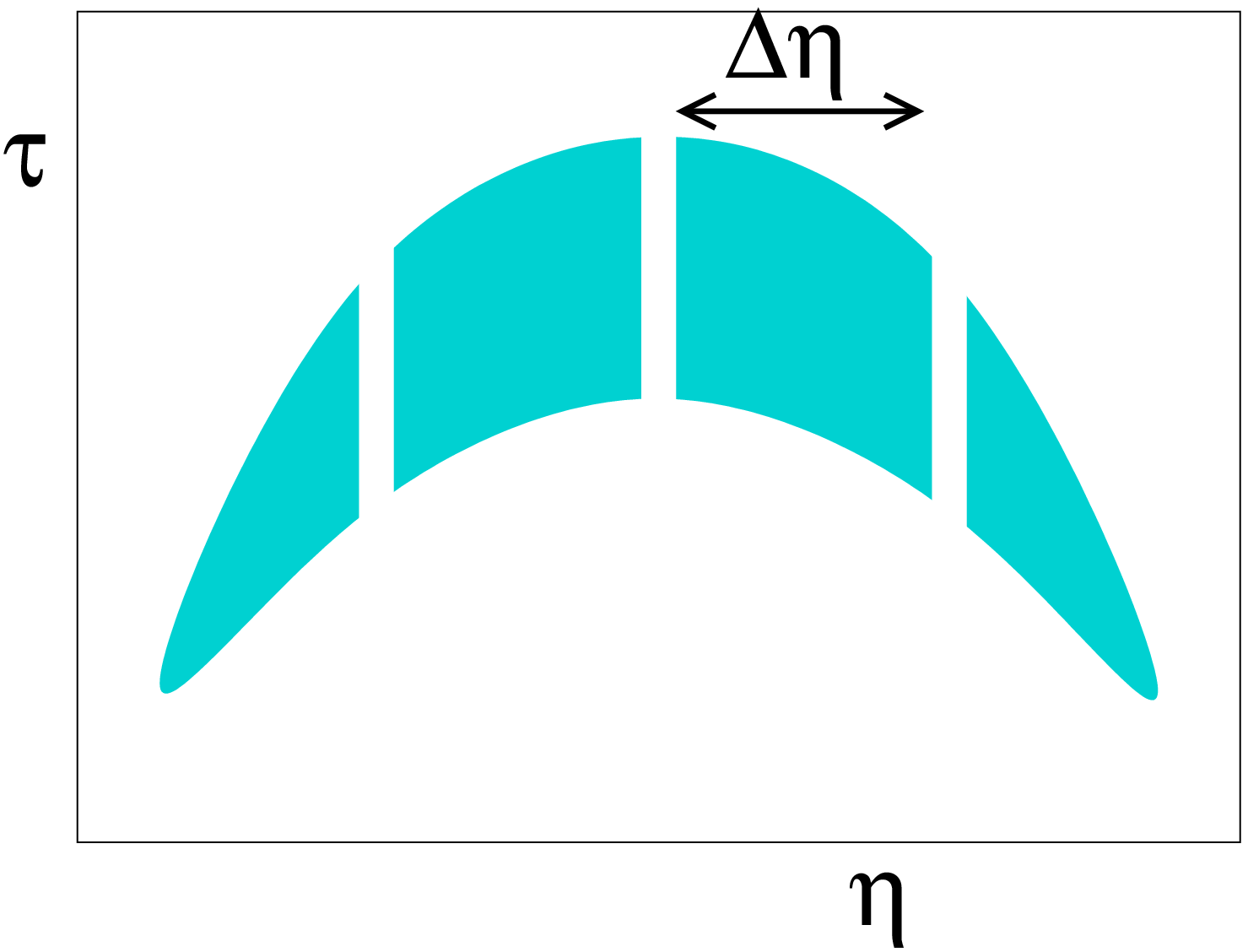} 
\caption{Splitting of the complete hypersurface (left) into pieces of width
$\Delta\eta$ (right).\label{splitting-of-the-complete} }
\end{figure}

In particular the important width in $\eta$ brings up (again) the
question of how to deal with the splitting of the complete hypersurface
into pieces of width $\Delta\eta$, as sketched in Fig. \ref{splitting-of-the-complete}.
For small systems (like pp scattering), the total effective masses
are not very big, in our three examples between 30 and 115 GeV, and
splitting these into small pieces will lead to very small effective
masses, with big effects concerning the production of heavy baryons.
As shown in Sec. \ref{-------Microcanonical-hadronization-of-plasma-droplets-------},
for masses at 25 GeV, one sees already deviations in the microcanonical
result compared with the grand canonical one. So the big question is:
does the system decay as single effective mass $M_{\mathrm{eff}}$
or as several independent objects of width $\Delta\eta$.

in Refs. \cite{Oliinychenko:2019,Oliinychenko:2020}, Oliinychenko et al.
presented a sampling method for the transition from relativistic hydrodynamics
to particle transport, which preserves the local event-by-event conservation
of energy, momentum, baryon number, strangeness, and electric charge,
for each sampled configuration in spatially compact regions (patches).
In our method, using finite $\Delta\eta$ amounts to cutting the hypersurface
into distinct regions, defined by $\eta$ ranges. In principle it
would be possible to further cut these regions into smaller pieces,
which indeed might be necessary to study local fluctuations. This
is an option for future development, but I do not follow this possibility
in this paper, since I mainly focus on the transition to small systems,
where already the regions defined by $\eta$ ranges are small.

\subsection{The role of the splitting parameter $\boldsymbol{\boldsymbol{\Delta\eta}}$
in core hadronization \label{-------the-role-of-the-splitting-------}}

Let me consider $\Delta\eta$ as a free parameter, in order to try
several choices of $\Delta\eta$ in the following, to learn how it
affects the results. One choice will be $\Delta\eta=\infty$, which
means one has just one single object as in Fig. \ref{splitting-of-the-complete}  (left).
In addition, one also investigates $\Delta\eta=4.5$ and $\Delta\eta=1.8$,
as in Fig. \ref{splitting-of-the-complete} (right).

One should keep in mind that the total width of the FO hypersurface
is around 10 fm/c, at LHC energies, see Fig. \ref{Hypersurface-element-positions},
and the total effective masses in pp scattering are between several
tens up to more than 100 GeV, depending on the Pomeron number.

\begin{figure}[h]
\centering{}\includegraphics[bb=20bp 60bp 642bp 570bp,clip,scale=0.33]
{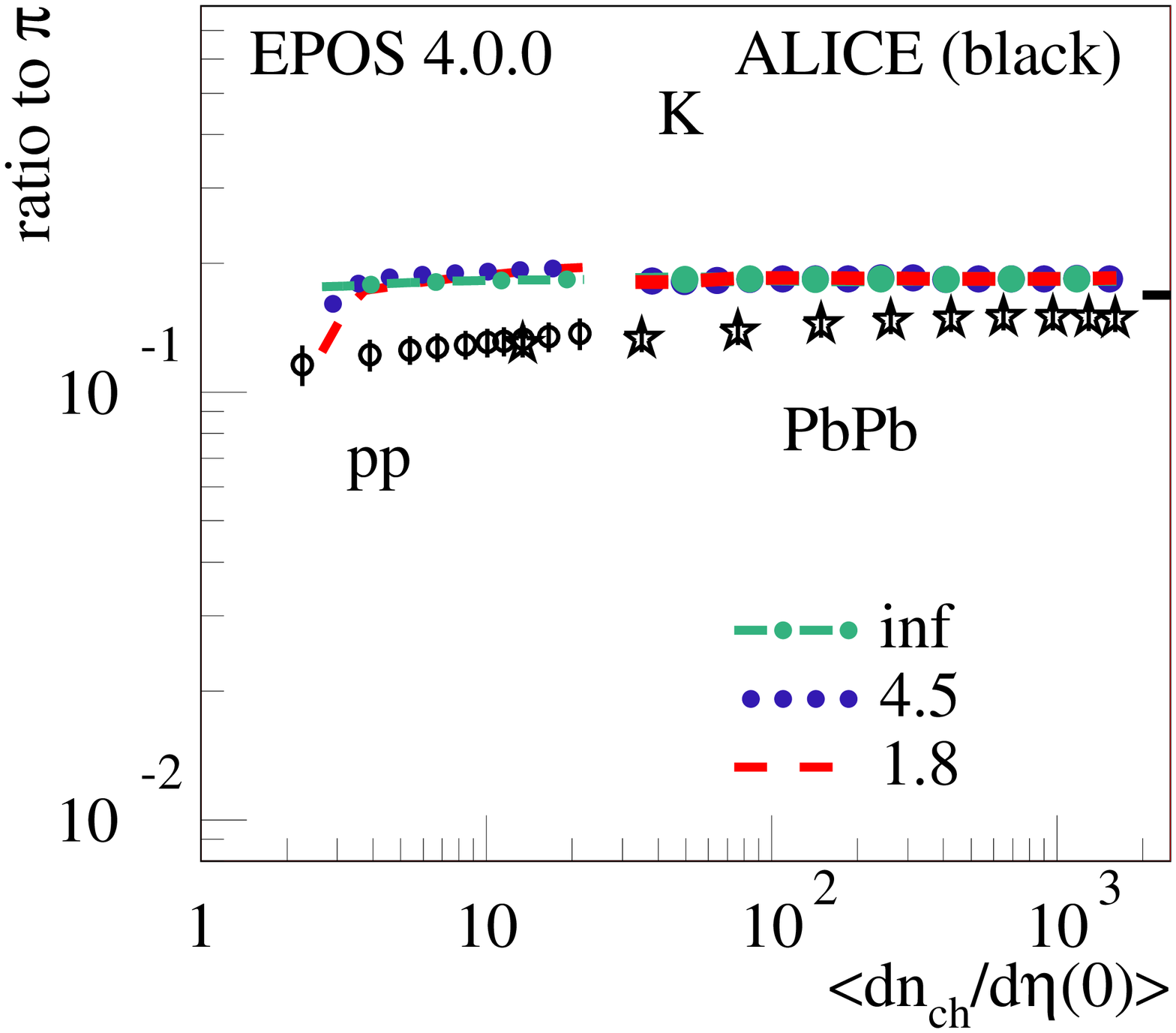}\\
 \includegraphics[bb=20bp 60bp 642bp 570bp,clip,scale=0.33]
{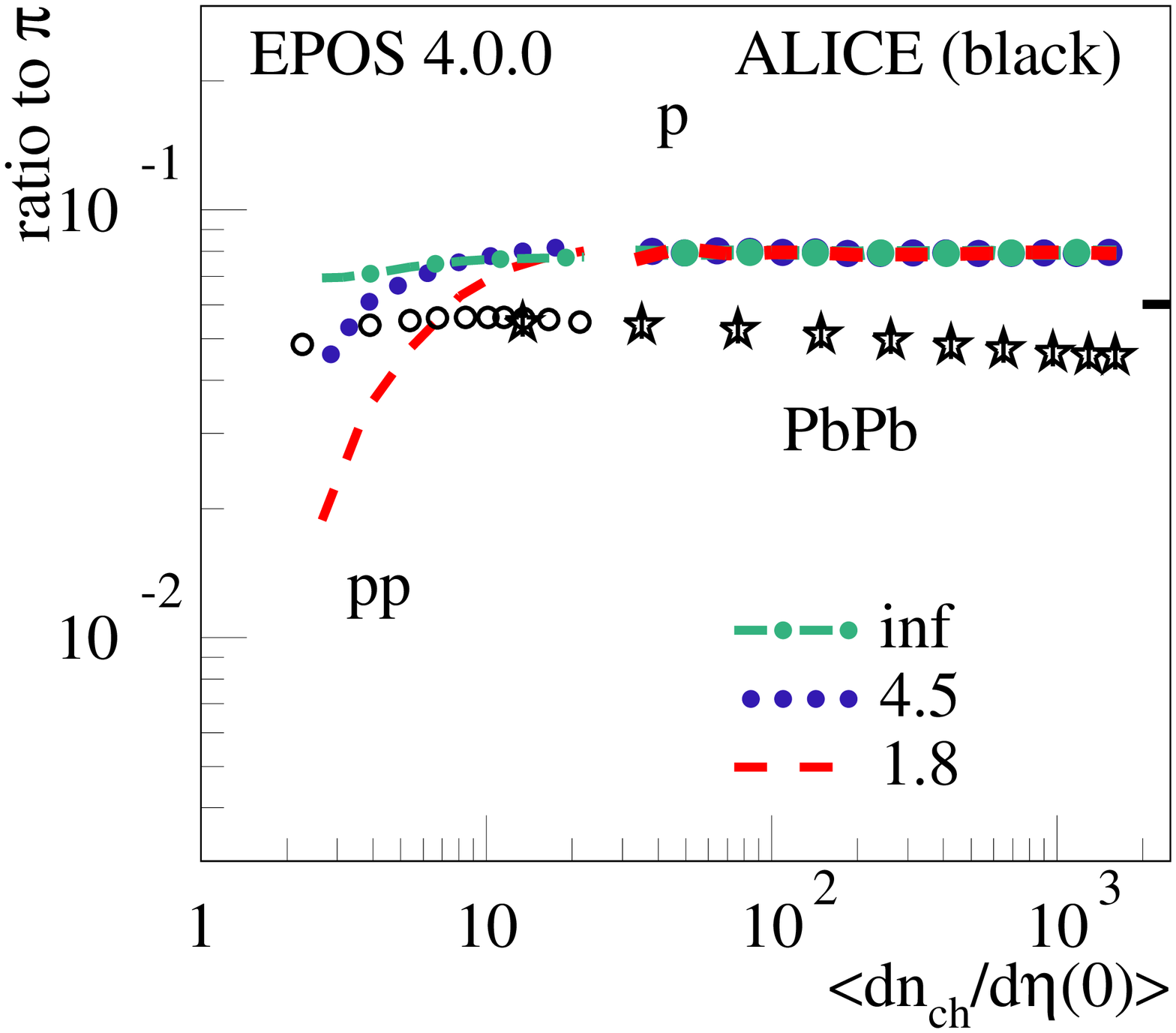}
\\
\includegraphics[bb=20bp 60bp 642bp 570bp,clip,scale=0.33]
{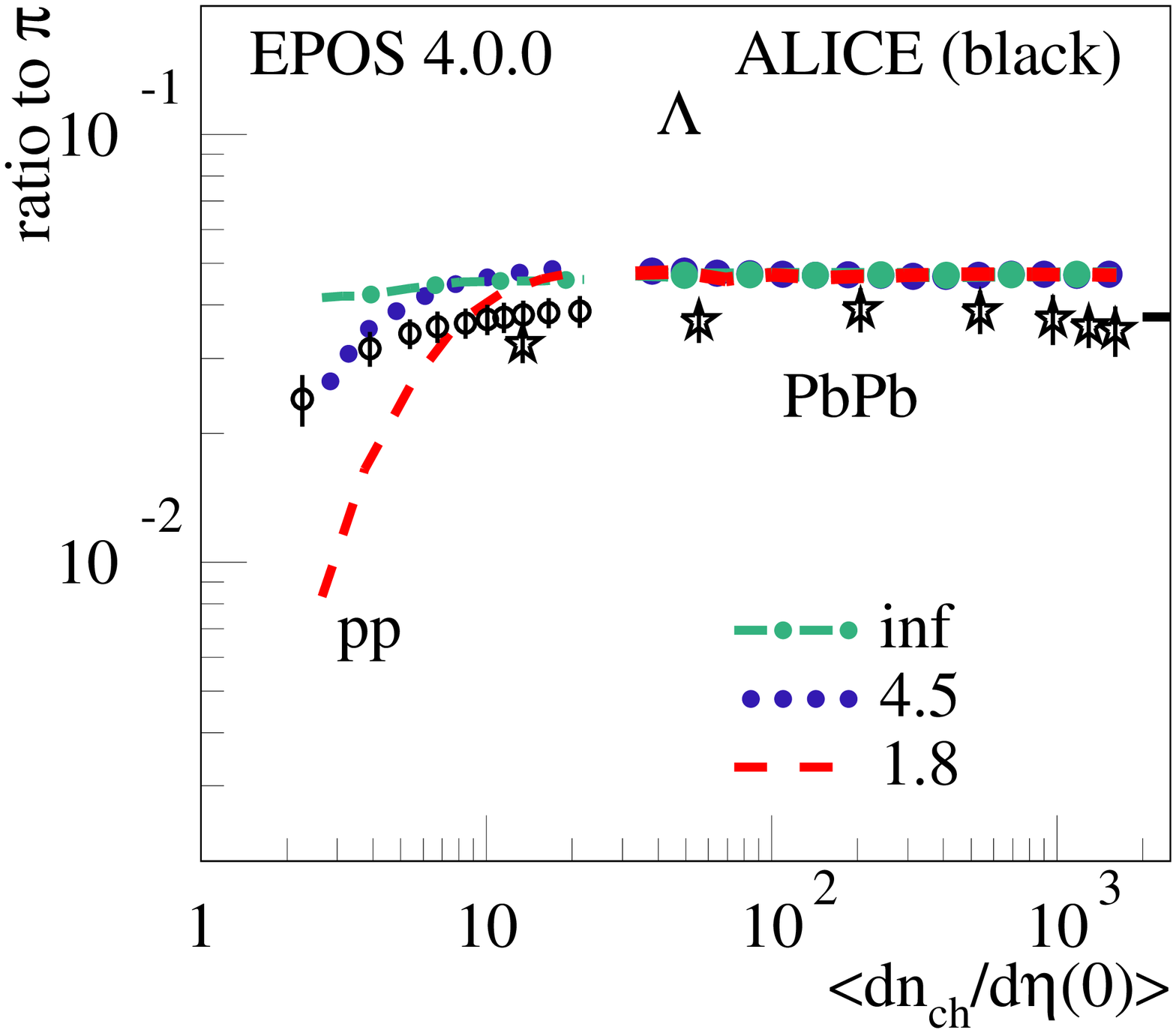}
\caption{Ratio $K/\pi$ (upper plot), $p/\pi$ (middle plot), and $\Lambda/\pi$
(lower plot), for pp at 7 TeV and PbPb at 2.76 TeV as a function of
the multiplicity $dn_{\mathrm{ch}}/d\eta(\eta=0)$, compared with ALICE
data.\label{fig: kaon to  pion ratio}}
\end{figure}

One considers only the core contributions for the moment, decayed according
to microcanonical hadronization, for pp and PbPb scattering. 
Particle ratios (with respect to pions always) are computed as a function of the
charged multiplicity $\left\langle dn_{\mathrm{ch}}/d\eta(0)\right\rangle $,
which allows one to put pp and PbPb results on the same plot. One considers
simulation results from EPOS4.0.0 for pp at 7 TeV and PbPb at 2.76
TeV, compared with data from ALICE \cite{ALICE:2013-PbPb-k-pi-p,ALICE:2013-PbPb-Ks-Lda,ALICE:2013-PbPb-Xi-Oga,ALICE:2015-pp-pi-K-p,ALICE:2016-pp-Ks-Lda-Xi-Oga}.
In Fig. \ref{fig: kaon to  pion ratio}, I show 
\begin{figure}[h]
\centering{} \includegraphics[bb=20bp 50bp 642bp 570bp,clip,scale=0.33]
{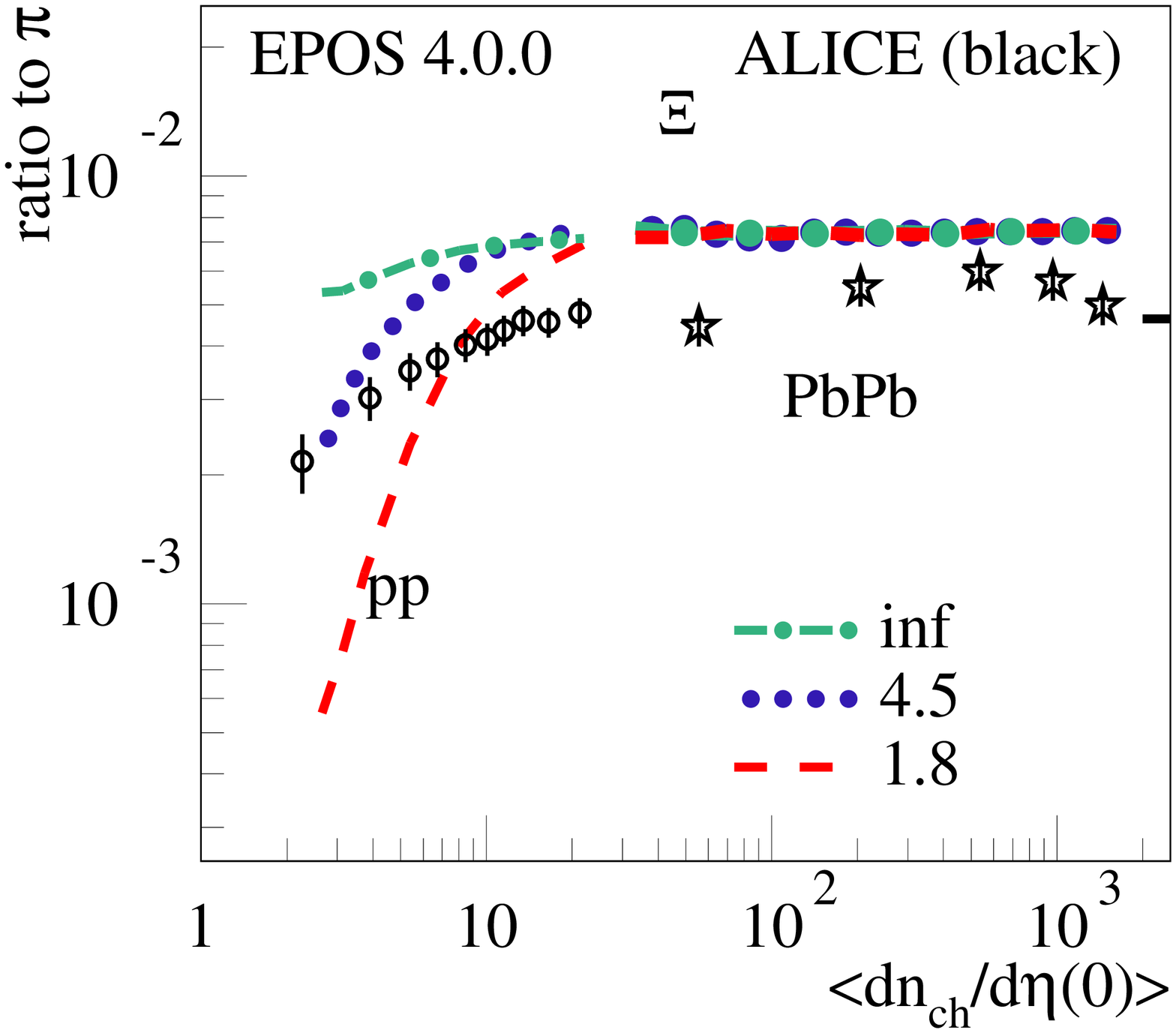}\\
 \includegraphics[bb=20bp 60bp 642bp 570bp,clip,scale=0.33]
{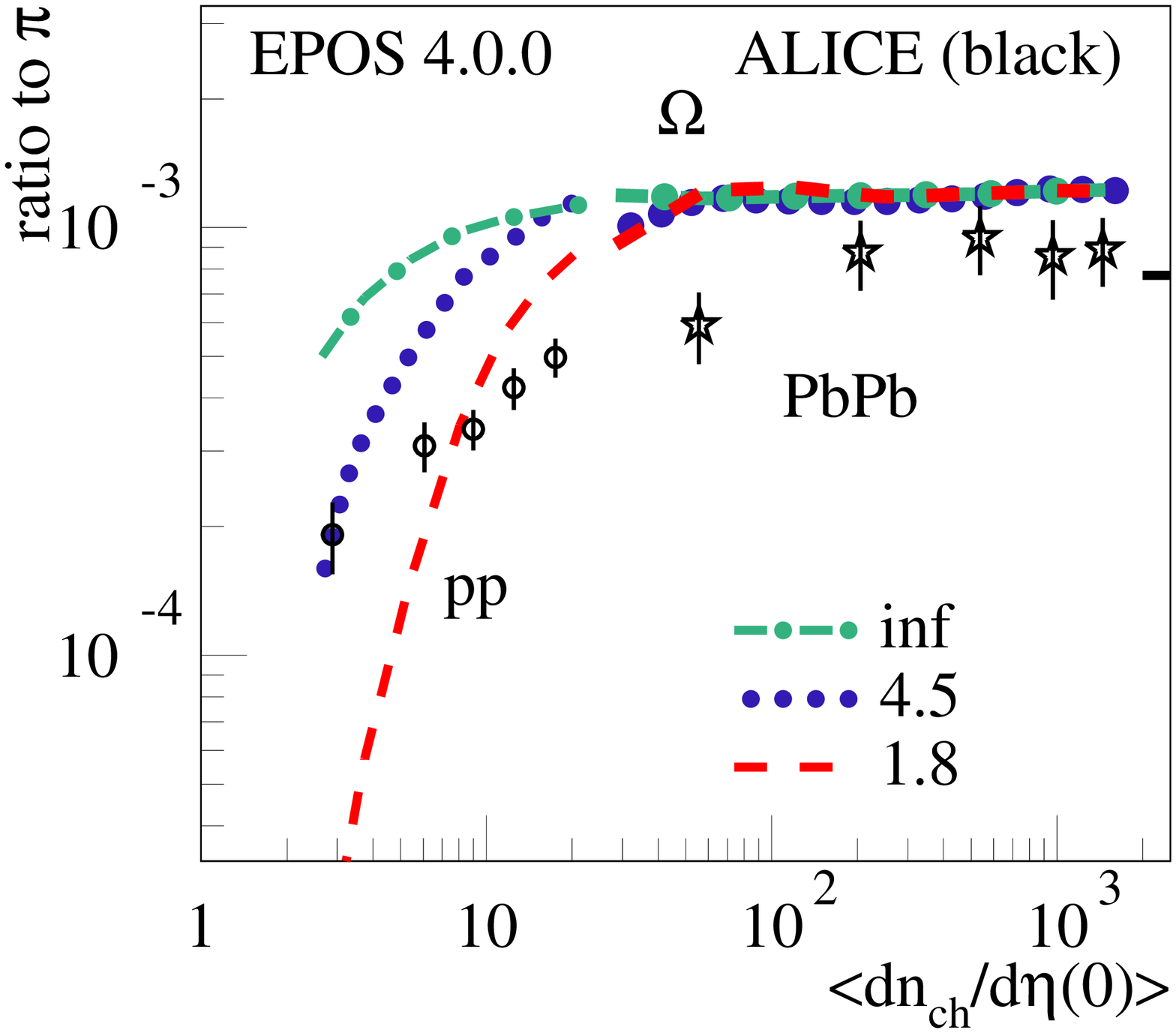}
\caption{Same as Fig. \ref{fig: kaon to  pion ratio}, but for $\Xi/\pi$ (upper
plot) and $\Omega/\pi$ (lower plot).\label{fig: lambda to pion ratio}}
\end{figure}
results for $K/\pi$ (upper plot), $p/\pi$ (middle plot), and $\Lambda/\pi$
(lower plot), and in Fig. \ref{fig: lambda to pion ratio} results
for $\Xi/\pi$ (upper plot) and $\Omega/\pi$ (lower plot). In all
cases, I show simulation results for $\Delta\eta=\infty$ (green
dashed-dotted lines), $\Delta\eta=4.5$ (blue dotted lines) and $\Delta\eta=1.8$
(red dahsed lines). For the simulation results, thin line style is
used for pp, and thick lines for PbPb, whereas for the data, circles
are used for pp and stars for PbPb. The short black horizontal lines
on the right side of the plots correspond to predictions from the
``thermal model'' \cite{Andronic}.

First of all, one sees that all the curves show a continuous behavior,
when passing from pp to PbPb. Concerning the PbPb results, there is
no difference between the the three $\Delta\eta$ choices, with the
only exception of $\Omega/\pi$ ratios, where the $\Delta\eta=\infty$
gives a flat ratio, whereas for $\Delta\eta=4.5$ and $\Delta\eta=1.8$
the curves drop slightly on the left end (peripheral collisions).
The situation is quite different for the pp curves. In all cases,
the curves drop when going to smaller multiplicities, and this drop
is more and more pronounced with decreasing $\Delta\eta$, and comparing
the curves with fixed $\Delta\eta$, the drop is increasing with particle
mass: the biggest effect is seen for $\Omega/\pi$ ratios. One should
also keep in mind that grand canonical particle production (what has been 
used in earlier EPOS version) leads to flat curves, the same for pp
and PbPb. So one sees a big effect of microcanonical hadronization compared
to the grand canonical one, and the deviation depends strongly on
$\Delta\eta$.

\begin{figure}[h]
\centering{}\includegraphics[bb=30bp 60bp 700bp 570bp,clip,scale=0.30]
{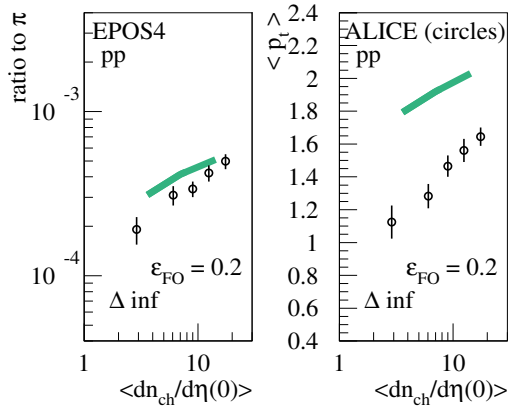}\\
 \caption{Ratio $\Omega/\pi$ (left plot) and average $p_{t}$ (right plot)
for pp at 7 TeV, for $\Delta\eta=\infty$ and $\varepsilon_{\mathrm{FO}}=0.2$.
EPOS simulations are compared with data from ALICE \cite{ALICE:2016-pp-Ks-Lda-Xi-Oga}.\label{epsilon-FO-0.20}}
\end{figure}

Concerning the comparison with experimental data, one sees that none
of the choices of $\Delta\eta$ can explain the data: for big values
of $\Delta\eta$, one stays clearly above the data, and for small
values, the ratios do drop sufficiently, but too fast. So it is clear
that even a ``tuning'' of $\Delta\eta$ will not help to reproduce
the data \textendash{} for a pure core approach. I will investigate
the full core-corona approach in the next section.

As a side remark: it is of course possible to reduce baryon production
by using a smaller value of the freeze-out energy density $\varepsilon_{\mathrm{FO}}$,
as shown in Fig. \ref{epsilon-FO-0.20} (left), where I plot the ratio
$\Omega/\pi$ for a pp simulation at 7 TeV, using $\Delta\eta=\infty$
and $\varepsilon_{\mathrm{FO}}=0.2$. But as seen in \ref{epsilon-FO-0.20} (right),
the average $p_{t}$ is much too high. Similar results are obtained
for other baryons. So it seems impossible to reproduce in this way
ratios and other important observables at the same time, as it will
be possble in the core-corona approach, where in addition, one 
employs a unique value of $\varepsilon_{\mathrm{FO}}$ for all systems.

\subsection{Results for the full core-corona picture in pp and PbPb scattering
\label{-------Core-corona-picture-------}}

In the following, results concerning the full core-corona
approach will be discussed. Let me briefly recall the procedure: as explained in detail
in Sec. \ref{=======the-role-of-core=======},
primary interactions lead to the production of prehadrons, a core-corona
procedure separates the core and corona prehadrons. The former constitute
the core, the latter are considered to be hadrons. The core is meant
to be matter, which evolves according to viscous hydrodynamics, which
allows one to compute the space-time dependence of the energy-momentum
tensor $T^{\mu\nu}$ (expressed in terms of energy density $\varepsilon$,
the flow vector $u^{\mu}$, the shear stress tensor, and bulk pressure)
and of the vector $J_{A}^{\mu}$ representing the current of conserved
quantities $A$.

As shown in Sec. \ref{-------Flow-through-hypersurface-elements-------},
based on the energy density $\varepsilon$, one defines a freeze-out
(FO) hypersurface via $\varepsilon(\tau,\eta,r,\varphi)=\varepsilon_{\mathrm{FO}}$,
which allows defining (after discretization) small hypersurface elements
$\Delta\Sigma_{\mu}$. Together with the knowledge of $T^{\mu\nu}$,
this allows one to compute energy-momentum flow four-vectors $\Delta P$
and invariant mass elements $\Delta M=\sqrt{\Delta P\cdot\Delta P}$,
which may be summed up to get effective invariant masses $\Delta M_{\mathrm{eff}}(\eta_{i})$,
covering space-time rapidity intervals of width $\Delta\eta$, which
are decayed microcanonically (this is called ``hadronization'').

As discussed in the previous section, the outcome of core hadronization
depends on $\Delta\eta$. I use in the following $\Delta\eta=\infty$,
which seems to be ``the best'' value when comparing with data, in
the sense of a comparison with very large set of data concerning very
different observables, for many different systems and energies, much
beyond what is shown in this paper. The aim of this work is not to
accommodate local fluctuations, in that case one needed in particular
for heavy ion collisions to introduce ``local patches'' and not
only finite $\Delta\eta$, which would be an easy extension of the
present work. The main focus here is the transition from big to small
systems, and for small systems the value of $\Delta\eta$ is important,
and (from our findings) the best global fit is obtained for $\Delta\eta=\infty$,
which is simply an empirical fact, somewhat counterintuitive. Anyway,
$\Delta\eta$ is one of the EPOS parameters the user may change.

After the hadronization of the fluid, the created hadrons as well
as the corona prehadrons (having been promoted to hadrons) may still
interact via hadronic scatterings, and here one uses UrQMD \cite{Bas98,Ble99}.

In the following, I want to study core and corona contributions to
hadrons production, for pp and PbPb collisons at LHC energies. I
will distinguish: 
\begin{description}
\item [{(A)}] The ``\textbf{core+corona}'' contribution: primary interactions
(S-matrix approach for parallel scatterings), plus core-corona separation,
hydrodynamic evolution, and microcanonical hadronization, but without
hadronic rescattering. 
\item [{(B)}] The ``\textbf{core}'' contribution: like (A), but considering
only core particles. 
\item [{(C)}] The ``\textbf{corona}'' contribution: like (A), but considering
only corona particles. 
\item [{(D)}] The ``\textbf{full}'' EPOS4 scheme: like (A), but in addition
hadronic rescattering. 
\end{description}
\noindent In cases (A), (B), and (C), one needs to exclude the hadronic
afterburner, because the latter affects both core and corona particles,
so in the full approach, the core and corona contributions are not
visible anymore.

\begin{figure}[h]
\centering{}\includegraphics[bb=20bp 20bp 700bp 600bp,clip,scale=0.33]
{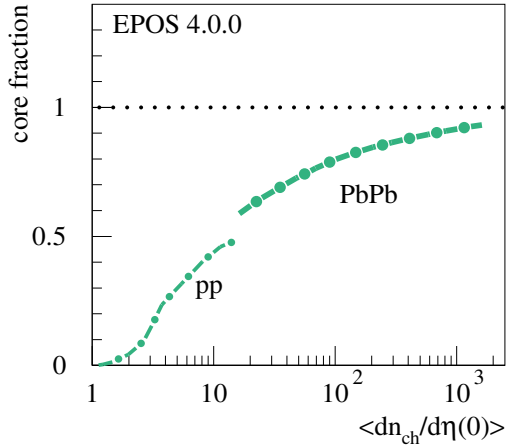}
\caption{The core fraction [core / (core+corona)] for pp at 7 TeV and PbPb
at 2.76 TeV as a function of the multiplicity $dn_{\mathrm{ch}}/d\eta(\eta=0)$.\label{fig: Core fraction}}
\end{figure}

In Fig. \ref{fig: Core fraction},
I show the core fraction [core / (core+corona)] in pp at 7 TeV and
PbPb at 2.76 TeV as a function of the multiplicity $dn_{\mathrm{ch}}/d\eta(\eta=0)$,
being defined as pion yield from core over core + corona contribution
(since pions are by far the most frequent species).

I will take core+corona as a reference, and plot ratios X /
core+corona versus $p_{t}$, with X being the corona contribution,
the core , and the full contribution, for four event classes
and four different particle species.

\begin{figure}[h]
\centering{}\includegraphics[bb=30bp 30bp 570bp 620bp,clip,scale=0.47]
{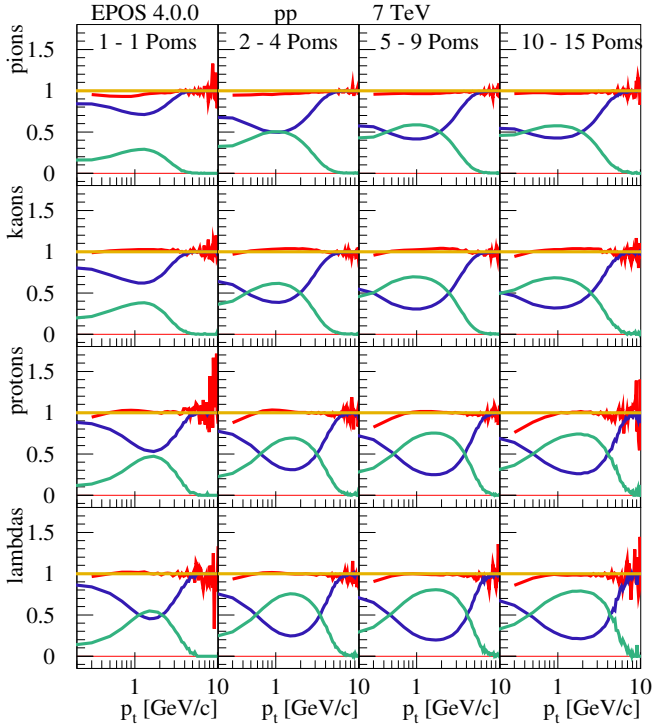}
\caption{The X / core+corona ratio, with X being the corona contribution
(blue), the core (green), and the full contribution (red),
for four event classes and four different particle species, for pp at
7 TeV.\label{core-corona-1}}
\end{figure}

In Fig. \ref{core-corona-1}, I show results for pp collisions at
7 TeV, for (from top to bottom) pions ($\pi^{\pm}$), kaons ($K^{\pm}$),
protons ($p$ and $\bar{p}$), and lambdas ($\Lambda$ and $\bar{\Lambda}$),
which correspond to hadrons with increasing masses.
\begin{figure}[h]
\centering{}\includegraphics[bb=30bp 30bp 570bp 620bp,clip,scale=0.47]
{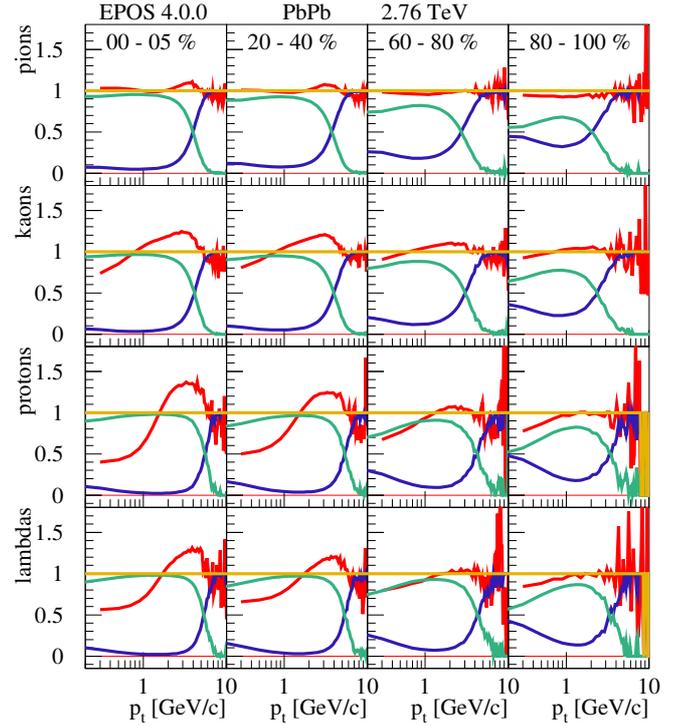}
\caption{The X / core+corona ratio, with X being the corona contribution
(blue), the core (green), and the full contribution (red),
for four centrality classes and four different particle species, for PbPb at 2.76 TeV.\label{core-corona-1-1}}
\end{figure}
The four columns represent four different event classes defined in
terms of the number $N_{\mathrm{Pom}}$ of Pomerons (from left to
right): 1, 2-4, 5-9, 10-15, which may be compared with the average number
of Pomerons in minimum bias pp at 7 TeV of around 2.

Looking at the green (core) and blue (corona) curves,
one observes that the core contribution increases with $N_{\mathrm{Pom}}$,
but it also increases with the hadron mass (from top to bottom). The
biggest core contribution is observed for lambdas for events with
10-15 Pomerons. There is an interesting $p_{t}$ dependence. In all
cases, the maximum of the green core curves is around 1-2$\,\mathrm{GeV/c}$,
with bigger values for the heavier particles. This is clearly a flow
effect, the core particles are produced from a radially expanding
fluid, which moves particles to higher $p_{t}$ values, as compared
with a decaying static droplet.

Whereas the core contribution goes down toward small $N_{\mathrm{Pom}}$,
one still observes a non-vanishing contribution even for $N_{\mathrm{Pom}}=1$,
which means even a very small number of strings may create a core
and flow effects. And this is actually needed to describe experimental
data, as I discuss later. It is often said that ``collective
effects'' show up in high multiplicity pp events, but it seems that
flow effects are present everywhere.

The red curves represent full over core+corona, the
difference between the two is the effect of the hadronic cascade in
the full case. Here, one sees only a small effect, essentially
some baryon-antibaryon annihilation, which suppresses the baryon yield
at small $p_{t}$.

In Fig. \ref{core-corona-1-1}, I show ratios X / core+corona versus
$p_{t}$, for PbPb collisions at 2.76 TeV, again for (from top to
bottom) pions ($\pi^{\pm}$), kaons ($K^{\pm}$), protons ($p$ and
$\bar{p}$), and lambdas ($\Lambda$ and $\bar{\Lambda}$). The four
columns represent four different centrality classes, namely 0-5\%,
20-40\%, 60-80\%, 80-100\%.

Looking at the green (core) and blue (corona) curves,
one observes that the core contribution increases with centrality, but
it also increases with the hadron mass (from top to bottom). Concerning
the $p_{t}$ dependence, one also observes a maximum of the green core
curves around 1-2$\,\mathrm{GeV/c}$, less pronounced compared with
the pp results, but still, at very low $p_{t}$ the core contribution
goes down, so even at very small $p_{t}$ values the corona contributes.
The crossing of the green core and the blue corona curves (core =
corona) occurs between around 2GeV/c (mesons, peripheral) and 5GeV/c
(baryons, central).

The red curve, full over core+corona, represents the
effect of the hadronic cascade in the full case. The pions are
not much affected, but for kaons and even more for protons and lambdas,
rescattering makes the spectra harder. One should keep in mind that
rescattering involves particles from fluid hadronization, but also
corona particles from hard processes. Concerning the baryons, rescattering
reduces (considerably) low $p_{t}$ yields, due to baryon-antibaryon
annihilation.

In the following, I show results of particle production in pp
scattering at 7 TeV and PbPb collisions at 2.76 TeV.
\begin{figure}
\centering{}\includegraphics[bb=150bp 70bp 600bp 770bp,clip,scale=0.9]
{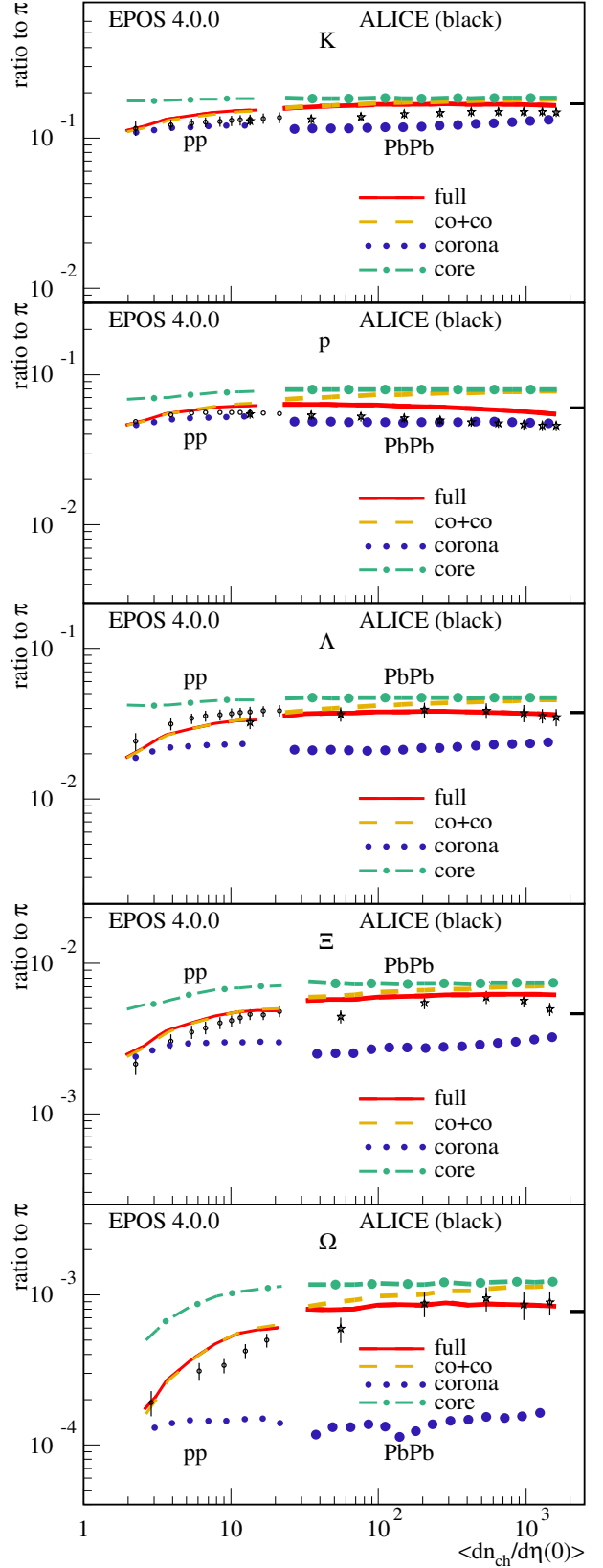}
\caption{Ratio to $\pi$ for (from top to bottom) $K$, $p$, $\Lambda$, $\Xi$,
$\Omega$ for pp at 7 TeV and PbPb at 2.76 TeV as a function of the
multiplicity $dn_{\mathrm{ch}}/d\eta(\eta=0)$. I show the different
contributions: core, corona, core+corona, and full (see text), compared
with ALICE data.\label{ratio5}}
\end{figure}
\begin{figure}
\centering{}\includegraphics[bb=150bp 30bp 600bp 750bp,clip,scale=0.9]
{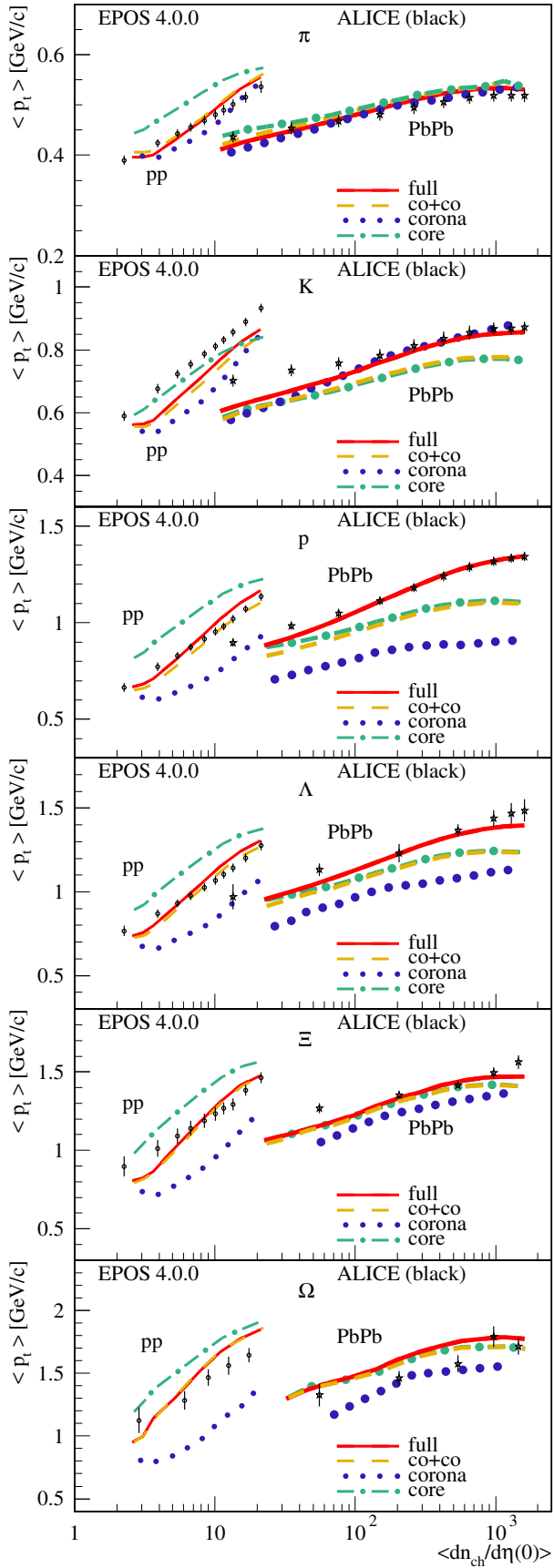}
\caption{Average transverse momentum of (from top to bottom) $\pi$, $K$,
$p$, $\Lambda$, $\Xi$, $\Omega$ for pp at 7 TeV and PbPb at 2.76
TeV as a function of the multiplicity $dn_{\mathrm{ch}}/d\eta(\eta=0)$.
I show the different contributions: core, corona, core+corona, and
full (see text), compared with ALICE data.\label{apt6}}
\end{figure}
In Fig. \ref{ratio5}, I show ratios of hadron yields over pion yields,
at rapidity zero, for proton-proton scattering at 7 TeV and PbPb at
2.76 TeV, as a function of the average multiplicity $\left\langle dn_{\mathrm{ch}}/d\eta(\eta=0)\right\rangle $.
Concerning the EPOS simulations, I show the different contributions:
core (green dashed-dotted lines), corona (blue dotted
lines), core+corona (co+co, yellow dashed lines), and full
(red full lines). Thin lines are used for pp and thick ones for PbPb.
I also show ALICE data \cite{ALICE:2013-PbPb-k-pi-p,ALICE:2013-PbPb-Ks-Lda,ALICE:2013-PbPb-Xi-Oga,ALICE:2015-pp-pi-K-p,ALICE:2016-pp-Ks-Lda-Xi-Oga},
using black symbols, stars for PbPb, and circles for pp. I consider
charged kaons, protons, lambdas, as well as $\Xi$ baryons and $\Omega$
baryons.

One sees almost flat lines for the corona contributions, similar for
pp and PbPb, which is understandable, since corona means particle
production from string fragmentation, which does not depend on the
system. One observes also flat curves for the core at high multiplicity,
which is again expected since the core hadronization is determined
by the freeze-out energy density, which is system independent. However,
when the system gets very small, one gets a reduction of heavy particle
production due to the microcanonical procedure (with its energy and
flavor conservation constraints), whereas a grand canonical treatment
would give a flat curve down to small multiplicities. This effect
increases with particle mass, it is biggest for Omega baryons, where
the reduction is about a factor of 2.

The yellow core+corona curves simply interpolate between the corona
and the core curves, with the core weight increasing continuously
with multiplicity. The increase is biggest for the $\Omega$. Here,
the core curve is far above the corona one, which simply
reflects the fact that $\Omega$ production is much more suppressed
in string decay, compared with statistical (``thermal'') production.
This explains why the core+corona contribution increases by
one order of magnitude from low to high multiplicity, because simply
the relative weight of the core fraction increases from zero to unity.
Both microcanonical decay and core-corona procedure contribute to
the decrease of the ratios toward small multiplicity, but it seems
that the core-corona mechanism is more important.

Finally, there is some effect from hadronic rescattering (difference
between full and co+co), mainly a suppression due to baryon-antibaryon
annihilation at large multiplicities.

Whereas the particle ratios are essentially smooth curves, from pp
to PbPb, the situation changes completely when looking at the average
transverse momentum $\left\langle p_{t}\right\rangle $ versus multiplicity,
as shown in Fig. \ref{apt6}, where I show simulation results for
pp (thin curves) and PbPb (thick curves) for charged $\pi$ mesons,
charged $K$ mesons, (anti)protons, $\Lambda$ (anti)baryons, $\Xi$
(anti)baryons, and $\Omega$ (anti)baryons. I again show the different
contributions: core (green dashed-dotted lines), corona
(blue dotted lines), core+corona (co+co, yellow dashed lines),
and full (red full lines). Simulation are compared with ALICE
data \cite{ALICE:2013-PbPb-k-pi-p,ALICE:2013-PbPb-Ks-Lda,ALICE:2013-PbPb-Xi-Oga,ALICE:2015-pp-pi-K-p,ALICE:2016-pp-Ks-Lda-Xi-Oga},
using black symbols, stars for PbPb, and circles for pp.

Here, one sees (for all curves) a significant discontinuity when going
from pp to PbPb. The corona contributions are not flat (as the
ratios), but they increase with multiplicity, in the case of pp even
more pronounced as for PbPb. This is a ``saturation effect'': the
saturation scale increases with multiplicity, which means that with
increasing multiplicity the events get harder, producing higher $p_{t}$.
The situation is different for PbPb, where an increase in multiplicity
is mainly due to an increase in the number of active nucleons, with
a more modest increase of the saturation scale with multiplicity.
Also, the core curves increase strongly with multiplicity, and
here as well more pronounced in the case of pp, due to the fact that
one gets for high-multiplicity pp high energy densities within a small
volume, leading to strong radial flow. Again, the core+corona contribution
is understood based on the continuous increase of the core fraction
from low to high multiplicity.

It is very useful (and necessary) to consider at the same time the
multiplicity dependence of particle ratios and of mean $p_{t}$ results,
since their behavior is completely different (the former is continuous,
the latter jumps). Despite these even qualitative differences between
the two observables, the physics issues behind these results is the
same, namely saturation, core-corona effects which mix flow (being
very strong) and non-flow, and microcanonical hadronization of the
core.

\begin{figure}[h]
\centering{}\includegraphics[bb=20bp 30bp 600bp 750bp,clip,scale=0.40]
{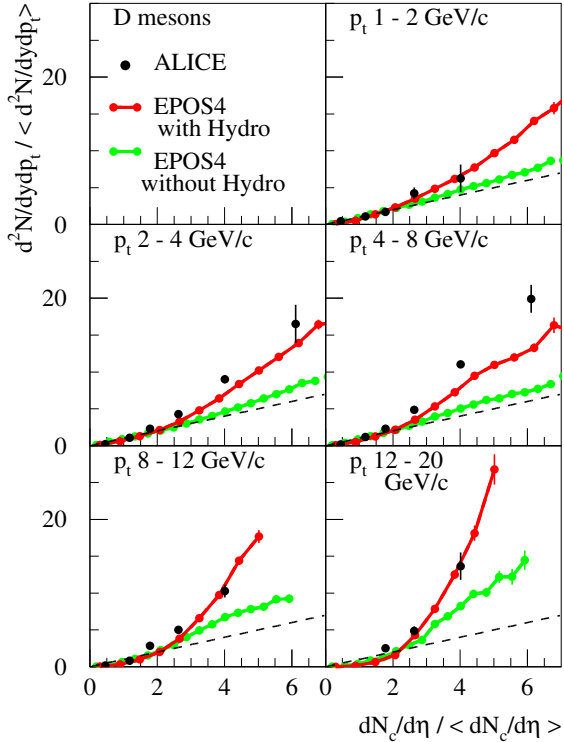}
\caption{Normalized $D$ meson multiplicity as a function of the normalized
charged particle multiplicity for different $p_{t}$ ranges in pp
scattering at 7 TeV. I show EPOS results with and without hydro,
compared with ALICE data. \label{charm-versus-charged}}
\end{figure}

Another very important and useful variable is the multiplicity dependence
of $D$ meson production, where ``D'' stands for the sum of $D^{0}$,
$D^{+}$, and $D^{*+}$. This is much more than just ``another particle'',
since the $D$ meson contains a charm quark, the latter being created
exclusively in the parton ladder and not during fragmentation or in
the plasma. In Fig. \ref{charm-versus-charged},
I plot the normalized $D$ meson multiplicity ($\frac{d^{2}N}{dydp_{t}}/<\!\!\frac{d^{2}N}{dydp_{t}}\!\!>$)
as a function of the normalized charged particle multiplicity ($\frac{dN_{c}}{dy}/<\!\!\frac{dN_{c}}{dy}\!\!>$)
for different $p_{t}$ ranges in pp scattering at 7 TeV, compared
with ALICE data \cite{ALICE:2015-pp-mult-charm}. It is interesting
to see in which way the simulations and the data deviate from the
reference curve, which is the dashed black line representing identical
multiplicity dependence for D mesons and charged particles. Considering
the EPOS results without hydro (green lines), for low $p_{t}$ (1-2
GeV/c) the curve is slightly above the reference, but with increasing
$p_{t}$ the green curves get steeper, which is due to the fact that
with increasing multiplicity the saturation scale increases, and the
events get harder, producing more easily both high-$p_{t}$ and charmed
particles. Considering EPOS with hydro (red curves), the increase
compared with the green curves is much stronger, which is due to the
fact that ``turning on hydro'' will reduce the multiplicity (the
available energy is partly transformed into flow). The red curves
are close to the experimental data, both showing a much stronger increase
compared with the reference curve, with the effect getting bigger with
increasing $p_{t}$. So one may conclude this paragraph: to get these
final results (the strong increase), two phenomena are crucial, namely,
saturation which makes high multiplicity events harder, and the ``hydro
effect'' which reduces multiplicity and ``compresses'' the multiplicity
axis.

\section{Summary}

After recalling briefly the EPOS4 parallel scattering approach, as
well as the core-corona method, I presented new developments concerning
the microcanonical decay of plasma droplets, providing very efficient
methods to decay droplets of all sizes, allowing the study of the transition
towards the grand canonical limit (the method usually employed). I
then discussed in detail new methods to construct effective (droplet)
masses from energy-momentum flow through freeze-out hypersurfaces
of expanding fluids in pp or AA collisions, and I discussed results
of microcanonical decay of these droplets. Finally, I investigated
the multiplicity dependence of multistrange hadron yields in proton-proton
and lead-lead collisions at several TeV, which allows the study of the
transition from very big to very small systems, in particular concerning
collective effects. Here, the ``full
model'' was employed: our core-corona approach, using new microcanonical
hadronization procedures, as well as the new methods allowing to transform
energy-momentum flow through freeze-out surfaces into invariant-mass
elements. It was tried to disentangle effects due to ``canonical suppression''
and ``core-corona effects'', which will both lead to a reduction
of the yields at low multiplicity.

\appendix
\vspace*{1cm}

\noindent \textbf{\huge{}{}APPENDIX}{\huge \par}

\section{The NRPS integral \label{=======Coordinate-transformations-for-NRPS=======}}

The NRPS integral Eq. (\ref{NRPS-integral}) may be written as (writing
$\Phi$ instead of $\Phi_{\mathrm{NRPS}}$) 
\begin{align}
\Phi & =(4\pi)^{n}\int_{m_{1}}^{\infty}dE_{1}\ldots\int_{m_{n}}^{\infty}dE_{n}\prod_{i=1}^{n}p_{i}\,E_{i}\\
 & \times\delta(M-\sum_{i=1}^{n}E_{i})\,W(p_{1},\ldots,p_{n}).
\end{align}
With $t_{i}:=E_{i}-m_{i}$ and $T:=M-\sum_{i=1}^{n}m_{i}$ one obtains
\begin{eqnarray}
\Phi(M,m_{1},\ldots,m_{n}) & = & (4\pi)^{n}\int_{0}^{\infty}dt_{1}\ldots\int_{0}^{\infty}dt_{n}\prod_{i=1}^{n}p_{i}\,\varepsilon_{i}\nonumber \\
 &  & \times\,\delta\big(T-\sum_{i=1}^{n}t_{i}\big)\,W(p_{1},\ldots,p_{n}).\label{26}
\end{eqnarray}
One now introduces ``accumulated'' kinetic energies $s_{i}$ via 
\begin{equation}
s_{i}:=\sum_{j=1}^{i}t_{j},\label{27}
\end{equation}
with the inverse $t_{i}=s_{i}-s_{i-1},\quad s_{0}=0$ and $dt_{i}=ds_{i}$.
One obtains 
\begin{align}
   & \Phi(M,m_{1},\ldots,m_{n}) = (4\pi)^{n}\int_{0}^{\infty}ds_{1}\int_{s_{1}}^{\infty}ds_{2}\ldots\int_{s_{n-1}}^{\infty}ds_{n}\nonumber \\
   & \quad\quad\quad\times\prod_{i=1}^{n}p_{i}\,\varepsilon_{i}\,\delta(T-s_{n})W(p_{1},\ldots,p_{n}).\label{30}
\end{align}
The integration over $s_{n}$ is trivial and may be performed, to
obtain 
\begin{align}
  & \Phi(M,m_{1},\ldots,m_{n})\nonumber \\
  & \quad\quad\quad=(4\pi)^{n}\int_{0}^{\infty}ds_{1}\int_{s_{1}}^{\infty}ds_{2}\ldots\int_{s_{n-3}}^{\infty}ds_{n-2}\int_{s_{n-2}}^{T}ds_{n-1}\nonumber \\
  & \quad\quad\quad\times\prod_{i=1}^{n}p_{i}\,\varepsilon_{i}\,W(p_{1},\ldots,p_{n}).\label{31}
\end{align}
All upper limits may be replaced by $T$. Introducing energy fractions,
$x_{i}:=s_{i}/T,$ one gets 
\begin{eqnarray}
 &  & \Phi(M,m_{1},\ldots,m_{n})=(4\pi)^{n}\,T^{n-1}\label{33}\\
 &  & \times\int_{0\leq x_{1}\leq\ldots\leq x_{n-1}\leq1}dx_{1}\ldots dx_{n-1}\prod_{i=1}^{n}p_{i}\,\varepsilon_{i}\,W(p_{1},\ldots,p_{n}).\nonumber 
\end{eqnarray}
Using the definition 
\begin{equation}
\psi(p_{1},\ldots,p_{n}):=\frac{(4\pi)^{n}\,T^{n-1}}{(n-1)!}\prod_{i=1}^{n}p_{i}\,\varepsilon_{i}\,W(p_{1},\ldots,p_{n}),\label{34}
\end{equation}
One may write 
\begin{eqnarray}
 &  & \Phi(M,m_{1},\ldots,m_{n})\label{35}\\
 &  & =(n\!-\!1)!\!\int_{0\leq x_{1}\leq\ldots\leq x_{n-1}\leq1}dx_{1}\ldots dx_{n-1}\psi(x_{1},\ldots,x_{n-1}),\nonumber 
\end{eqnarray}
where $\psi(x_{1},\ldots,x_{n-1})$ is meant to be $\psi(p_{1},\ldots,p_{n})$
with $p_{i}$ and $\varepsilon_{i}$ expressed in terms of $x_{1},\ldots,x_{n-1}$.
This may be solved via Monte Carlo as 
\begin{equation}
\Phi(M,m_{1},\ldots,m_{n})=\frac{1}{N}\sum_{\beta=1}^{N}\psi(x_{1}^{(\beta)}\ldots x_{n-1}^{(\beta)}),\label{36}
\end{equation}
where the $x_{i}^{(\beta)}$ are ordered random numbers, 
\begin{equation}
0\leq x_{1}^{(\beta)}\leq x_{2}^{(\beta)}\leq\ldots\leq x_{n-1}^{(\beta)}\leq1.\label{37}
\end{equation}
So for each Monte Carlo step, $n-1$ random numbers have to be generated,
ordered according to size, and then used to evaluate $\psi(x_{1}^{(\beta)},\ldots,x_{n-1}^{(\beta)})$.
To avoid ordering, one may introduce the variables 
\begin{equation}
z_{i}:=\frac{x_{i}}{x_{i+1}},\label{38}
\end{equation}
using the definition $x_{n}:=1$. One gets 
\begin{equation}
dx_{i}=dz_{i}\,x_{i+1}=dz_{i}\prod_{j=i+1}^{n-1}z_{j},\label{39}
\end{equation}
the last equation holding for $i<n-1$; so one has 
\begin{eqnarray}
\prod_{i=1}^{n-1}dx_{i} & = & \prod_{i=1}^{n-1}dz_{i}\prod_{i=1}^{n-2}\prod_{j=i+1}^{n-1}z_{j}\nonumber \\
 & = & \prod_{i=1}^{n-1}dz_{i}\prod_{i=1}^{n-1}z_{i}^{i-1}.\label{40}
\end{eqnarray}
From Eq. (\ref{35}), one gets 
\begin{eqnarray}
 &  & \Phi(M,m_{1},\ldots,m_{n})\label{41}\\
 &  & =\int_{0}^{1}dz_{1}\ldots\int_{0}^{1}dz_{n-1}\prod_{i=1}^{n-1}i\,z_{i}^{i-1}\,\psi(z_{1},\ldots,z_{n-1}),\nonumber 
\end{eqnarray}
where obviously $\psi(z_{1},\ldots,z_{n-1})$ is meant to be $\psi(p_{1},\ldots,p_{n})$
with $p_{i}$ expressed in terms of $z_{i}$. One now introduces 
\begin{equation}
r_{i}:=\int_{0}^{z_{i}}i\,\xi^{i-1}\,d\xi=z_{i}^{i}\label{42}
\end{equation}
to obtain 
\begin{equation}
\Phi(M,m_{1},\ldots,m_{n})=\int_{0}^{1}dr_{1}\ldots\int_{0}^{1}dr_{n-1}\,\psi(r_{1},\ldots,r_{n-1}).\label{43}
\end{equation}
The $r_{i}$ are now uncorrelated, no ordering is required. The final
result is $\Phi=\int d\Phi$ with 
\begin{equation}
d\Phi=dr_{1}...dr_{n-1}\frac{(4\pi)^{n}\,T^{n-1}}{(n-1)!}\prod_{i=1}^{n}p_{i}\,E_{i}\,W(p_{1},...,p_{n}),\label{dPhi-NRPS-rvariables}
\end{equation}
with $r_{i}\in[0,1]$, and with $z_{i}=r_{i}^{1/i}$, ~$x_{i}=z_{i}x_{i+1}$,
~$s_{i}=x_{i}T$, ~$t_{i}=s_{i}-s_{i-1}$, ~$E_{i}=t_{i}+m_{i}$,
~$T=M-\sum_{i=1}^{n}m_{i}$.

\section{The LIPS integral \label{=======LIPS-method-for-small-n=======}}

\begin{figure}
\centering{}\includegraphics[scale=0.3]
{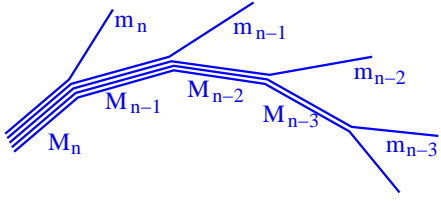}
\caption{Successive two-body decays.\label{fig:Successive-two-body-decays}}
\end{figure}

For given mass $M$, for given $n$, and for given masses $m_{1},...,m_{n}$,
one defines the LIPS phase-space integral $\Phi_{\mathrm{LIPS}}=\int d\Phi_{\mathrm{LIPS}}$
as

\begin{equation}
\Phi_{\mathrm{LIPS}}=\int\delta(M-\Sigma E_{i})\,\delta^{3}(\Sigma\vec{p}_{i})\,\prod_{i=1}^{n}\frac{d^{3}p_{i}}{\,2E_{i}},
\end{equation}
which may be written as \cite{James:1968} 
\begin{equation}
\Phi_{\mathrm{LIPS}}=\int\delta^{4}(P-\Sigma p_{i})\,\prod_{i=1}^{n}\delta(p_{i}^{2}-m_{i}^{2})d^{4}p_{i}\qquad\qquad\qquad
\end{equation}
\[
\quad=\int\left\{ \int\delta^{4}\left(P\!-\!P_{l}\!-\!\sum_{i>l}p_{i}\right)\,\qquad\qquad\qquad\qquad\qquad\qquad\right.
\]
\begin{equation}
\times\prod_{i>l}\delta(p_{i}^{2}-m_{i}^{2})d^{4}p_{i}\,\delta(P_{l}^{2}-M_{l}^{2})d^{4}P_{l}
\end{equation}
\[
\quad\quad\left.\times\int\delta^{4}\left(P_{l}-\sum_{i\leq l}p_{i}\right)\,\prod_{i\leq l}\delta(p_{i}^{2}-m_{i}^{2})d^{4}p_{i}\right\} dM_{l}^{2}
\]
which leads to 
\begin{equation}
\Phi_{\mathrm{LIPS}}=\int\,\Phi_{\mathrm{LIPS}}(M;M_{l},m_{l+1},...,m_{n})\,
\end{equation}
\[
\qquad\qquad\times\Phi_{\mathrm{LIPS}}(M_{l};m_{1},...,m_{l})\,dM_{l}^{2}.
\]
Iterating this equation, one gets (using $M_{n}=M$, $M_{1}=m_{1}$)
\[
\Phi_{\mathrm{LIPS}}=\frac{1}{2m_{1}}\int\,\prod_{i=1}^{n-1}2\,M_{i}\,\Phi_{\mathrm{LIPS}}(M_{i+1};M_{i},m_{i+1})\,\prod_{i=2}^{n-1}dM_{i}\,,
\]
where the two-body LIPS factor may be written as 
\[
\Phi_{\mathrm{LIPS}}(M;m_{a},m_{b})\qquad\qquad\qquad\qquad\qquad\qquad\qquad\qquad
\]
\begin{equation}
=\int\delta^{4}(P-p_{a}-p_{b})\,\delta(p_{b}^{2}-m_{b}^{2})\theta(E_{b})\,\,d^{4}p_{b}\,\frac{d^{3}p_{a}}{\,2E_{a}}
\end{equation}
\begin{equation}
\qquad=\int\delta\left((P-p_{a})^{2}-m_{b}^{2}\right)\theta\left((P-p_{a})_{0}\right)\,\frac{d^{3}p_{a}}{\,2E_{a}}.
\end{equation}
Evaluating this expression in the center-of-mass system, one gets 
\[
P=(M,\vec{0}),\:p_{a}=(E_{a},\vec{p}),\;p_{b}=(E_{b},-\vec{p}),
\]
and so 
\begin{equation}
(P-p_{a})^{2}-m_{b}^{2}=P^{2}-2Pp_{a}+p_{a}^{2}-m_{b}^{2}
\end{equation}
\begin{equation}
=M^{2}-2ME_{a}+m_{a}^{2}-m_{b}^{2},
\end{equation}
which gives 
\begin{align}
 & \Phi_{\mathrm{LIPS}}(M;m_{a},m_{b})\\
 & =\int\delta\left(M^{2}-2ME_{a}+m_{a}^{2}-m_{b}^{2}\right)\theta\left(M-E_{a}\right)\,\frac{p_{a}dE_{a}d\Omega}{\,2}\\
 & =\frac{p_{a}}{4M}\,\int d\Omega.
\end{align}
So one gets 
\begin{equation}
\Phi_{\mathrm{LIPS}}(M;m_{a},m_{b})=\frac{\text{\ensuremath{\pi}}}{M}\,p(M;m_{a},m_{b}),
\end{equation}
where $p$ is the momentum of the two-body decay in the rest frame,
given as 
\begin{equation}
p(M;m_{a},m_{b})=\sqrt{E_{a}^{2}-m_{a}^{2}},
\end{equation}
with (from integrating the mass-shell delta function) 
\begin{equation}
E_{a}=\frac{1}{2M}\left(M^{2}+m_{a}^{2}-m_{b}^{2}\right).
\end{equation}
The n-body phase-space element is then 
\begin{equation}
\Phi_{\mathrm{LIPS}}=\frac{\pi(2\pi)^{n-2}}{M}\:\prod_{i=1}^{n-1}p(M_{i+1};M_{i},m_{i+1})\,\prod_{i=2}^{n-1}dM_{i}\,,
\end{equation}
which amounts to successive two-body decays (F. James, 1968 ), see
Fig. \ref{fig:Successive-two-body-decays}, as 
$M_{n}\to M_{n-1}+m_{n},$ $M_{n-1}\to M_{n-2}+m_{n-1}$ etc. As a
consequence, the integration limits are 
\begin{equation}
M_{i-1}+m_{i}\leq M_{i}.
\end{equation}
Defining variables $x_{i}$ via 
\begin{equation}
M_{i}=\sum_{j=1}^{i}m_{j}+x_{i}\left(M-\sum_{j=1}^{n}m_{j}\right),
\end{equation}
with 
\begin{equation}
0\leq x_{2}\leq...\leq x_{i-1}\leq x_{i}...\leq x_{n-1}\leq1\text{ ,}
\end{equation}
one recovers indeed %
{} $M_{i-1}+m_{i}\leq M_{i}.$ So one gets 
\begin{align}
d\Phi_{\mathrm{LIPS}}&=\frac{\pi(2\pi)^{n-2}}{M}\:\prod_{i=1}^{n-1}p(M_{i+1};M_{i},m_{i+1})  \nonumber \\
&\qquad\,\Big(M-\sum_{j=1}^{n}m_{j}\Big)^{n-2}\,\prod_{i=2}^{n-1}dx_{i},
\end{align}
with ordered $x_{i}$. With $x_{i+1}=x'_{i}$, $x'_{i}=z_{i}x'_{i+1}$,
and $(z_{i})^{i}=r_{i},$ one gets 
\begin{equation}
\prod_{i=2}^{n-1}dx_{i}=\frac{1}{(n-2)!}\prod_{i=1}^{n-2}dr_{i}.
\end{equation}
with independent variables $r_{i}\in[0,1]$, and so 
\begin{align}
&d\Phi_{\mathrm{LIPS}}(M;m_{1},...,m_{n}) \nonumber \\
&\qquad\qquad=\frac{\pi(2\pi)^{n-2}}{M}\:\prod_{i=1}^{n-1}p(M_{i+1};M_{i},m_{i+1})  \nonumber \\
&\qquad\qquad\,\Big(M-\sum_{j=1}^{n}m_{j}\Big)^{n-2}\,\frac{1}{(n-2)!}\prod_{i=1}^{n-2}dr_{i},\label{dPhi-LIPS-rvariables}
\end{align}
with $x_{i+1}=x'_{i}$, $x'_{i}=z_{i}x'_{i+1}$, and $(z_{i})^{i}=r_{i},$
and 
\begin{equation}
M_{i}=\sum_{j=1}^{i}m_{j}+x_{i}\left(M-\sum_{j=1}^{n}m_{j}\right),
\end{equation}
The successive two-body decays are done in the center-of-mass of the
decaying object. In the Monte Carlo procedure, the particles have
therefore to be boosted into the frame of the object they are originating
from, in an iterative fashion, starting with the last decay.

\section{Sampling via Markov chains \label{=======Sampling-via-Markov=======}}

One wants to generate randomly hadron configurations $\{H,P\}$, with
$H=\{h_{1},\ldots,h_{n}\}$ specifying the hadron species and $P=\{\vec{p}_{1},...,\vec{p}_{n}\}$
their momenta, according to the weight given in Eq. (\ref{microcanonical-law-1}),
which can be written as
[see Eqs. (\ref{microcanonical-law-2},\ref{microcanonical-law-3},\ref{dPhi-NRPS-rvariables-1},\ref{dPhi-LIPS-rvariables-0},\ref{dPhi-LIPS-rvariables-1})] 
\begin{equation}
dQ\:W(H,Q),
\end{equation}
with $Q=\{q_{1},...q_{k}\}$ representing $k$ independent variables
with $q_{i}\in[0,1]$, which characterize $P$. Depending on the method
one uses, one has $k=3n-1$ (improved Cerulus-Hagedorn method) or $k=3n-4$
(LIPS method). The expression for $W$ can be found by comparing
with Eqs. (\ref{microcanonical-law-1},\ref{microcanonical-law-2},\ref{microcanonical-law-3},\ref{dPhi-NRPS-rvariables-1},\ref{dPhi-LIPS-rvariables-0},\ref{dPhi-LIPS-rvariables-1}),
keeping in mind that there one only writes the non-trivial variables
$r_{i}$, the variables $u$ and $w$ are simply uniform random variables
in $[0,1]$). So to have complete expressions, one needs to add terms
like $\prod f(u_{j})\prod f(w_{j})$ with $f$ being the function
defined as $f=1$ in $[0,1]$ and zero elsewhere.

There is one technical point that needs to be discussed. Rather than
considering configurations (like the two hadron configuration ``one
pion and one kaon''), one considers ordered sequences of hadrons (like
``first hadron = pion, second hadron = kaon''). In addition, one
allows in the sequence $\{h_{1},...h_{L}\}$ also ``holes'', $h_{i}=0$,
and one uses a fixed length $L$. The number of hadrons is then the
number of nonzero places in the sequence. Consequently, one adds a
factor 
\begin{equation}
{\frac{1}{n!}}\,\left\{ \prod_{\alpha\in\mathcal{S}}\,n_{\alpha}!\right\} \,\frac{n!\,(L-n)!}{L!}
\end{equation}
to our probability distribution $W$.

To simplify the notation, I will use in the following 
\begin{equation}
K=\{H,Q\}
\end{equation}
for a configuration, and $\sum_{K}...$ means by definition $\sum_{H}\int...dQ$.
Finally, one defines the normalized distribution $\Omega(K)=W(K)/\sum_{k}W(K)$,
so one has 
\begin{equation}
\sum_{K}\Omega(K)=1.
\end{equation}s

To generate $K$ according to $\Omega(K)$, one considers a sequence
of random configurations 
\begin{equation}
K_{0}\,,\:K_{1}\,,\,\,K_{2},...
\end{equation}
with some $\Omega_{t}$ being the law for $K_{t}$. Let $A$ and $B$
be possible configurations. One defines an operator $T$ as $T\Omega_{t}(B)=\sum_{A}\Omega_{t}(A)\,p(A\to B)$,
with $p(A\to B)$ being called transition probability (or matrix).
Normalization: $\sum_{B}p(A\to B)=1.$ Considering homogeneous Markov
chains, the law for $\Omega_{t+1}$ is by definition given as $T\Omega_{t}$.
A law is called stationary if $T\Omega=\Omega.$ According to the
fundamental theorem of Markov chains, one knows, if a stationary law
$T\Omega=\Omega$ exists, then $T^{k}\Omega_{0}$ converges in a unique
fashion towards $\Omega$, for any $\Omega_{0}$. A sufficient condition
for $T\Omega=\Omega$ is detailed balance, defined as 
\begin{equation}
\Omega(A)\,p(A\to B)=\Omega(B)\,p(B\to A)\,,
\end{equation}
and ergodicity, which means that for any $A,B$ there must exist some
$r$ with the probability to get from $A$ to $B$ in $r$ steps being
nonzero. Henceforth, one uses the abbreviations 
\begin{equation}
\Omega_{A}:=\Omega(A);\,\,\,\,\,\,p_{AB}:=p(A\rightarrow B).
\end{equation}
Following Metropolis-Hastings \cite{Metropolis49,Metropolis53,Hastings70},
one makes the ansatz 
\begin{equation}
p_{AB}=w_{AB}\,u_{AB}\,\,,\label{pab-as-product}
\end{equation}
with a so-called proposal matrix $w$ and an acceptance matrix $u$.
Detailed balance now reads 
\begin{equation}
\frac{u_{AB}}{u_{BA}}=\frac{\Omega_{B}}{\Omega_{A}}\frac{w_{BA}}{w_{AB}}\,\,,\label{detailes-balance}
\end{equation}
which is obviously fulfilled for 
\begin{equation}
u_{AB}=F\left(\frac{\Omega_{B}}{\Omega_{A}}\frac{w_{BA}}{w_{AB}}\right)\,\,,\label{u-equal-f}
\end{equation}
with some function $F$ fulfilling $F(z)\,/\,F(z^{-1})=z.$ One takes
\begin{equation}
F(z)=\min(z,1)\,\,.\label{function-F}
\end{equation}
The power of the method is due to the fact that an arbitrary $w$
may be chosen, in connection with $u$ being given by Eq. (\ref{u-equal-f}).
So the task is twofold: one needs an efficient algorithm to calculate,
for given $K$, the weight $\Omega(K)$, and one needs to find an
appropriate proposal matrix $w$ which leads to \textbf{fast} convergence
(small $I_{\textrm{eq}}$), which is not trivial. The first task can
be solved, as shown in section \ref{-------Microcanonical-hadronization-of-plasma-droplets-------}.
\\

In the following, I discuss how to construct an appropriate proposal
matrix $w_{AB}$. Let $n_{\mathrm{specs}}$ be the number of hadron
species considered (the latter being the list of hadrons according
to the particle data group PDG, without charm, bottom, top). The index
$n_{\mathrm{specs}}+1$ is used for the ``hole'' (missing particle),
which is formally considered as particle species. One defines weights
for the hadron species $h$ as 
\begin{equation}
e(h)=\left\{ \begin{array}{cc}
f_{h}/\{2\sum f_{h}\} & \mathrm{for}\,\mathrm{hadrons\,}h\\
1/2 & \mathrm{for}\,\mathrm{the\,hole}
\end{array}\right.,
\end{equation}
with $f_{h}$ being the grand canonical yields, see Eq. (\ref{GraCan distribution}).
One defines the proposal matrix $w_{AB}$ in terms of an algorithm which
constructs $B$ starting form some configuration $A$. As discussed
above, a configuration has the structure $\{\{h_{1},\ldots,h_{n}\},\{q_{1},...q_{k}\}\}$,
where $h_{i}$ refers to the particle species of hadron $i$, and
the $q_{i}$ are independent variables (defined in $[0,1]$) which
define the momenta of the hadrons (the $q_{i}$ are not associated
to individual hadrons). Starting from $A=$ $\{\{h_{1},\ldots,h_{n}\},\{q_{1},...q_{k}\}\}$: 
\begin{description}
\item [{A1}] chose randomly two integer numbers $i$ and $j\ne i$ between
1 and $n$, and replace $h_{i}$ and $h_{j}$ in $A$ by $h'_{i}$
and $h'_{j}$, the latter having been generated with weights $e(h'_{i})$
and $e(h'_{j})$. Let me name the new configuration $A'$ 
\item [{A2}] chose randomly two more integer numbers $k$ and $l\ne k$
between 1 and $n$, different from $i$ and $j$ 
\item [{A3}] establish a list of pairs of particle species $h'_{a}$ and
$h''_{a}$, $a=1,2,3...N$, considering those which conserve flavor,
if $h_{i}$, $h_{j}$, $h_{k}$, and $h_{l}$ are replaced by $h'_{i}$,
$h'_{j}$, $h'_{a}$, and $h''_{a}$. Associate a weight $c_{a}=e(h'_{a})$$e(h''_{a})$
to pair $a$. Choose a pair index $a$ with weight $c_{a}/\sum_{a=1}^{N}c_{a}$. 
\item [{A4}] replace $h_{k}$ and $h_{l}$ in $A'$ by $h'_{a}$ and $h''_{a}$,
with the $a$ chosen in the previous step, which gives the new configuration
$B$. 
\item [{A5}] In case of change 'hadron to hole' or vice versa, replace
one of the $q_{i}$ by $q'_{i}$, chosen randomly in $[0,1]$ (note
that $q_{i}$ is not associated directly to hadron $i$). 
\end{description}
The asymmetry $w_{AB}/w_{BA}$ is given as 
\begin{equation}
\frac{w_{AB}}{w_{BA}}=\frac{e(h'_{i})e(h'_{j})e(h'_{a})e(h''_{a})}{e(h_{i})e(h{}_{j})e(h{}_{k})e(h_{l})}.
\end{equation}
Finally, one computes $\Omega_{B}$, and with $\Omega_{A}$ already
known (computed in the step before), this allows one to compute 
\begin{equation}
u_{AB}=F\left(\frac{\Omega_{B}}{\Omega_{A}}\frac{w_{BA}}{w_{AB}}\right)\,\,.\label{u-equal-F-2}
\end{equation}
The proposal will be accepted with this weight, otherwise one continues
with configuration $A$.

With this choice (algorithm A1-5) of a proposal matrix, fast convergence
can be achieved. Concerning in A3 the ``list of pairs that conserve
flavor'', one predefines for all possible flavor numbers $N_{u}$,
$N_{d},$$N_{s}$ (each one between $-6$ and 6) tables $\mathrm{idpairs_{i}}(N_{u},N_{d},$$N_{s};K$)
and $\mathrm{wgtpairs}(N_{u},N_{d},$$N_{s};K$) with the hadron ids
($i=1$ and $i=2$) and the weights of the $K^{th}$ pair, with $K=1,2,...,$
which allows a very fast generation of pairs (even with a ``complete''
set of hadron species, being close to 400).

$\qquad\qquad\qquad\qquad$%

\section{Hypersurfaces and Milne coordinates\label{=======Hyper-surfaces-and-Milne=======}}

\noindent One considers a hadronization hypersurface parametrized in
Minkowski space as $x^{\mu}=x^{\mu}(\tau,\varphi,\eta)$, with 
\begin{equation}
x^{0}=\tau\cosh\eta,\;x^{1}=r\cos\varphi,\;x^{2}=r\sin\varphi,\;x^{3}=\tau\sinh\eta,
\end{equation}
where $r=r(\tau,\varphi,\eta)$ is some function of the three parameters
$\tau,\:\varphi,\:\eta$. One allows for several sheets in the sense
that for given $\tau,\varphi,\eta$, there are several values of $r$
satisfying the freeze-out condition. For each sheet, for given values
of the three parameters $\tau,\:\varphi,\:\eta$, the hypersurface
element is 
\begin{equation}
d\Sigma_{\mu}=\varepsilon_{\mu\nu\kappa\lambda}\frac{\partial x^{\nu}}{\partial\tau}\frac{\partial x^{\kappa}}{\partial\varphi}\frac{\partial x^{\lambda}}{\partial\eta}\,d\tau\,d\varphi\,d\eta,
\end{equation}
with $\varepsilon^{0123}=-\varepsilon_{0123}=1$. Computing the partial
derivatives $\partial x^{\mu}/d\alpha$, with $\alpha=\tau,$ $\varphi$,
$\eta$, one gets, still in Minkowski space, 
\begin{align}
 & d\Sigma_{0}=\left\{ -r\frac{\partial r}{\partial\tau}\tau\cosh\eta+r\frac{\partial r}{\partial\eta}\sinh\eta\right\} d\tau\,d\varphi\,d\eta,\\
 & d\Sigma_{1}=\left\{ \quad\quad\frac{\partial r}{\partial\varphi}\tau\sin\varphi+r\,\tau\cos\varphi\;\right\} d\tau\,d\varphi\,d\eta,\\
 & d\Sigma_{2}=\left\{ \quad\,-\frac{\partial r}{\partial\varphi}\tau\cos\varphi+r\,\tau\sin\varphi\;\right\} d\tau\,d\varphi\,d\eta,\\
 & d\Sigma_{3}=\left\{ \quad r\frac{\partial r}{\partial\tau}\tau\sinh\eta-r\frac{\partial r}{\partial\eta}\cosh\eta\right\} d\tau\,d\varphi\,d\eta.
\end{align}

It is useful to choose Milne coordinates $\{x'^{\mu}\}=$ $\{\tau,x,y,\eta\}$,
which are expressed in terms of Minkowski coordinates $\{x^{\mu}\}=$
$\{t,x,y,z\}$ as $\tau=\sqrt{t^{2}-z^{2}}$ and $\eta=\frac{1}{2}\mathrm{ln(}(t+z)/(t-z))$,
or the other way round $t=\tau\cosh\eta$ and $z=\tau\sinh\eta$.
The definitions of $x$ and $y$ coordinates are unchanged. One chooses
$(+,-,-,-)$ signature of $g_{\mu\nu}$ in Minkowski space. Let $e_{\mu}$
be the natural basis with respect to Minkowski coordinates.%
The natural basis vectors $e'_{\lambda}$ with respect to Milne coordinates
are 
\begin{equation}
e'_{\lambda}=e_{\mu}\,M_{\lambda}^{\mu}\;,
\end{equation}
with $M_{\lambda}^{\mu}=\partial x^{\mu}/\partial x'^{\lambda}$.
One gets 
\begin{equation}
M=\left(\begin{array}{c}
\cosh\eta\\
0\\
0\\
\sinh\eta
\end{array}\quad\begin{array}{c}
0\\
1\\
0\\
0
\end{array}\quad\begin{array}{c}
0\\
0\\
1\\
0
\end{array}\begin{array}{c}
\tau\,\sinh\eta\\
0\\
0\\
\tau\,\cosh\eta
\end{array}\right).\;
\end{equation}
The corresponding metric is $\left\{ g_{\mu\nu}\right\} =\{e'_{\mu}\cdot e'_{\nu}\}$$=\mathrm{diag}(1,-1,-1,-\tau^{2})$.
The transformation for contravariant coordinates is 
\begin{equation}
x'^{\,\lambda}=(M^{-1})_{\mu}^{\lambda}\,x^{\mu}\;,
\end{equation}
with 
\begin{equation}
M^{-1}=\left(\begin{array}{c}
\cosh\eta\\
0\\
0\\
-\frac{1}{\tau}\sinh\eta
\end{array}\quad\begin{array}{c}
0\\
1\\
0\\
0
\end{array}\quad\begin{array}{c}
0\\
0\\
1\\
0
\end{array}\begin{array}{c}
-\sinh\eta\\
0\\
0\\
\frac{1}{\tau}\cosh\eta
\end{array}\right).
\end{equation}

\noindent The inverse transformation (Milne to Minkowski) is 
\begin{equation}
x^{\lambda}=M_{\mu}^{\lambda}\,x'^{\mu}\;.
\end{equation}
For covariant components (Minkowski to Milne) one has 
\begin{equation}
A'_{\lambda}=A_{\mu}\,M_{\lambda}^{\mu}\,,
\end{equation}
so for $d\Sigma_{\mu}$ one gets in Milne coordinates %
\begin{eqnarray}
d\Sigma'_{\tau} & = & \left\{ \,\,\,-r\,\frac{\partial r}{\partial\tau}\,\tau\,\,\,\right\} d\tau\,d\varphi\,d\eta,\\
d\Sigma'_{x} & = & \left\{ \frac{\partial r}{\partial\varphi}\tau\sin\varphi+r\,\tau\cos\varphi\right\} d\tau\,d\varphi\,d\eta,\\
d\Sigma'_{y} & = & \left\{ -\frac{\partial r}{\partial\varphi}\tau\cos\varphi+r\,\tau\sin\varphi\right\} d\tau\,d\varphi\,d\eta,\\
d\Sigma'_{\eta} & = & \left\{ \,\,\,-r\,\frac{\partial r}{\partial\eta}\tau\,\,\,\,\right\} \,d\tau\,d\varphi\,d\eta.
\end{eqnarray}

\noindent Considering a longitudinal velocity vector $u$, given as
$\left(\cosh y,0,0,\sinh y\right)$ in Minkowski space, one gets in
Milne space $u'^{\,\lambda}=(M^{-1})_{\mu}^{\lambda}\,u^{\mu}$, i.e.
$\left(\cosh(y-\eta),0,0,\tau^{-1}\sinh(y-\eta)\right)$. It is useful
here and actually for any vector to define $\tilde{A}'^{\mu}=h_{\alpha}^{\mu}A'^{\alpha}$
and $\tilde{A}'_{\mu}=A'_{\alpha}k_{\mu}^{\alpha}$ with 
\begin{align}
h & =\mathrm{diag}(1,1,1,\tau)\,\,\,,\\
k & =\mathrm{diag}(1,1,1,\tau^{-1})\:.
\end{align}
For scalar products, one has $\tilde{A}'_{\mu}\tilde{B}'^{\mu}=A'_{\mu}B'^{\mu}.$
One uses the same ``tilde'' definition for tensors, for example~:
$\tilde{T}'^{\mu\nu}=h_{\alpha}^{\mu}h_{\beta}^{\nu}T'^{\alpha\beta}$.
For the velocity vector, one gets 
\begin{equation}
\tilde{u'}=\left(\cosh(y-\eta),0,0,\sinh(y-\eta)\right),
\end{equation}
which is identical to the Minkowski expression, in a frame which moves
with rapidity $\eta$. Furthermore, one gets for the hypersurface element

\noindent 
\begin{eqnarray}
d\tilde{\Sigma}'_{0} & = & \left\{ \,\,\,-r\,\frac{\partial r}{\partial\tau}\,\tau\,\,\,\right\} d\tau\,d\varphi\,d\eta,\\
d\tilde{\Sigma}'_{1} & = & \left\{ \frac{\partial r}{\partial\varphi}\tau\sin\varphi+r\,\tau\cos\varphi\right\} d\tau\,d\varphi\,d\eta,\\
d\tilde{\Sigma}'_{2} & = & \left\{ -\frac{\partial r}{\partial\varphi}\tau\cos\varphi+r\,\tau\sin\varphi\right\} d\tau\,d\varphi\,d\eta,\\
d\tilde{\Sigma}'_{3} & = & \left\{ \,\,\,-r\,\frac{\partial r}{\partial\eta}\,\,\,\,\right\} \,d\tau\,d\varphi\,d\eta.
\end{eqnarray}
which is identical to the Minkowski expression, in a frame which moves
with rapidity $\eta$. %
One also defines ``tilde'' quantities for the transformation matrices,
as $(\widetilde{M^{-1}})_{\beta}^{\alpha}=h_{\mu}^{\alpha}(M^{-1})_{\beta}^{\mu}$
and $\widetilde{M}{}_{\beta}^{\alpha}=M{}_{\mu}^{\alpha}k{}_{\beta}^{\mu}$,
which gives 
\begin{equation}
\tilde{M}=\left(\begin{array}{cccc}
\cosh\eta &  &  & \sinh\eta\\
 & \,\,\,1\,\,\,\\
 &  & \,\,\,1\,\,\,\\
\sinh\eta &  &  & \cosh\eta
\end{array}\right),
\end{equation}
\begin{equation}
\widetilde{M^{-1}}=\left(\begin{array}{cccc}
\cosh\eta &  &  & -\sinh\eta\\
 & \,\,\,1\,\,\,\\
 &  & \,\,\,1\,\,\,\\
-\sinh\eta &  &  & \cosh\eta
\end{array}\right).
\end{equation}
For contravariant vector components $A^{\mu}$ and $A'^{\mu}$, covariant
vector components $B_{\mu}$ and $B'_{\mu}$, and for tensors $T^{\mu\nu}$
and $T'^{\mu\nu}$of rank 2 (with prime ``~'~'' refers to Milne,
without to Minkowski), one has%
\begin{align}
A^{\lambda}
&=M_{\mu}^{\lambda}\,A'^{\mu}=M_{\alpha}^{\lambda}\,\delta_{\mu}^{\alpha}\,A'^{\mu}=M_{\alpha}^{\lambda}\,k_{\nu}^{\alpha}\,h_{\mu}^{\nu}\,A'^{\mu} \nonumber \\
&=\widetilde{M}{}_{\nu}^{\lambda}\,\tilde{A}'^{\nu},  \\
B_{\mu} & =B'_{\lambda}(M^{-1})_{\mu}^{\lambda}=B'_{\lambda}\delta_{\alpha}^{\lambda}(M^{-1})_{\mu}^{\alpha}=B'_{\lambda}\,k_{\nu}^{\lambda}\,h_{\alpha}^{\nu}\,(M^{-1})_{\mu}^{\alpha}\nonumber \\
 & =\tilde{B}'_{\nu}\,(\widetilde{M^{-1}})_{\mu}^{\nu}\:,  \\
T^{\mu\nu}& =M_{\alpha}^{\mu}M_{\beta}^{\nu}T'^{\alpha\beta}=\widetilde{M}{}_{\kappa}^{\mu}h_{\alpha}^{\kappa}\,\widetilde{M}{}_{\lambda}^{\nu}h_{\beta}^{\lambda}\,k_{\omega}^{\alpha}k_{\rho}^{\beta}\tilde{T}'^{\omega\rho} \nonumber \\
&=\widetilde{M}{}_{\kappa}^{\mu}\widetilde{M}{}_{\lambda}^{\nu}\tilde{T}'^{\kappa\lambda},
\end{align}
so vector and tensor components in Minkowski space can be entirely
expressed in terms of (Milne) ``tilde'' quantities, as 
\begin{align}
A^{\lambda} & =\widetilde{M}{}_{\nu}^{\lambda}\,\tilde{A}'^{\nu},\\
B_{\mu} & =\tilde{B}'_{\nu}\,(\widetilde{M^{-1}})_{\mu}^{\nu}\:,\\
T^{\mu\nu} & =\widetilde{M}{}_{\kappa}^{\mu}\widetilde{M}{}_{\lambda}^{\nu}\tilde{T}'^{\kappa\lambda}\:.
\end{align}
In particular, one has 
\begin{align}
B_{\mu}A^{\mu}&=\tilde{B}'_{\nu}\tilde{A}'^{\nu},  \\
T^{\mu\nu}B_{\nu}&=\widetilde{M}{}_{\kappa}^{\mu}\,\tilde{T}'^{\kappa\lambda}\:\tilde{B}'_{\lambda}\:.
\end{align}
This means: 
\begin{itemize}
\item One considers vector and tensor components in Milne coordinates 
\item One performs trivial transformations to get ``tilde'' quantities,
essentially removing factors of $\tau$ or $1/\tau$. The hydro calculations
are actually done based on these tilde quantities, no need to explicitly
do any transformation. 
\item One then gets the vector components in Minkowski space (lab frame) by
simply employing the transformation matrices $\tilde{M}$ (contravariant)
or $\widetilde{M^{-1}}$ (covariant components). Scalar products of
vectors in Minkowski space are equal to the corresponding product
of ``tilde'' quantities. 
\end{itemize}
This procedure is understandable, since ``tilde transformations''
of vector and tensor components in Milne coordinates correspond exactly
to Minkowski components, but in a frame boosted with $y=\eta$. The
transformation $\tilde{M}$ (or $\widetilde{M^{-1}}$) is nothing
but the Lorentz boost back to the laboratory frame.

I apply this procedure for computing the energy-momentum flow
$T^{\mu\nu}d\Sigma_{\nu}$ as well as charge flows $J_{A}^{\nu}d\Sigma_{\nu}$
through hyper-surface elements.%

\end{document}